\documentclass[prd,aps,twocolumn,a4paper,showkeys,nofootinbib,showpacs]{revtex4-2}

\usepackage[colorlinks=true,linkcolor=blue,linktocpage=true,anchorcolor=black,citecolor=blue,filecolor=black,menucolor=black,urlcolor=blue]{hyperref}

\usepackage{graphicx,psfrag}
\usepackage{mathrsfs}
\usepackage{amsmath,amsfonts,amssymb}
\usepackage{comment}
\usepackage{ulem}
\usepackage[top=2cm,bottom=2cm,left=1.4cm,right=1.4cm]{geometry}

\usepackage[T1]{fontenc}

\newcommand{\be}{\begin{equation}}
\newcommand{\ee}{\end{equation}}
\newcommand{\bea}{\begin{eqnarray}}
\newcommand{\eea}{\end{eqnarray}}
\newcommand{\bel}{\begin{align}}
\newcommand{\eel}{\end{align}}

\def\Msun{M_{\odot}}
\def\GMc2{G M_{\odot} c^{-2}}

\def\F{{\cal F}}
\def\lm{{\ell m}}

\def\lm{{\ell m}}

\def\de{\partial}
\def\lm{{\ell m}}

\def\F{{\cal F}}

\def\Msun{M_\odot}

\def\TEOB{\texttt{TEOBResumS}}
\def\TEOBResumS{\texttt{TEOBResumS}}

\def\nrsurqeight{{\texttt{NRHybSur3dq8}}}
\def\nrsurqfifteen{{\texttt{NRHybSur2dq15}}}
\def\TEOBd{{\texttt{TEOBResumS-Dal\'i}}}
\def\TEOBg{{\texttt{TEOBResumS-GIOTTO}}}
\def\daliAN{{\tt{Dal\'i$_{\tt 4PN-analytic}$}}}
\def\daliNR{{\tt{Dal\'i$_{\tt 4PN-NRtuned}$}}}

\DeclareSymbolFontAlphabet{\mathrsfs}{rsfs}
\DeclareMathAlphabet{\mathcal}{OMS}{cmsy}{m}{n}
\DeclareSymbolFontAlphabet{\mathrsfs}{rsfs}
\DeclareMathAlphabet\mathbfcal{OMS}{cmsy}{b}{n}

\usepackage{color}
\definecolor{cyan}{rgb}{0,0.9,0.9}
\definecolor{orange}{rgb}{0.9,0.5,0}
\definecolor{magenta}{rgb}{1,0,1}
\definecolor{purple}{rgb}{0.8,0.4,0.8}
\definecolor{gray}{rgb}{0.8242,0.8242,0.8242}
\definecolor{dodgerblue}{rgb}{0.12, 0.56, 1.0}

\begin{document}
        
\title{Effective-one-body waveform model for non-circularized, planar, coalescing black hole binaries:
the importance of radiation reaction}
\author{Alessandro \surname{Nagar}${}^{1,2}$}
\author{Rossella \surname{Gamba}${}^{3,4}$}
\author{Piero \surname{Rettegno}${}^{1}$}
\author{Veronica \surname{Fantini}${}^{2}$}
\author{Sebastiano \surname{Bernuzzi}${}^{5}$}

\affiliation{${}^1$INFN Sezione di Torino, Via P. Giuria 1, 10125 Torino, Italy}
\affiliation{${}^2$Institut des Hautes Etudes Scientifiques, 91440 Bures-sur-Yvette, France}
\affiliation{${}^3$Institute for Gravitation \& the Cosmos, The Pennsylvania State University, University Park PA 16802, USA}
\affiliation{${}^4$Department of Physics, University of California, Berkeley, CA 94720, USA}
\affiliation{${}^5$Theoretisch-Physikalisches Institut, Friedrich-Schiller-Universit{\"a}t Jena, 07743, Jena, Germany}  

\begin{abstract}
We present an updated version of the {\tt TEOBResumS-Dal\'i} effective one body (EOB) waveform model for 
spin-aligned binaries on noncircularized orbits. Recently computed 4PN (nonspinning) terms are incorporated
in the waveform and radiation reaction. The model is informed by a restricted sample ($\sim60$) of spin-aligned, 
quasicircular, Numerical Relativity (NR) simulations. In the quasicircular limit, the model displays EOB/NR maximal 
unfaithfulness ${\bar{\cal F}}^{\rm max}_{\rm EOBNR}\lesssim 10^{-2}$ (with median~ $1.06\times 10^{-3}$)
(with Advanced LIGO noise and in the total mass range $10-200M_\odot$) for the dominant 
$\ell=m=2$ mode all over the 534 spin-aligned configurations available through the Simulating 
eXtreme Spacetime (SXS) catalog of NR waveforms. Similar figures are also obtained with the 28 public 
eccentric SXS simulations as well as good compatibility between EOB and NR scattering angles.
The quasi-circular limit of  {\tt TEOBResumS-Dal\'i} is  highly consistent with the {\tt TEOBResumS-GIOTTO} 
quasi-circular model. We then systematically
explore the importance of NR-tuning {\it also} the radiation reaction of the system. When this is done,
the median of the distribution of quasi-circular ${\bar{\cal F}}^{\rm max}_{\rm EOBNR}$ is lowered 
to $3.92\times 10^{-4}$, though balanced by a tail up to $\sim 0.1$ for large, positive spins.
The same is true for the eccentric-inspiral datasets. We conclude that an improvement of
the analytical description of the spin-dependent flux (and its interplay with the conservative part) 
is likely to be the cornerstone to lower the EOB/NR unfaithfulness below the $10^{-4}$ level 
all over the parameter space, thus grazing the current NR uncertainties as well as the expected 
needs for next generation of GW detector like Einstein Telescope.
\end{abstract}

\maketitle

\section{Introduction}
Prompted by the desire of obtaining models able to include a large class of physical effects, 
the last few years have seen an increasing interest from the gravitational waves (GW) community in the 
construction of accurate waveform models incorporating orbital eccentricity
and in general configurations that go beyond the standard quasi-circular case. These efforts have been particularly vibrant
within the Effective-One-Body (EOB) framework, with many studies~\cite{Hinderer:2017jcs,Chiaramello:2020ehz,Ramos-Buades:2021adz}
proposing different techniques to model non-circularized binaries.
In particular, the \TEOBd{} model~\cite{Chiaramello:2020ehz} immediately proved to be
sufficiently mature to pioneer several parameter estimation studies involving both bound configurations 
(i.e. eccentric inspirals)~\cite{Bonino:2022hkj} and unbound ones (i.e. scattering or dynamical capture)~\cite{Gamba:2021gap}.
This model is built upon the crucial understanding that the factorized and resummed EOB quasi-circular waveform 
and radiation reaction~\cite{Damour:2008gu} can be generalized to the case of eccentric binaries by simply 
considering {\it generic} Newtonian prefactors in the waveform and fluxes~\cite{Chiaramello:2020ehz,Nagar:2021gss}. 
Although this procedure neglects some (high-order) physical effect, it proved sufficiently accurate in several
context. The idea, technically complemented by the analytical implementation of (high-order) time derivatives via an iterative 
procedure~\cite{Chiaramello:2020ehz,Damour:2012ky}, was thoroughly tested versus a large amount of numerical data 
both in the comparable mass~\cite{Nagar:2020xsk,Gamba:2021gap,Nagar:2021gss,Nagar:2021xnh,Hopper:2022rwo,Andrade:2023trh,Nagar:2023zxh} 
and in the large mass ratio limit~\cite{Albanesi:2021rby,Albanesi:2022ywx}, notably also exploring the effect of 
higher-order PN terms in radiation reaction and waveform~\cite{Placidi:2021rkh,Albanesi:2022ywx,Albanesi:2022xge}.
Among the many findings of this lineage of work, Refs.~\cite{Chiaramello:2020ehz,Albanesi:2022ywx} 
clearly proved that the Newton-factorized azimuthal part of the radiation reaction is more accurate 
than the 2PN-accurate one proposed in Ref.~\cite{Bini:2012ji} (see Ref.~\cite{Khalil:2021txt} for the 3PN calculation).
We note that the approach of Ref.~\cite{Chiaramello:2020ehz} and subsequent works was not adopted 
in a different lineage of eccentric EOB-based models, 
dubbed {\tt SEOBNRv4EHM}~\cite{Ramos-Buades:2021adz,Khalil:2021txt,Ramos-Buades:2023yhy}.
In this respect, while \TEOBd{} was proven to be quantitatively accurate also for dynamical capture 
configurations as well as scattering ones~\cite{Gamba:2021gap,Hopper:2022rwo,Andrade:2023trh}, the corresponding 
studies involving {\tt SEOBNRv4EHM} in this regime were at most qualitative~\cite{Ramos-Buades:2021adz}.
The Achilles' heel of \TEOBd{} was however hidden in its quasi-circular limit, where
the model was found to perform not as well as the quasi-circular {\tt TEOBResumS-GIOTTO} version, 
especially for large, positive spins~\cite{Nagar:2021gss,Bonino:2022hkj}. This problem, 
related to the strong-field behavior of the radial part of the radiation reaction, $\F_r$ was solved, 
in the nonspinning case, in Ref.~\cite{Nagar:2023zxh} adopting a different analytical expression for
it (see discussion in Sec.IV of Ref.~\cite{Nagar:2023zxh} and in particular Fig.~12 therein). 
Note in this respect that Ref.~\cite{Nagar:2023zxh} did not consider, on purpose, the eccentric 
spin case, that deserved more dedicated understanding and work.

Here we build upon the knowledge acquired in Ref.~\cite{Nagar:2023zxh} and present an improved
version of the \TEOBd{} model in its avatar introduced in Ref.~\cite{Nagar:2021xnh} 
(that also deals with spin-aligned binaries). The quasi-circular limit of this new version yields 
an excellent consistency with {\tt TEOBResumS-GIOTTO} as well as with the Simulating eXtreme Spacetimes 
(SXS)~\cite{Boyle:2019kee} quasi-circular Numerical Relativity (NR) datasets.
The model incorporates some new analytical information, namely the 4PN term in the 
quadrupolar waveform (and flux) recently computed in Refs.~\cite{Blanchet:2023bwj,Blanchet:2023sbv,Blanchet:2023soy}.
The availability of this new information enables a detailed investigation of the 
effect of minimal changes in the radiation reaction and their nonnegligible impact on the phasing. 
In this respect, we explore the possibility of tuning the radiation reaction to the NR data; we
conclude that this will likely be needed to obtain waveform templates highly faithful to NR data 
(say, $\sim 10^{-4}$ level) as they are expected to be needed for Third Generation (3G) detectors.

The paper is organized as follows. In Sec.~\ref{sec:eob} we recall the main elements of the 
\TEOBd{} model of Refs.~\cite{Nagar:2021xnh,Nagar:2023zxh} and highlight 
the modifications introduced in this work. In particular, Sec.~\ref{sec:4PN} is dedicated to the  
factorization and resummation of the 4PN waveform of Ref.~\cite{Blanchet:2023bwj} following the standard 
EOB approach~\cite{Damour:2008gu}, while Sec.~\ref{sec:eob_dyn} discusses the dynamics
and more generally the spin sector. In Sec.~\ref{sec:ham_rho} we present the new 
spin-aligned model, discussing in detail quasi-circular configurations, eccentric configurations 
as well as scattering. In Sec.~\ref{sec:tuning_rho22} we break new ground with respect to previous
work by investigating various improvements in the model that can be obtained by NR-informing
{\it also} the radiation reaction. Concluding remarks are collected in Sec.~\ref{sec:conclusions}.
The main text is complemented by a few appendices. In particular, Appendix~\ref{sec:alter_resum} 
identifies some analytical systematics related to the Pad\'e resummation of the waveform and 
discusses their solution; Appendix~\ref{sec:giotto4PN} explores the impact of the 4PN-accurate waveform
(and radiation reaction) on the \TEOBg{} quasi-circular model in the nonspinning case; 
Appendix~\ref{sec:eccentric_ics} presents the implementation of the initial conditions
for eccentric inspirals using eccentricity and mean anomaly instead of using eccentricity and
frequency at the apastron  as it was done in previous work.

We adopt the following notations and conventions. The black hole masses
are denoted $(m_1,m_2)$, the mass ratio $q=m_1/m_2\geq 1$, the total mass $M\equiv m_1+m_2$, 
the symmetric mass ratio $\nu\equiv m_1 m_2/M^2$
and the mass fractions $X_i\equiv m_i/M$ with $i=1,2$. 
The dimensionless spin magnitudes are $\chi_i\equiv S_i/m_1^2$ with $i=1,2$, 
and we indicate with $\tilde{a}_0\equiv \tilde{a}_1+\tilde{a}_2\equiv X_1\chi_1+X_2\chi_2$
the effective spin, usually called $\chi_{\rm eff}$ in the literature. 
Unless otherwise stated, we use geometric units with $c=G=1$.

\section{Analytic EOB structure: waveform and dynamics}
\label{sec:eob}

As previously mentioned, we build upon the spin-aligned, eccentric \TEOBd{} model
discussed extensively in Ref.~\cite{Nagar:2021xnh} and Sec.~IIIB.2 of Ref.~\cite{Nagar:2023zxh}, 
improving few key aspects of it. In this section, we discuss the analytical structure of the model.
First, we focus on the pure-orbital sector, and incorporate 4PN waveform information in the $\ell=m=2$ contribution
to waveform and radiation reaction. Then, we remove the next-to-next-to-leading order (NNLO) spin-square effects 
that were first introduced in a factorized and resummed form in Ref.~\cite{Nagar:2021xnh}.
This will prompt a new determination of the $(a_6^c,c_3)$ EOB flexibility parameters, that will be discussed 
in the following section (see Sec.~\ref{sec:ham_rho}).

\subsection{The 4PN factorized and resummed nonspinning waveform}
\label{sec:4PN}
\begin{figure*}[t]
	\center
	\includegraphics[width=0.31\textwidth]{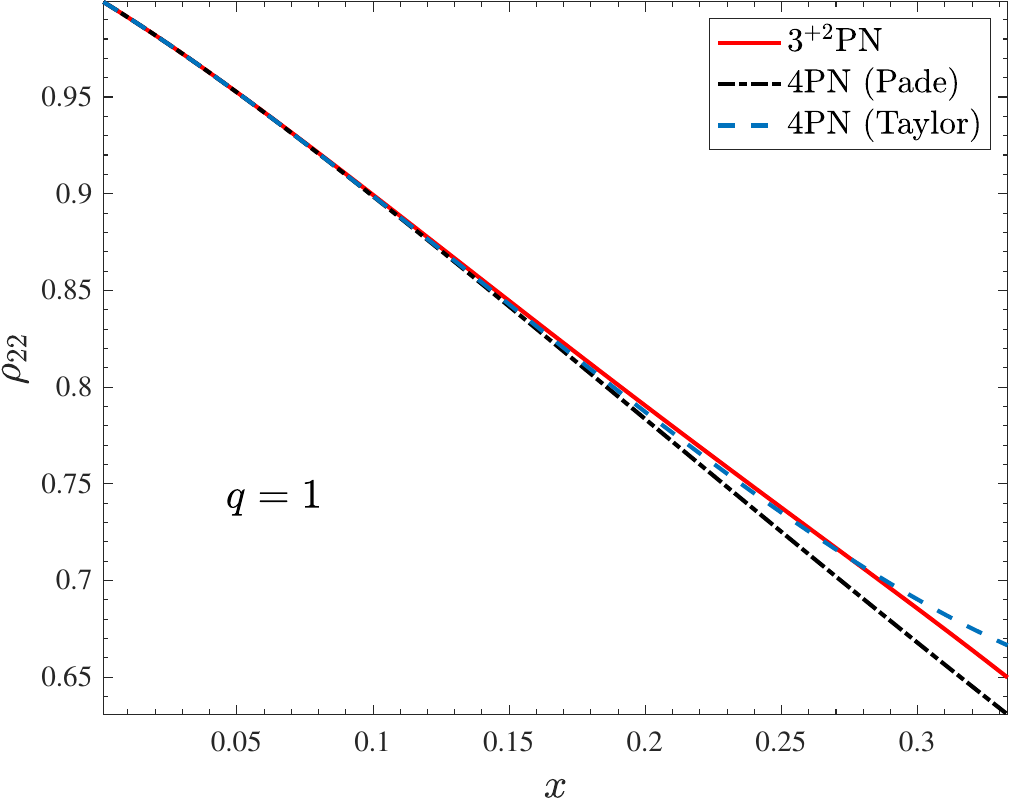}
	\includegraphics[width=0.31\textwidth]{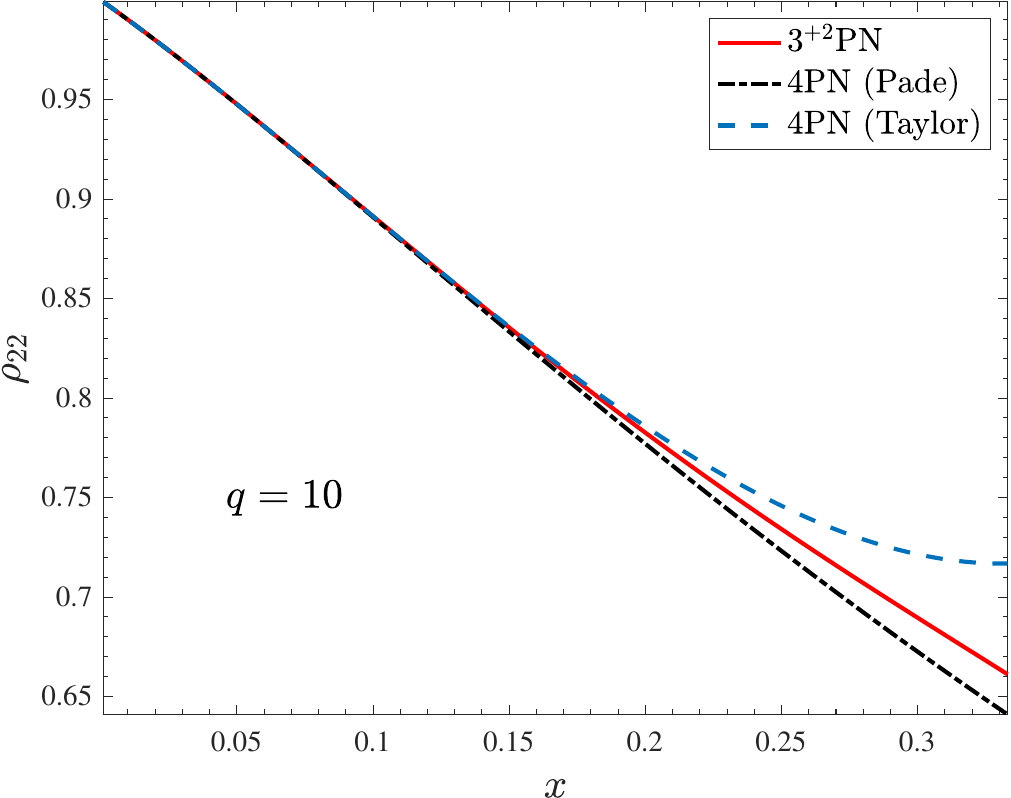}
	\includegraphics[width=0.31\textwidth]{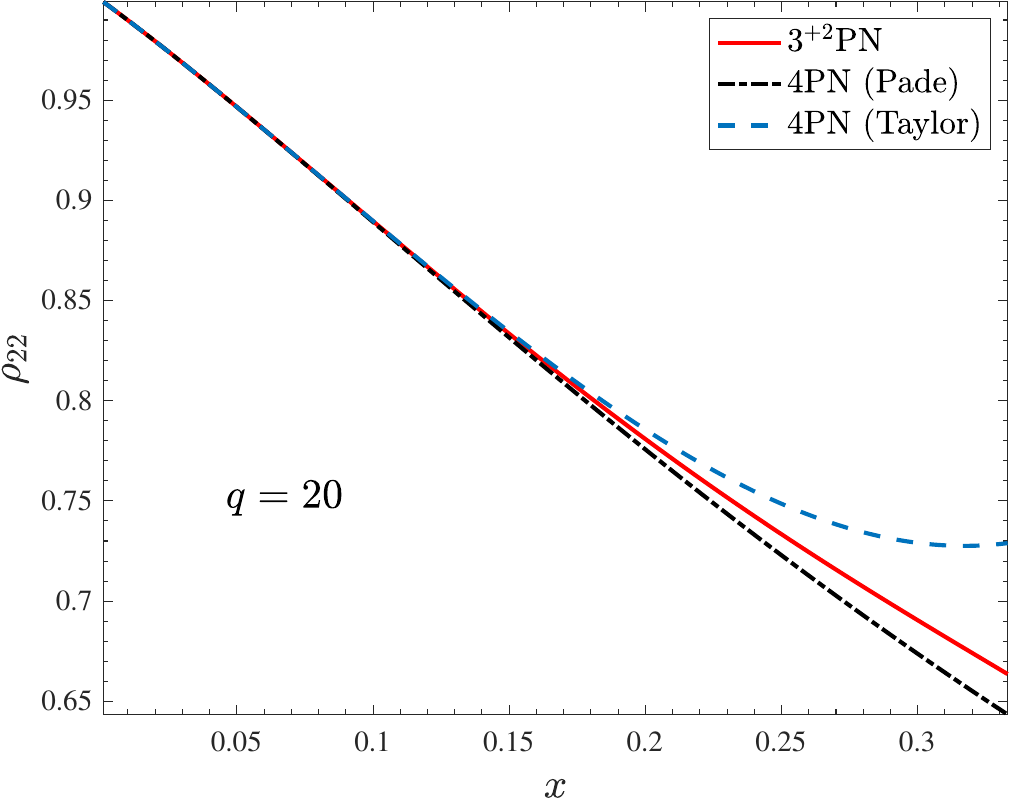}
	\caption{\label{fig:rho22}Comparing various approximations of the function $\rho_{22}$ for mass ratios $q=(1,10,20)$: 
	the $3^{+2}$~PN one used in the standard implementation of \TEOB{}; the $\rho_{22}$ at 4PN accuracy, 
	resummed with the $(2,2)$ Pad\'e approximant; the 4PN, Taylor-expanded, function. Note the consistency between 
	the $3^{+2}$ and the 4PN resummed as well as their weak dependence on the mass ratio.}
\end{figure*}
In order to be employed in EOB models, PN expression typically need to be recast in 
factorized and resummed form~\cite{Damour:1997ub,Damour:2008gu}. This is particularly 
important for the radiation reaction, where the factorization and resummation of the fluxes
is crucial to obtain a faithful description of the dynamics.
Here, we start from the 4PN accurate waveform obtained in Refs.~\cite{Blanchet:2023soy,Blanchet:2023sbv,Blanchet:2023bwj}
and recast it in the desired form of~\cite{Damour:2008gu}, following the procedure
of Ref.~\cite{Faye:2014fra}.

Let us first recall our notation. The multipolar expansion of the strain waveform is
\be
h_+-ih_\times = \dfrac{1}{D_L}\sum_{\ell=2}^{\infty}\sum_{m=-\ell}^\ell h_{\lm}{}_{-2}Y_{\lm} \, ,
\ee
where $D_L$ is the luminosity distance and ${}_{-2}Y_{\lm}$ are the $s=-2$ 
spin-weighted spherical harmonics. For each multipolar mode, the circular waveform is factorized as 
\be
h_\lm = h_\lm^N\hat{h}_\lm \ ,
\ee
where $h_\lm^N$ is the Newtonian prefactor (given in closed form e.g. in Ref.~\cite{Damour:2008gu})
and $\hat{h}_\lm$ is the PN correction. Following~\cite{Damour:2008gu}, this latter is factorized as
\be
\hat{h}_\lm = \hat{S}_{\rm eff}T_\lm e^{i\delta_\lm}(\rho_\lm)^\ell \ ,
\ee
where $\hat{S}_{\rm eff}$ is the effective source\footnote{$\hat{S}_{\rm eff}$ is the 
effective EOB Hamiltonian when $\ell+m={\rm even}$ and the Newton-normalized angular momentum when $\ell+m={\rm odd}$.}, 
$T_\lm$ is the tail factor~\cite{Damour:2008gu}, while $\rho_\lm$ and $\delta_\lm$ are the residual amplitude and phase corrections.
The tail factor explicitly reads
\be
\label{eq:Tlm}
T_\lm=\dfrac{\Gamma\left(\ell+1-2{\rm i}\hat{\hat{k}}\right)}{\Gamma(\ell+1)}e^{\pi \hat{\hat{k}}}e^{2{\rm i}\hat{\hat{k}}\log(2 k r_0)} \ .
\ee
Indicating with $E$ the energy along a circular orbit of frequency $\Omega$, we have
$\hat{\hat{k}}\equiv m E \Omega$, $k\equiv m\Omega$ and $r_0=2/\sqrt{e}$~\cite{Fujita:2010xj}.
The formula above is specified to the $\ell=m=2$ case starting from Eq.~(11) of Ref.~\cite{Blanchet:2023bwj}
(where $\hat{h}_{22}\equiv H_{22}$ therein) and $E$, at 4PN accuracy, given by Eq.~(3) therein. Note that
$x\equiv (M\Omega)^{2/3}$. The factorization (following the procedure and conventions of 
Ref.~\cite{Faye:2014fra} for consistency with the results given in~\cite{Blanchet:2023bwj}) 
yields the following 4PN-accurate $\rho_{22}$ function:
\begin{widetext}
\begin{align}
\label{eq:rho22}
\rho_{22}^{\rm 4PN}(x)&=1+\left(-\dfrac{43}{42}+\dfrac{55}{84}\nu\right)x+\left(-\dfrac{20555}{10584}-\dfrac{33025}{21168}\nu+\dfrac{19583}{42336}\nu^2\right)x^2\nonumber\\
&+\bigg[\dfrac{1556919113}{122245200}-\frac{428}{105}\text{eulerlog}_2(x)
+\left(\frac{41 \pi^2}{192}-\frac{48993925}{9779616}\right)\nu-\frac{6292061}{3259872}\nu^2+\frac{10620745}{39118464}\nu^3\bigg]x^3 \nonumber\\
&+\bigg[-\frac{387216563023}{160190110080}+\text{eulerlog}_2(x) \left(\frac{9202}{2205}+\frac{8819}{441}\nu\right)+\left(-\frac{6718432743163}{145627372800}-\frac{9953 \pi ^2}{21504}\right)\nu\nonumber\\
&+\left(\frac{10815863492353}{640760440320}-\frac{3485 \pi ^2}{5376}\right)\nu^2 -\frac{2088847783}{11650189824}\nu^3+\frac{70134663541}{512608352256} \nu ^4\bigg]x^4 \ ,
\end{align}
\end{widetext}
where $\text{eulerlog}_m(x)\equiv \gamma_E + \log(2m\sqrt{x})$. The residual phase, instead, reads:
\begin{align}
\delta_{22}&=\dfrac{7}{3}y^{3/2}-24\nu y^{5/2} + \dfrac{428}{105}\pi y^3 \nonumber\\
                 &+ \left(\dfrac{30995}{1134}\nu+\dfrac{962}{135}\nu^2\right)y^{7/2} -\dfrac{5536}{105}\pi\nu y^4 \ ,
\end{align}
with $y = (E\Omega)^{2/3}$. 

Once the first factorization is performed, the residual functions need to be resummed.
Phase and amplitude are considered separately, and their behaviors in the high-velocity 
limit studied. 
Let us first discuss the 4PN correction to $\delta_{22}$.
The analytical expression for $\delta_{22}$ implemented in \TEOB{} dates back to to Ref.~\cite{Damour:2012ky} 
(see Sec.~IIB.1 and Fig.~1 therein). There, it was obtained by factorizing the leading-order (LO)
part of $\delta_{22}$, $\delta_{22}^{\rm LO}=7/3y^{3/2}$, and resumming the remaining factor,
$\hat{\delta}_{22}$, with a Pad\'e $(2,2)$ approximant in the variable $v_y=\sqrt{y}$. 
In this respect, Fig.~1 of Ref.~\cite{Damour:2012ky} illustrates that the chosen 
Pad\'e approximant is effective in averaging the various PN-truncations of $\delta_{22}$. 
This fact by itself indicates that the resummed expression should give a representation of the function 
$\delta_{22}$ more robust than the truncated Taylor expansion and, as such, should be extended 
at the next available PN order. 
Attempting to follow this procedure, we compute the $\hat{\delta}_{22}$ factor, which at 4PN reads:
\begin{align}
\hat{\delta}_{22}&=1-\dfrac{72}{7}\nu v_y^2 + \dfrac{428\pi}{245} v_y^3 \nonumber\\
                        &+ \nu\left(\dfrac{30995}{2646}+\dfrac{962}{315}\nu\right)v_y^4-\dfrac{5536\pi}{245}\nu v_y^5 \ .
\end{align}
We have explored several ways of treating this expression analytically. First, one considers the straightforward,
Taylor-expanded expression. If for $q=1$ it is close to the former Pad\'e $(2,2)$ one, as $\nu$ decreases the
function is found to abruptly grow as $v_y \to 0.3$. When moving to Pad\'e approximants, it is natural to consider
the near-diagonal ones, i.e. $P^2_3$ and $P^3_2$. However, one finds that the $P^2_3$ develops a spurious
pole, while the $P^3_2$ increases again for $v_y \to 0.3$ when $\nu$ decreases. By contrast, the $(2,2)$
approximant remains robust and keeps the same functional shape for any choice of $\nu$. In view of these
results, for robustness, we decided to neglect the new 4PN contribution to $\hat{\delta}_{22}$ and just keep
using the Pad\'e $(2,2)$ approximant.

The $\rho_{22}(x)$ function, Eq.~\eqref{eq:rho22}, is similarly resummed using a Pad\'e $(2,2)$ approximant.
Following standard practice within the EOB framework~\cite{Damour:2009kr}, the $\log(x)$ functions 
appearing Eq.~\eqref{eq:rho22} above are treated as constant when computing the Pad\'e approximant~\cite{Nagar:2016ayt}. 
The $\log(x)$ are then replaced in the resulting rational function. Note that this approach is implemented
for all higher order modes, as suggested in Refs.~\cite{Nagar:2016ayt,Messina:2018ghh,Nagar:2020pcj}.
This choice, though simple and consistent with the low-order PN expansion, eventually introduces some 
qualitative incorrectness in the high-order terms as guessed by the resummation procedure.
For consistency with previous work we pursue this approach in the main text of the paper. However, in 
Appendix~\ref{sec:alter_resum} we revisit this standard choice and propose a different 
(though eventually more accurate) resummation strategy.
To appreciate the importance of the resummation , let us compare the Pad\'e resummed 
function with its Taylor-expanded expression as well as with the $\rho_{22}$ at $3^{+2}$~PN accuracy used 
in all implementations of \TEOBResumS{} so far, starting from Ref.~\cite{Damour:2009kr}. Let us remind the
reader that the notation $3^{+2}$~PN means that the function, dubbed $\rho_{22}^{3^{+2}{\rm PN}}$
hereafter, is obtained by hybridizing the 3PN-accurate one (with the complete $\nu$-dependence) with
4PN and 5PN test-mass terms~\cite{Damour:2008gu}. It explicitly reads:
\begin{widetext}
\begin{align}
\rho_{22}^{3^{+2}{\rm PN}}&=1+\left(-\dfrac{43}{42}+\dfrac{55}{84}\nu\right)x+\left(-\dfrac{20555}{10584}-\dfrac{33025}{21168}\nu+\dfrac{19583}{42336}\nu^2\right)x^2\nonumber\\
&+\bigg[\dfrac{1556919113}{122245200}-\frac{428}{105}\text{eulerlog}_2(x)
+\left(\frac{41 \pi^2}{192}-\frac{48993925}{9779616}\right)\nu-\frac{6292061}{3259872}\nu^2+\frac{10620745}{39118464}\nu^3\bigg]x^3 \nonumber\\
&+\bigg(-\frac{387216563023}{160190110080}+\frac{9202}{2205}\text{eulerlog}_2(x)\bigg)x^4 +\left(-\dfrac{16094530514677}{533967033600}+ \dfrac{439877}{55566}\text{eulerlog}_2(x)\right)x^5 \ .
\end{align}
\end{widetext}
Figure~\ref{fig:rho22} compares $\rho_{22}^{3^{+2}{\rm PN}}$ with $P^2_2(\rho_{22}^{\rm 4PN})$ and the 
Taylor-expanded $\rho_{22}^{\rm 4PN}$. The figure illustrates that, while $\rho_{22}^{\rm 4PN}$ shows a
strong dependence on $\nu$, both $\rho_{22}^{3^{+2}{\rm PN}}$ and $P^2_2(\rho_{22}^{\rm 4PN})$ are
weakly dependent on it and in addition are semi-quantitatively consistent among themselves.
As it will be shown below, this guarantees the robustness of the model all over the 
parameter space even if $\rho_{22}^{3^{+2}\rm PN}$ is replaced by  $P^2_2(\rho_{22}^{\rm 4PN})$, 
though this entails some changes in the value of the NR-informed effective 5PN parameter $a_6^c$.
In the following main text we will only focus on including the complete 4PN function in the \TEOBd{}
model, where, as we will see, will yield improvements with respect to previous work. For completeness, 
we have also explored the impact of the 4PN waveform (and flux) correction on \TEOBg{}, finding
however that it does not improve\footnote{This might be due to the combinaton of the iteration on NQC corrections
needed for the \TEOBg{} model together with the lower PN order of the (resummed) $\bar{D}$ 
and $Q$ functions, that yield quantitative differences towards merger, see discussion in Ref.~\cite{Nagar:2023zxh}.} 
the current state-of-the-art quasi-circular model (see Appendix~\ref{sec:giotto4PN}).
From now on, we will thus consider $P^2_2(\rho_{22}^{\rm 4PN})$ and the Pad\'e resummed 3.5PN 
$\delta_{22}$ as our default choices for the waveform and radiation reaction.
Evidently, when implemented in the complete EOB model, the energy along circular orbits $E$ in Eq.~\eqref{eq:Tlm}
will be replaced by the actual energy during the EOB evolution. Similarly, the argument of the function along
circular orbits, that is now $x=\Omega^{2/3}$, will become $x=(r_\omega \Omega)^2$, where $r_\Omega$ is a
Kepler's law correct orbital radius~\cite{Damour:2007yf,Damour:2009kr,Damour:2012ky,Damour:2014sva}.

\subsection{Spin-aligned EOB dynamics: centrifugal radius and waveform}
\label{sec:eob_dyn}

The conservative part of the model, i.e. the Hamiltonian, is based on the one discussed
extensively in Sec.~II of Ref.~\cite{Nagar:2021xnh}, with a few differences highlighted below. 
The EOB orbital dynamics is encoded within three potentials $(A,D,Q)$ while the spin-orbit 
sector is determined by the two gyro-gravitomagnetic functions $(G_S,G_{S_*})$. 
The real EOB Hamiltonian $H_{\rm EOB}$ is related to the effective 
one $\hat{H}_{\rm eff}\equiv H_{\rm eff}/\mu$ as~\cite{Buonanno:1998gg}
\be
H_{\rm EOB}=M\sqrt{1+2\nu\left(\hat{H}_{\rm eff}-1\right)} \ ,
\ee
where $\hat{H}_{\rm eff}$ reads:
\be
\hat{H}_{\rm eff}=\hat{H}_{\rm eff}^{\rm orb}+\tilde{G}p_{\varphi} \ ,
\ee
with
\be
\label{eq:Gtilde}
\tilde{G}\equiv G_S\hat{S}+G_{S_*}\hat{S}_{*} \ ,
\ee
where we defined
\begin{align}
\hat{S}&\equiv (S_1+S_2)M^{-2} \ ,\\
\hat{S}_{*}&\equiv \left(\dfrac{m_2}{m_1}S_1 + \dfrac{m_1}{m_2}S_2\right)M^{-2} \ .
\end{align}
The functions $(A,D)$ are taken at formal 
5PN order (see e.g.~\cite{Bini:2019nra}), with two free (yet uncalculated)
5PN coefficients $a_6^c$ and $d_5^{\nu^2}$, see Eqs.~(2) and~(3) in Ref.~\cite{Nagar:2021xnh}.
Then we fix $d_5^{\nu^2}=0$, while $a_6^c$ is informed using NR data. Both functions are
resummed, $A$ using a $(3,3)$ Pad\'e approximant and $D$ using a $(3,2)$ Pad\'e approximant
(see Eqs.~(6) and~(7) of Ref.~\cite{Nagar:2021xnh}). The $Q$ function includes only the local
part and is taken in Taylor-expanded form as in Eq.~(5) of Ref.~\cite{Nagar:2021xnh}.

Concerning the spin sector, the $(G_S,G_{S_*})$ functions also follow Ref.~\cite{Nagar:2021xnh}
and~\cite{Damour:2014sva} at next-to-next-to-leading order (NNLO) with the NR-informed 
next-to-next-to-next-to-leading (N$^3$LO) parameter $c_3$ (see Eqs.~(20)-(21) in~\cite{Nagar:2021xnh}).
Note that Ref.~\cite{Nagar:2021xnh} also explored the effect of using the analytical N$^3$LO 
results obtained in Ref.~\cite{Antonelli:2020aeb,Antonelli:2020ybz} (see also~\cite{Placidi:2024yld})
but here we only focus on the NR-informed approach to the spin-orbit sector.
Concerning instead the differences with respect to~\cite{Nagar:2021xnh}, 
here we modify: (i) the PN-order of even-in-spin effects incorporated
in the Hamiltonian through the centrifugal radius $r_c$, see Ref.~\cite{Damour:2014sva};
(ii) the PN order of spin-dependent terms entering the waveform.
Let us focus first on $r_c$, as introduced in Ref.~\cite{Nagar:2021xnh}  
to incorporate quadratic-in-spin corrections at NLO.  This is still the state-of-the-art implementation
in \TEOBg{}, even if corrections are actually available up to NNLO 
(see in particular Ref.~\cite{Nagar:2018plt} and references therein). 
As an exploratory study, Ref.~\cite{Nagar:2021xnh}  attempted to incorporate
NNLO effects in a special factorized and resummed form that eventually turned out to be unsatisfactory because
of the limited flexibility for large, positive, spins (see in particular Sec.~IIB.3 of ~\cite{Nagar:2021xnh}).
Here we thus go back to using the standard expression of $r_c$ at NLO. 
More precisely, using for consistency the notation of Sec.~IIB.3 of ~\cite{Nagar:2021xnh}, 
the centrifugal radius reads:
\be
r_c^2 = (r_c)^{\rm LO}\hat{r}_c^2 \ ,
\ee
with
\be
\hat{r}_c^2 = 1+ \dfrac{\delta a^2_{\rm NLO}}{r (r_c^{\rm LO})^2} \ ,
\ee
and
\begin{align}
(r_c^{\rm LO})^2&=r^2 + \tilde{a}_0^2\left(1+\dfrac{2}{r}\right)\ , \\
\delta a^2_{\rm NLO}&=-\dfrac{9}{8}\tilde{a}_0^2 - \dfrac{1}{8}(1-4\nu)\tilde{a}_{12}+\dfrac{5}{4}\tilde{a}_0\tilde{a}_{12} \ .
\end{align}
For what concerns the spin-dependent content of the waveform (and radiation reaction)
we adopt the results of Ref.~\cite{Nagar:2020pcj} outlined in Sec.~IIB therein
except for the $\ell=m=2$ mode that includes the N$^3$LO  and N$^4$LO spin-orbit corrections 
obtained by hybridizing the known $\nu$-dependent term up to NNLO with those
coming from the case of a spinning particle around a spinning black hole following the approach 
outlined in Sec.VB of Ref.~\cite{Messina:2018ghh}.
For the $\ell+m=\text{even}$ modes the residual waveform amplitudes
are written as 
\be
\rho_\lm = \rho_\lm^{\rm orb}+\rho_\lm^{\rm S} \ ,
\ee
and in particular for the $\rho_{22}^{\rm S}$ we formally have
\begin{align}
\rho_{22}^{\rm S}&= c_{\rm SO}^{\rm LO} x^{3/2}+c_{\rm SS}^{\rm LO}x^2 + c_{\rm SO}^{\rm NLO}x^{5/2}+c_{\rm SS}^{\rm NLO}x^3\nonumber\\
                           &+ (c_{\rm SO}^{\rm NNLO}+c^{\rm LO}_{\rm S^3})x^{7/2} +c_{\rm SO}^{\rm N^3LO} x^{9/2}+c_{\rm SO}^{\rm N^4LO} x^{11/2} \ ,
\end{align}
where the coefficients explicitly read
\begin{align}
c_{\rm SO}^{\rm LO} &=-\dfrac{\tilde{a}_0}{2} - \dfrac{1}{6}X_{12}\tilde{a}_{12} \ , \\
c_{\rm SS}^{\rm LO} & = \dfrac{1}{2}\tilde{a}_0^2 \ , \\
c_{\rm SO}^{\rm NLO} & = \left(-\dfrac{52}{63}-\dfrac{19}{504}\nu\right)\tilde{a}_0 - \left(\dfrac{50}{63}+\dfrac{209}{504}\nu\right)\tilde{a}_{12}X_{12} \ , \\
c_{\rm SS}^{\rm NLO} & =  \dfrac{221}{252}\tilde{a}_0\tilde{a}_{12}X_{12} + (\tilde{a}_1^2 + \tilde{a}_2^2)\left(-\dfrac{11}{21} + \dfrac{103}{504}\nu\right)\nonumber\\
            &+ \left(-\dfrac{85}{63} + \dfrac{383}{252}\nu\right)\tilde{a}_1\tilde{a}_2\ ,\\
c_{\rm S^3}^{\rm LO}& = \dfrac{7}{12}\tilde{a}_0^3 - \dfrac{1}{4}\tilde{a}_0^2\tilde{a}_{12}X_{12} \ ,\\
c_{\rm SO}^{\rm NNLO}&=\tilde{a}_0\left(\dfrac{32873}{21168} + \dfrac{477563}{42336}\nu  + \dfrac{147421}{84672}\nu^2\right)\nonumber\\
            &-\tilde{a}_{12}X_{12}\left(\dfrac{23687}{63504} -\dfrac{171791}{127008}\nu + \dfrac{50803}{254016}\nu^2\right) \ , \\
c_{\rm SO}^{\rm N^3LO}&=  c^+_{\rm N^3LO}\tilde{a}_0 + c^-_{\rm N^3LO}\tilde{a}_{12}X_{12} \\
c_{\rm SO}^{\rm N^4LO}&=  c^+_{\rm N^4LO}\tilde{a}_0 + c^-_{\rm N^4LO}\tilde{a}_{12}X_{12}\ .
\end{align}
where $c^{\rm \pm}_{\rm N^nLO}\equiv(c_a^{\rm N^nLO}\pm c^{\rm N^nLO}_\sigma)$ with
\begin{align}
c_a^{\rm N^3LO}           &= -\dfrac{8494939}{467775}+ \dfrac{2536}{315}{\rm eulerlog}_2(x) \ , \\
c_\sigma^{\rm N^3LO}  & =-\dfrac{14661629}{8731800} + \dfrac{214}{315}{\rm eulerlog}_2(x) \ , \\
c_a^{\rm N^4LO}           & =-\dfrac{890245226581}{26698351680} + \dfrac{328}{6615}{\rm eulerlog_(x)}\ , \\
c_\sigma^{\rm N^4LO}  & =-\dfrac{90273995723}{88994505600}  + \dfrac{428}{6615}{\rm eulerlog}_2(x) \ . 
\end{align}
We remind the reader that $(c_{\rm SO}^{\rm NNLO},c_{\rm SO}^{\rm N^3LO},c_{\rm SO}^{\rm N^4LO})$ 
are omitted from the quasi-circular {\tt TEOBResumS-GIOTTO} implementation.
Also note that the $c_{\rm SO}^{\rm NNLO}$ term is just one of the currently known 3.5PN-accurate contributions to the spin-dependent part
of the waveform recently obtained in Ref.~\cite{Henry:2022dzx}. In particular, these result correct some approximate expressions, e.g. for
the functions $\rho_{32}^{\rm S}$ or $\rho_{44}^S$, used in the current implementation. We have implemented these new amplitude 
corrections (after rewriting) and verified that the effect is so small that could be degenerate with the NR-informed parameter.
As a consequence, for simplicity, in this work we are not considering any of the new waveform terms of Ref.~\cite{Henry:2022dzx}.

\subsection{Radiation reaction forces}
\label{sec:FrFphi}
The residual multipolar amplitudes discussed above are then combined together to yield the
flux of angular momentum and the related radiation reaction force. Let us recall that in
the \TEOBd{} one deals with two forces, one taking care of the flux of angular momentum, 
$\hat{\F}_\varphi$ and the other one of the flux of radial momentum $\hat{\F}_{r_*}$. These
functions were detailed in previous works~\cite{Nagar:2021xnh,Nagar:2023zxh}, but we find 
it useful to briefly review here some information. See also Appendix~A of Ref.~\cite{Nagar:2024oyk} 
for technical details about the implementation. The two forces enter two Hamilton's equations as
\begin{align}
\dot{p}_\varphi &= \hat{\F}_\varphi \ , \\
\dot{p}_{r_*} &= -\sqrt{\dfrac{A}{B}}\de_r\hat{H}_{\rm EOB} +\hat{\F}_{r_*} \ ,
\end{align}
where $\hat{H}_{\rm EOB}\equiv H_{\rm EOB}/\mu$, $A B=D$ (see Eq.~(32) of Ref.~\cite{Damour:2014yha} 
and Sec.~II of Ref.~\cite{Nagar:2021xnh}), $p_\varphi\equiv P_\varphi/\mu$ is the orbital 
angular momentum, $p_{r_*}\equiv P_{R_*}/\mu$ is the radial momentum and 
$(\hat{\F}_\varphi,\hat{\F}_{r_*})$ follow respectively from 
Eqs.~(36)-(38) of Ref.~\cite{Nagar:2021xnh} and Eqs.~(6)-(7) of Ref.~\cite{Nagar:2023zxh}.

\section{Noncircularized waveform model with radiation reaction at 4PN}
\label{sec:ham_rho}
In this Section we complete the model by presenting the NR-informed
parameters and the performance all over the BBHs parameter space. 
The validation over the parameter space is performed -- as usual -- via EOB/NR comparisons
with various type of NR data. 
In particular: (i) for the quasi-circular limit, we compare with either the full
SXS catalog of NR quasi-circular (spin-aligned) waveform or with NR surrogates computing EOB/NR 
unfaithfulness (see below); (ii) for eccentric inspiral we perform the same analysis 
using the 28 SXS waveforms publicly available~\cite{Hinder:2017sxy}; (iii) for scattering configurations, 
we compare with the scattering angles of Refs.~\cite{Hopper:2022rwo,Damour:2014afa} 
and~\cite{Rettegno:2023ghr}. For pedagogic reasons we focus first on the quasi-circular 
nonspinning case and then gradually move on to considering quasi-circular, aligned spins
systems, eccentric inspirals and scattering configurations.
Before entering the discussion, let us recall that the above mentioned unfaithfulness $\bar{\F}$
is defined as follows. Given two waveforms $(h_1,h_2)$,  $\bar{\F}$ is a function of the total 
mass $M$ of the binary:
\be
\label{eq:barF}
\bar{\cal F}(M) \equiv 1-{\cal F}=1 -\max_{t_0,\phi_0}\dfrac{\langle h_1,h_2\rangle}{||h_1||||h_2||},
\ee
where $(t_0,\phi_0)$ are the initial time and phase. We used $||h||\equiv \sqrt{\langle h,h\rangle}$,
and the inner product between two waveforms is defined as 
$\langle h_1,h_2\rangle\equiv 4\Re \int_{f_{\rm min}^{\rm NR}(M)}^\infty \tilde{h}_1(f)\tilde{h}_2^*(f)/S_n(f)\, df$,
where $\tilde{h}(f)$ denotes the Fourier transform of $h(t)$, $S_n(f)$ is the detector power spectral density (PSD),
and $f_{\rm min}^{\rm NR}(M)=\hat{f}^{\rm NR}_{\rm min}/M$ is the initial frequency of the
NR waveform at highest resolution, i.e. the frequency measured after the junk-radiation
initial transient.
For $S_n$, in our comparisons we use either the zero-detuned, high-power noise spectral density of 
Advanced LIGO~\cite{aLIGODesign_PSD} or the predicted sensitivity of Einstein 
Telescope~\cite{Hild:2009ns, Hild:2010id}. Waveforms are tapered in the time-domain to reduce 
high-frequency oscillations in the corresponding Fourier transforms. 

\subsection{Nonspinning case: interplay between conservative and dissipative contributions}
\label{sec:Ham_vs_rr}
\begin{figure}[t]
	\center	
	\includegraphics[width=0.42\textwidth]{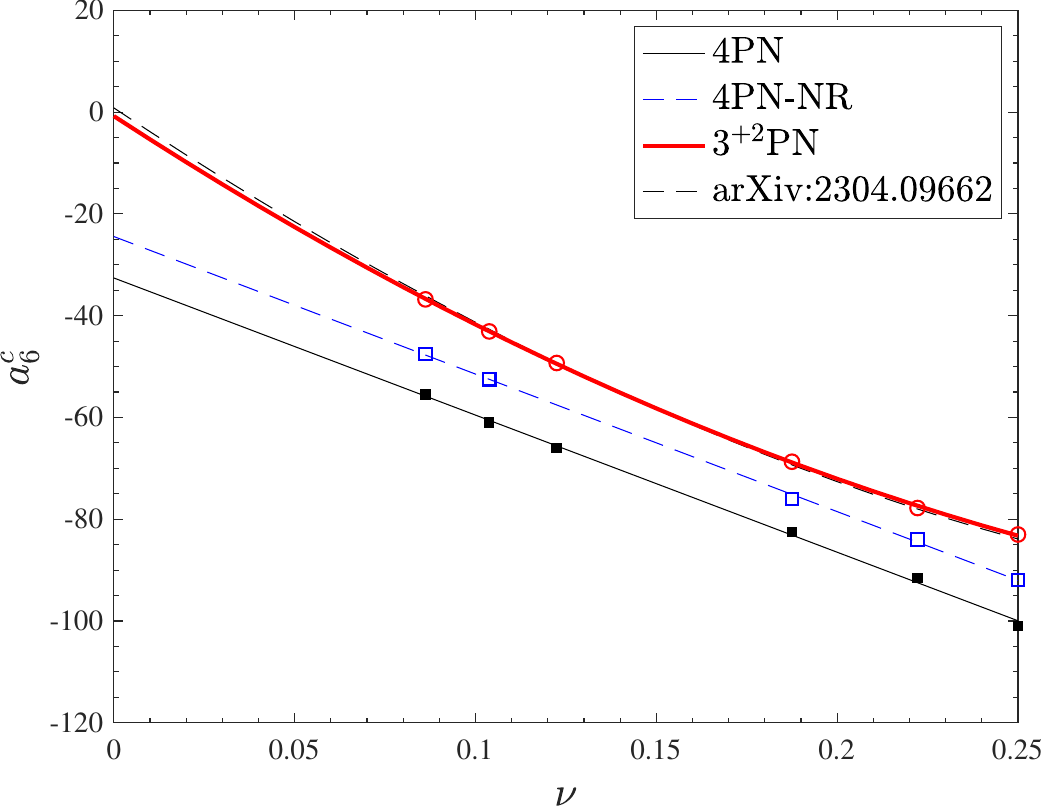}
	\caption{\label{fig:a6_fits}Behavior of $a_6^c$ for different approximations to the $\rho_{22}$ function. 
	The 4PN-accurate function (in Pad\'e-resummed form) entails values of $a_6^c$ that are smaller than 
	the case with $\rho_{22}$ at $3^{+2}$~PN accuracy in Taylor-expanded form. This mirrors a {\it more attractive} 
	conservative dynamics (the radius of the LSO is larger, see Table~\ref{tab:LSO}) that compensates for the weaker 
	action of radiation reaction due to $P^2_2(\rho_{22}^{\rm 4PN})<\rho_{22}^{\rm 3^{+2}PN}$, as shown 
	in Fig.~\ref{fig:rho22}. The 4PN-NR line, that lies between the other two, refers to the case where the 
	analytical $\nu$-dependence in the 4PN is replaced by a suitably NR-informed one, consistently 
	with Fig.~\ref{fig:rho22_NR_q1} below. See text for additional details.}
\end{figure}
%
\begin{table}[t]
	\caption{\label{tab:a6c}Data used to NR-inform the nonspinning section of the model(s) with various choices
	for radiation reaction. From left to right: the Taylor-expanded $\rho_{22}$ function at $3^{+2}$~PN accuracy 
	($\rho_{22}^{3^{+2}\rm PN}$); the $(2,2$) Pad\'e-resummed $\rho_{22}$ function at 4PN accuracy ($\rho_{22}^{\rm 4PN}$); 
	the (2,2) Pad\'e resummed $\rho_{22}$ at formal 4PN accuracy with the NR-informed $\nu$-dependence of the 4PN coefficient $c_4$. The corresponding
	values of $c_4^\nu$ are listed in the last column of the table. The fits of the various $a_6^c$ points are illustrated
	 in Fig.~\ref{fig:a6_fits}. Note that the $q=6$ data are used only when $\rho_{22}^{\rm 3^{+2}PN}$ is used.}
	\begin{center}
  \begin{ruledtabular}
	\begin{tabular}{lllc|c||ccc }
	  $\#$ & ID & $q$ & $a_6^{c,3^{+2}\rm PN}$ &$a_6^{\rm c, 4PN}$  & $a_6^{\rm c,4PN-NR}$ & $c_4^\nu$\\
	  \hline
  1 & SXS:BBH:0180 & $1$ & $-83$ & $-101$   & $-91.9$ & $-13.5$\\ 
 2 & SXS:BBH:0169 & $2$ & $-77.8$ & $-91.5$ & $-84$ & $-11.4$\\ 
 3 & SXS:BBH:0168 & $3$ & $-68.7$ & $-82.5$ & $-76$& $-8.5$\\ 
 4 & SXS:BBH:0166 & $6$ & $-49.3$ & $-66$   & $\dots$ & $\dots$\\ 
 5 & SXS:BBH:0299 & $7.5$ & $-43.1$ & $-61$  & $-52.5$& $-2.5$\\ 
 6 & SXS:BBH:0302& $9.5$ & $-36.8$ & $-55.5$ & $-47.5$& $-1.1$
 \end{tabular}
  \end{ruledtabular}
  \end{center}
  \end{table}
\begin{table}[t]
   \caption{\label{tab:LSO}Properties of the last stable orbit (LSO) obtained with the NR-informed $a_6^c$ 
   using $\rho_{22}^{\rm 3^{+ 2}PN}$ or $P^2_2(\rho_{22}^{\rm 4PN})$ with the analytical or the NR-informed
   4PN coefficient (4PN-NR case). Lowering $a_6^c$ as is needed  when using $P^2_2(\rho_{22}^{\rm 4PN})$ 
   entails a larger value of $r_{\rm LSO}$ and thus a faster plunge, so to compensate that
   $P^2_2(\rho_{22}^{\rm 4PN})<\rho_{22}^{\rm 3^{+2}PN}$ during the late inspiral as shown in Fig.~\ref{fig:rho22}.
   The 4PN-NR value is similarly understood by comparing with the NR-tuned 
   $P^2_2(\rho_{22}^{\rm 4PN})$ in Fig.~\ref{fig:rho22_NR_q1}.}
   \begin{center}
 \begin{ruledtabular}
   \begin{tabular}{cccc}
   model & $p_\varphi^{\rm LSO}$ & $r_{\rm LSO}$ & $u_{\rm LSO}$ \\
     \hline
$3^{+2}$PN &  3.034  & 2.72 & 0.367 \\ 
$4$PN    &  3.191  & 4.092 & 0.244 \\ 
4PN-NR & 3.167 & 3.631 &  0.275\\
\hline
\TEOBg{} & 3.225 & 4.517 & 0.221 \\
Schwarzschild & 3.464 & 6.0 & $0.1\bar{6}$ \\
\end{tabular}
 \end{ruledtabular}
 \end{center}
 \end{table}

\begin{figure}[t]
	\center	
	\includegraphics[width=0.21\textwidth]{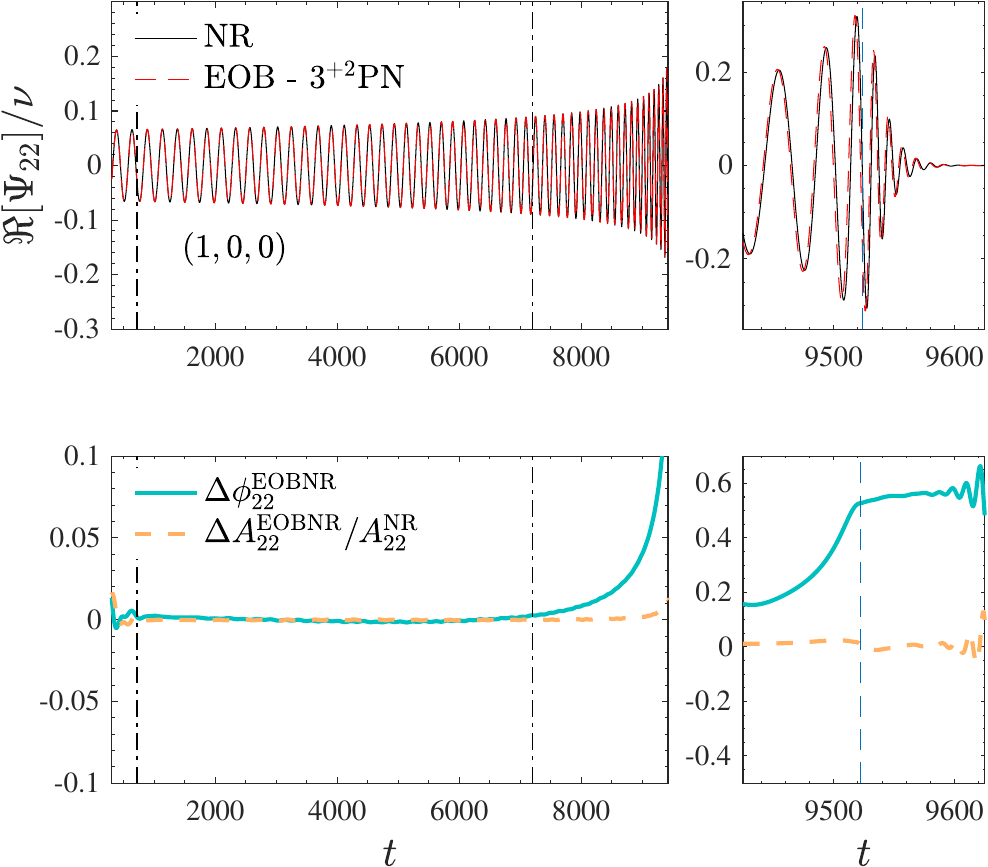}
	\qquad
	\includegraphics[width=0.21\textwidth]{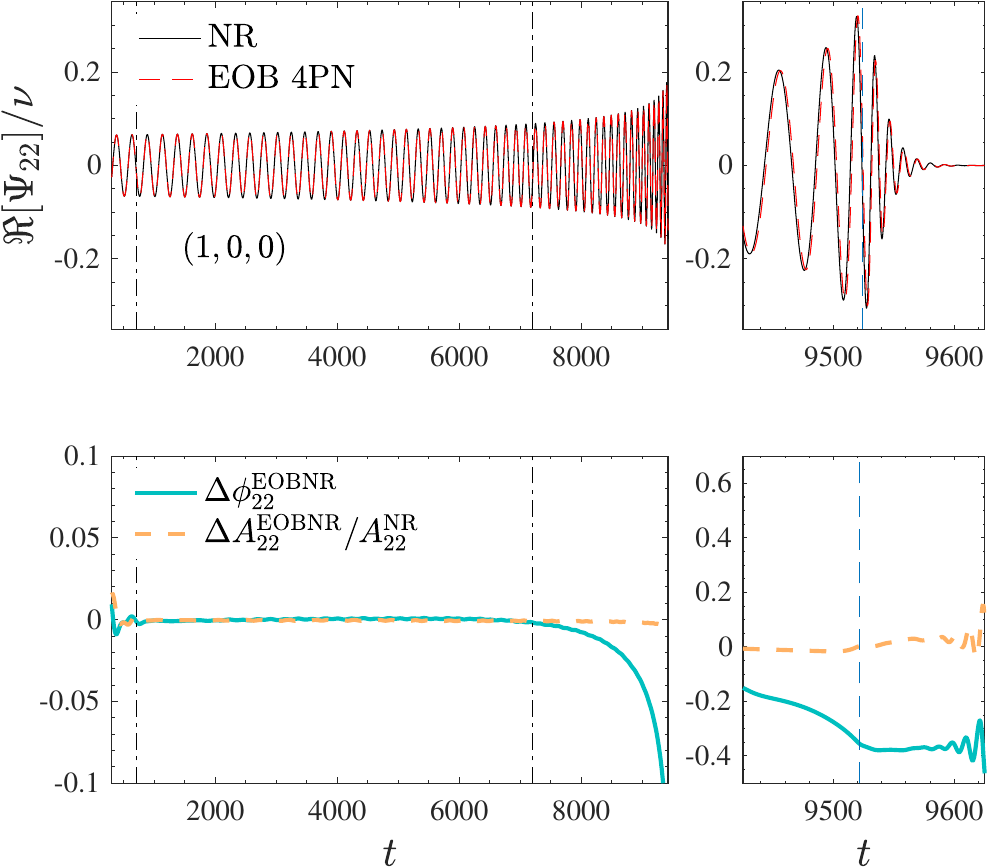}\\
	\vspace{5mm}
	\includegraphics[width=0.21\textwidth]{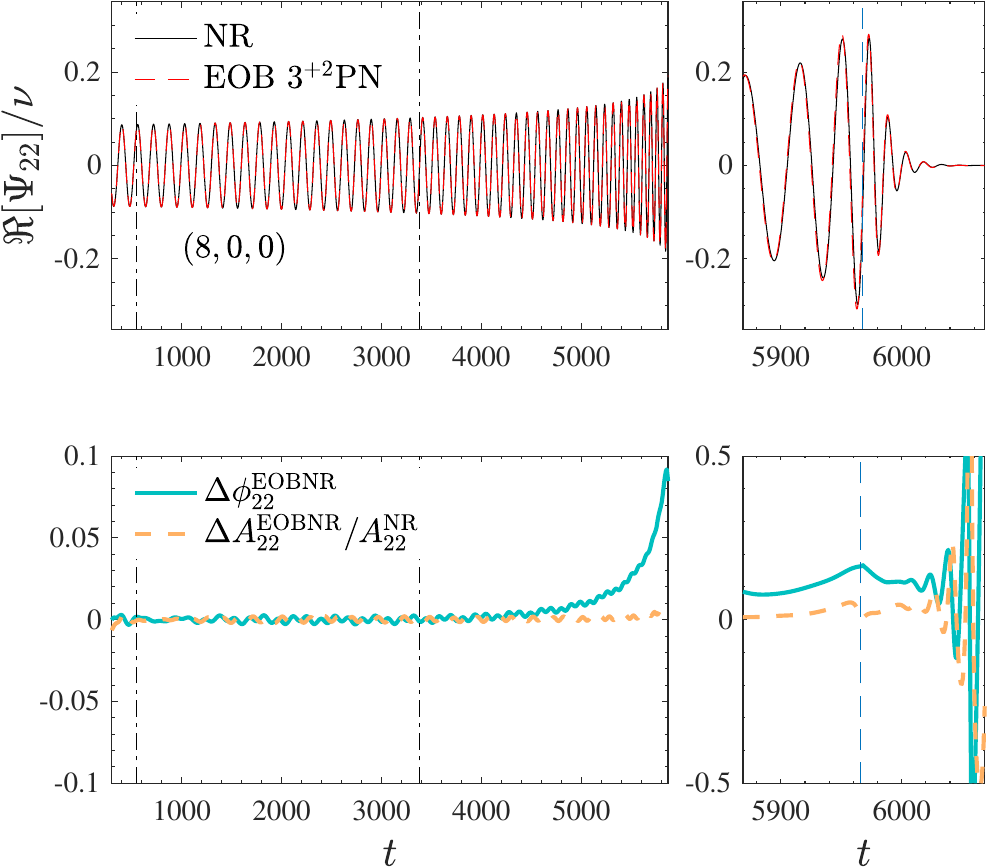}
	\qquad
	\includegraphics[width=0.21\textwidth]{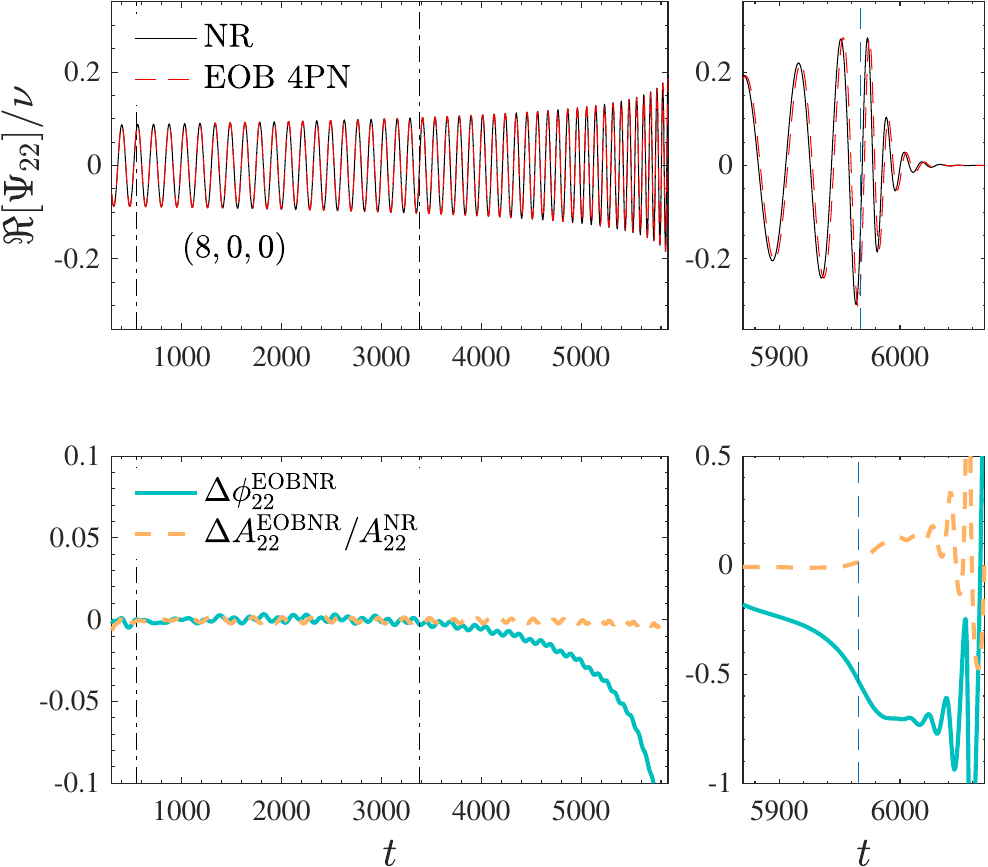}
	\caption{\label{fig:test_phasings}Left panels: EOB/NR phasings obtained with $\rho_{22}^{3^{+2}\rm PN}$ and $a_6^c$ given
	by Eq.~\eqref{eq:3p2PN}. Right panels: EOB/NR phasings obtained with $\rho_{22}^{4\rm PN}$ and $a_6^c$ given 
	by Eq.~\eqref{eq:4PN}. The dash-dotted vertical lines in the left part of each panel indicate the alignment window. The 
	dashed line in the right part of each panel marks the NR merger location.}
\end{figure}

In the nonspinning case, Ref.~\cite{Nagar:2023zxh} first introduced the model using the 
$\rho_{22}^{3^{+2}\rm PN}$ waveform and radiation reaction. Its performance was 
evaluated in the quasi-circular, eccentric and scattering case, with explicit comparisons 
of the scattering angle (see Figs.~12 and ~14 as well as Table~III of Ref.~\cite{Nagar:2023zxh}).
To start with, we need then to compare the performance of this model with the
new one obtained using the 4PN-resummed radiation reaction (and a newly determined $a_6^c$).
While doing so, we realized the presence of a small bug in the implementation of $F_r$ in
Ref.~\cite{Nagar:2023zxh}. Although this has  minimal quantitative effects, we redo here 
the full analysis of Sec.~IV of Ref.~\cite{Nagar:2023zxh}, while also NR-completing the 4PN-resummed
model. To start with, we determine $a_6^c$ by EOB/NR phasing comparisons, with the requirement,
clearly pointed out in~\cite{Nagar:2023zxh}, that the EOB/NR phase difference grows  monotonically,
so to have the smallest values of the EOB/NR unfaithfulness. To inform $a_6^c$ we use only six NR datasets,
that are listed in Table~\ref{tab:a6c}. The points are visualized in Fig.~\ref{fig:a6_fits}
They are easily representable by the following fits. For  $\rho_{22}$ at $3^{+2}$~PN accuracy the 
values are consistent with those of Ref.~\cite{Nagar:2023zxh} and can be fitted with a quadratic 
function\footnote{This is consistent with, but replaces, the function 
$a_6^c=175.5440\nu^3 + 487.6862\nu^2 -471.7141\nu+0.8178$ of~\cite{Nagar:2023zxh}, that is also
represented as a dashed line in Fig.~\ref{fig:a6_fits} for completeness.}
\be
\label{eq:3p2PN}
a_6^{c,{\rm 3^{+2}PN}} = 530.9514\nu^2 -462.5404\nu -0.78979 \ .
\ee
For $P^2_2(\rho_{22}^{\rm 4PN})$, the functional behavior of the $a_6^c$ points is simpler, as they 
can be accurately fitted by the following linear regression
\be
\label{eq:4PN}
a_6^{c,{\rm 4PN}}=-32.5953-269.4331\nu \ .
\ee
The fact that for the 4PN case $a_6^c$ is always smaller than for the $3^{+2}$PN case is the consequence
of $P^2_2(\rho_{22}^{\rm 4PN})<\rho_{22}^{\rm 3^{+2}PN}$. From the physical point of view, this follows from the 
fact that the radiation reaction (i.e., mainly the flux of angular momentum) is smaller in one case than in the other. 
As a result, to have the EOB waveform NR faithful one must tune the conservative dynamics 
(through $a_6^c$) so as to compensate this effect. In practice, as we will see below, lowering the value of $a_6^c$
means increasing the value of $r_{\rm LSO}$, which prompts a faster transition from the radiation-reaction driven
inspiral to plunge. 
In Figure~\ref{fig:test_phasings} we show four illustrative EOB/NR phasings for $q=1$ and $q=8$ obtained with either
the $3^{+2}$~PN prescription (left-panels) or the 4PN prescription (right panels).
Note that the EOB/NR phase difference is (essentially) monotonic in both cases, but its sign is different. 
The corresponding values of the LSO for $q=1$ are listed in Table~\ref{tab:LSO}. 
One sees that the fact that the lowering of $a_6^c$ needed when using $P^2_2(\rho_{22}^{\rm 4PN})$ 
entails a larger value of $r_{\rm LSO}$ and thus a faster plunge, so to compensate for 
$P^2_2(\rho_{22}^{\rm 4PN})<\rho_{22}^{\rm 3^{+2}PN}$ during the late inspiral.
On top of this, it is remarkable to note that when $\rho_{22}^{\rm 3^{+2}PN}$ is used,
the good, NR-informed, value of the LSO is rather small, $r_{\rm LSO}=2.72$, notably
a $25\%$ smaller than the value for $P^2_2(\rho_{22}^{\rm 4PN})$. This is needed to compensate
for what seems to be an incorrectly large radiation reaction during the inspiral.
With this vision in mind, one can better understand the left panels of Fig.~\ref{fig:test_phasings}
and in particular the meaning of the fact that the phase difference $\Delta\phi^{\rm EOBNR}_{22}\equiv \phi^{\rm EOB}_{22}-\phi^{\rm NR}_{22}$
is positive: the radiation-reaction-dominated inspiral progresses {\it faster} than the NR one, so that
$\phi_{22}^{\rm EOB}>\phi_{22}^{\rm NR}$ and thus  $\Delta\phi^{\rm EOBNR}_{22}>0$. This effect is 
compensated by the repulsive character of the EOB dynamics that is magnified by tuning $a_6^c$ so that the
LSO occurs at a rather small value of $r$.
With the same rationale in mind, it is similarly easy to interpret the right panels of Fig.~\ref{fig:test_phasings},
that exhibit a negative phase difference that begins to grow already during the late inspiral.
This indicates that the effect of radiation reaction (mainly related to the amplitude of $\rho_{22}$ being too small) 
is insufficient (with respect to the NR benchmark) and thus the transition from inspiral to plunge, merger and 
ringdown is delayed with respect to the NR case.
\begin{figure}[t]
	\center	
	\includegraphics[width=0.22\textwidth]{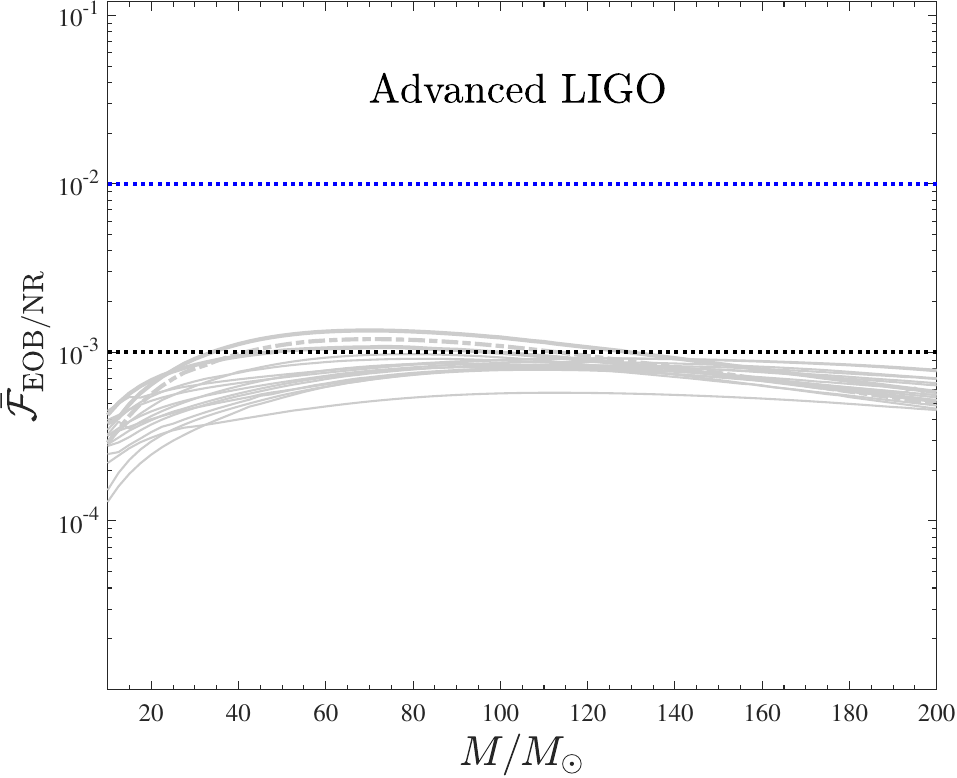}
	\includegraphics[width=0.22\textwidth]{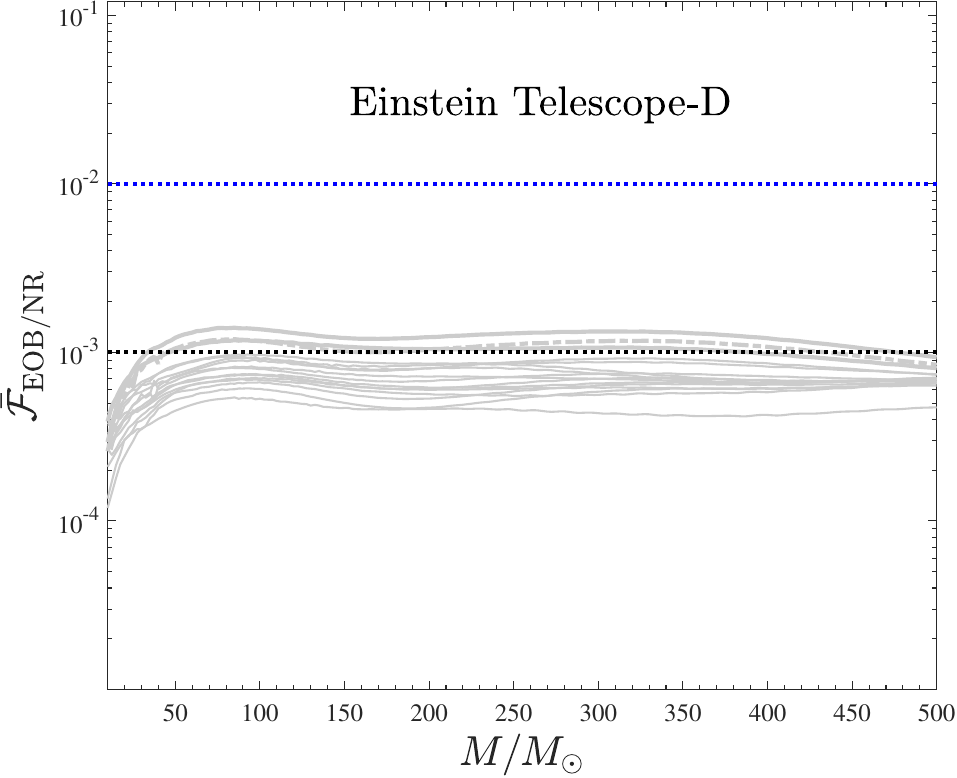}
	\caption{\label{fig:barFs04PN}EOB/NR unfaithfulness for the $\ell=m=2$ mode in the nonspinning  for all SXS nonspinning 
	datasets available (with $1\leq q\leq 10$ and $q=15$) using the 4PN-accurate (resummed) $\rho_{22}$ function.
	The performance is similar to, though slightly better than, the one obtained 
	for $\rho_{22}$ in Taylor-expanded form at $3^{+2}$~PN accuracy, Fig.~12 in Ref.~\cite{Nagar:2023zxh}.}
\end{figure}
The quantitative assessment of the quality of our new EOB model is finally completed by
computing the EOB/NR unfaithfulness. Since this quantity was computed in Ref.~\cite{Nagar:2023zxh} 
using the $3^{+2}$~PN expression of $\rho_{22}$, it is also pedagogically 
useful to compute it here with the 4PN-resummed model.
Figure~\ref{fig:barFs04PN} reports the values of $\bar{\F}$ for a sample of nonspinning binaries with
$1\leq q\leq 10$ stepped by 0.5. The performance of the 4PN model is substantially comparable to
that of the $3^{+2}$~PN one, although one has a small gain for high masses (cf. Fig.~12 in~\cite{Nagar:2023zxh}).
With this so well under control, we are ready to move to discussing the spin sector of the model.

\subsection{Spin-aligned and EOB/NR performance in the quasi-circular case}
\label{sec:eobnr}
\begin{table*}[t]
 \caption{\label{tab:c3_coeff}Coefficients for the fit of $c_3$ given by Eq.~\eqref{eq:c3fit} for the two NR-informed model discussed in the main text. 
 The {\tt Dal\'i}$_{\tt 4PN-analytic}$ one incorporates the 4PN information in the nonspinning $\rho_{22}^{\rm orb}$ function; 
 in the {\tt Dal\'i$_{\tt 4PN-NRTuned}$} model the same term is instead informed using nonspinning NR waveforms.}
   \begin{center}
     \begin{ruledtabular}
\begin{tabular}{c |c c c c c c | c c c c c c} 
Model   & \multicolumn{11}{c}{\hspace{-28mm}$c_3^{=}\equiv p_0\left(1 + n_1\tilde{a}_0 + n_2\tilde{a}_0^2 + n_3\tilde{a}_0^3 + n_4\tilde{a}_0^4\right)/\left(1 + d_1\tilde{a}_0\right)$}\\
            &  \multicolumn{11}{c}{$c_3^{\neq}\equiv \left(p_1\tilde{a}_0 + p_2\tilde{a}_0^2  + p_3\tilde{a}_0^3\right)\sqrt{1-4\nu}+ p_4\tilde{a}_0\nu \sqrt{1-4\nu} + \left(p_5\tilde{a}_{12}+ p_6\tilde{a}_{12}^2\right)\nu^2$}\\
            \hline
 {\tt TEOBResumS*}      & $p_0$ & $n_1$ & $n_2$   &  $n_3$   &     $n_4$  & $d_1$  & $p_1$ & $p_2$ & $p_3$ & $p_4$ & $p_5$ & $p_6$\\
\hline
{\tt Dal\'i}$_{\tt 4PN-analytic}$  &38.625 & $-0.105187$ & $-0.758427$  & $0.183613$ & $0.057817$ &   $0.905420$  &   $23.058$  & $12.544$ &  $-0.0157$ &  $-119.2596$ & $64.4709$ & $54.6568$   \\
{\tt Dal\'i$_{\tt 4PN-NRTuned}$}  & 44.616 & $-1.609364$ & 0.807277  & $-0.220357$ & $0.045408$ & $-0.7704$   &5.67453   &$-1.214$   
& 12.3433  & $-16.5925$  & $-0.7784$ & $-55.1691$   \\
\end{tabular}
\end{ruledtabular}
\end{center}
\end{table*}
To complete the spin sector, we need to NR-inform the N$^3$LO 
effective spin-orbit parameter $c_3$ introduced above (see~\cite{Damour:2014sva}). 
This procedure was already implemented in previous versions of the \TEOBd{} 
model~\cite{Nagar:2021gss,Nagar:2021xnh}, but it was always found complicated to reduce the 
EOB/NR unfaithfulness for large, positive, values of the spins, as discussed extensively in Ref.~\cite{Nagar:2021xnh}.
We find that the new analytical setup finally allows us to overcome this problem.
The NR-informed analytical expression for $c_3$ is obtained using the same functional form and 
the same  set of SXS NR data of Ref.~\cite{Nagar:2023zxh}. It reads:
\begin{align}
  \label{eq:c3fit}
c_3(\nu,\tilde{a}_0,\tilde{a}_{12})= c^{=}_{3}+c_3^{\neq} \ ,
\end{align}
where
\begin{align}
c_3^{=}&\equiv p_0\dfrac{1 + n_1\tilde{a}_0 + n_2\tilde{a}_0^2 + n_3\tilde{a}_0^3 + n_4\tilde{a}_0^4}{1 + d_1\tilde{a}_0}\\
c_3^{\neq}&\equiv \left(p_1\tilde{a}_0 + p_2\tilde{a}_0^2  + p_3\tilde{a}_0^3\right)\sqrt{1-4\nu} \nonumber\\
                 &+ p_4\tilde{a}_0\nu \sqrt{1-4\nu} + \left(p_5\tilde{a}_{12}+ p_6\tilde{a}_{12}^2\right)\nu^2 \ ,
\end{align}
The NR configurations we used to inform $c_3$ are listed in Tables~\ref{tab:c3_eqmass}-\ref{tab:c3_uneqmass} 
in Appendix~\ref{sec:nrdata}. For each configuration, we determine the best-guess value of $c_3$ via time-domain 
phasing comparison. Then, the resulting values are fitted using the above functional form. 
The coefficients of the fit are reported in Table~\ref{tab:c3_coeff}. Since this model relies on 4PN analytical information, 
we will refer to it as ${\tt TEOBResumS-Dal\'i _{4PN-analytic}}$ or just \daliAN{} for simplicity.

We focus first on the $\ell=m=2$ waveform mode
and estimate the EOB/NR unfaithfulness $\bar{\F}_{\rm EOBNR}$ with the Advanced LIGO PSD.
\begin{figure*}[t]
	\center	
	\includegraphics[width=0.42\textwidth]{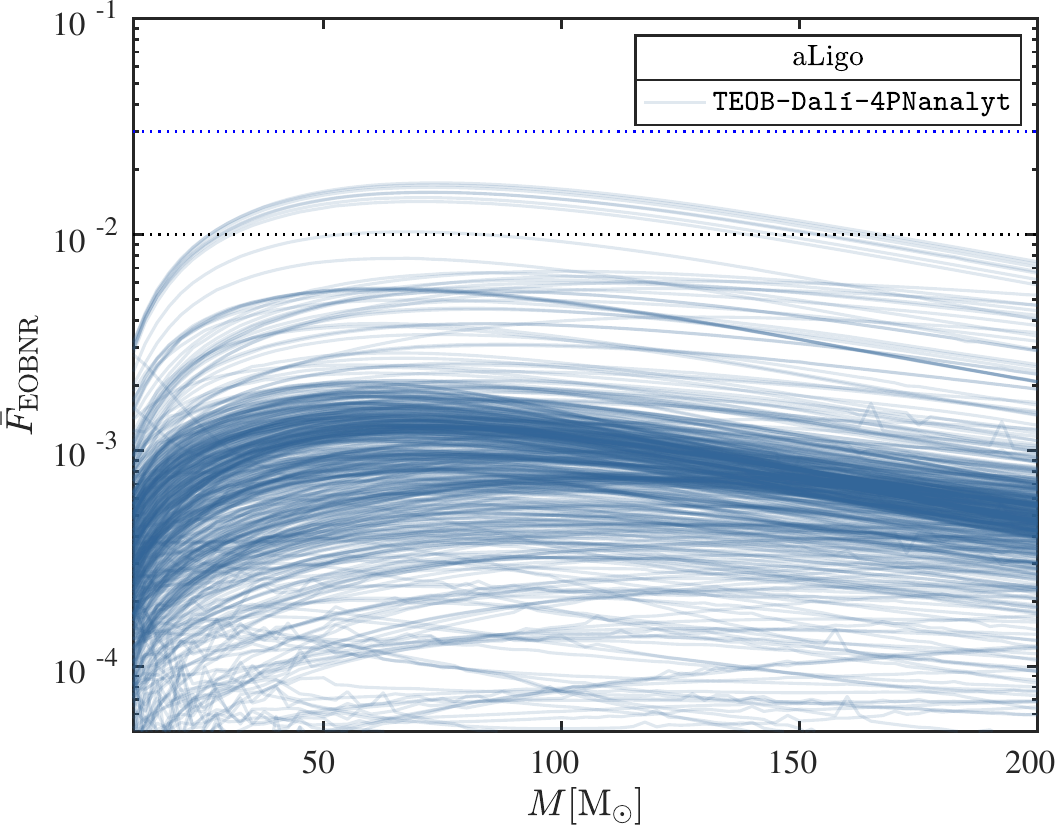}
	\includegraphics[width=0.45\textwidth]{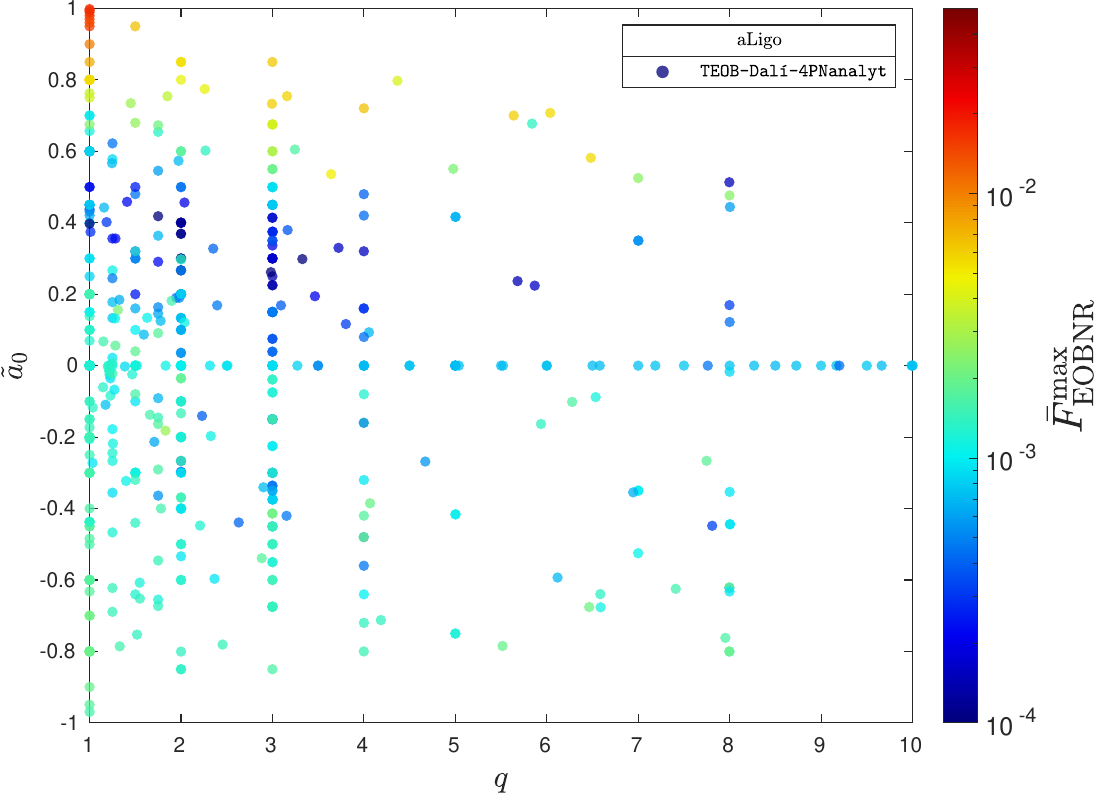}
	\caption{\label{fig:barF_circ_eobnr}EOB/NR unfaithfulness for the $\ell=m=2$ mode. The performance of the 
	model gets progressively worst in the equal-mass, high-spin corner. This is consistent with a previous version of
	the model, but less pronounced.}
\end{figure*}
Figure~\ref{fig:barF_circ_eobnr} shows $\bar{\F}_{\rm EOBNR}$ computed all over the 534 spin-aligned 
datasets of the SXS catalog. The left panel of the figure shows $\bar{\F}_{\rm EOBNR}$ versus the total mass, 
while the right panel the maximum value for each configurations, $\bar{\F}^{\rm max}_{\rm EOBNR}$. 
We see that the unfaithfulness always lies below the $1\%$ threshold except for a few outliers 
in the equal-mass, high (positive) spin corner, that in any case do not exceed the $2\%$ level. 
This result alone represents an improvement with respect to previous 
work~\cite{Nagar:2021gss}. It is also useful to provide a direct comparison with waveforms generated with the NR 
surrogates \nrsurqeight{}~\cite{Varma:2018mmi} and  \nrsurqfifteen{}~\cite{Yoo:2022erv}.
We generate 1000 randomly sampled configurations with $q \in [1, 8]$, total mass $M \in [40, 140] \Msun$ 
and dimensionless spins $|\chi_i|<0.8$ and compute mismatches in the frequency interval between $[20, 2048]$~Hz.
Similarly, when working with \nrsurqfifteen{}, we consider another 1000 randomly sampled configurations with 
$q\in[8,15]$, and dimensionless spins $|\chi_{1}|\leq 0.5$, $\chi_2=0$, corresponding to the validity range of the surrogate model.
The values of $\bar{\F}$ are reported in Fig.~\ref{fig:barF_circ_sur}, together with those corresponding
to the quasi-circular version of the model, {\tt TEOBResumS-GIOTTO}, calculated in Ref.~\cite{Nagar:2021xnh}.
\begin{figure}[t]
	\center	
	\includegraphics[width=0.42\textwidth]{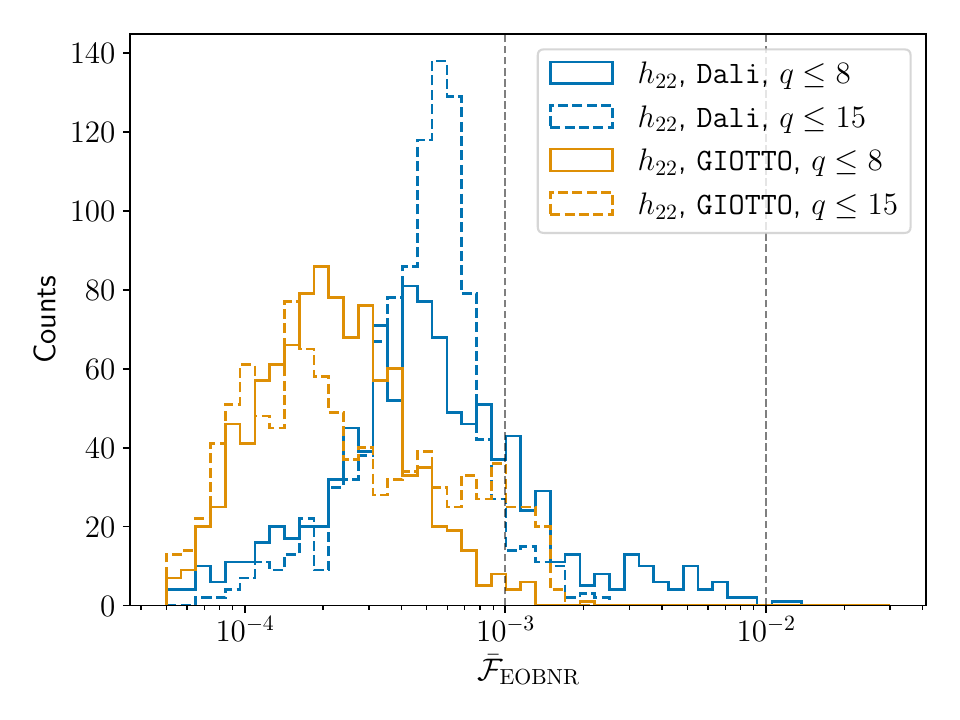}
	\caption{\label{fig:barF_circ_sur}EOB/NR unfaithfulness for the $\ell=m=2$ model against the NR surrogate 
	{\tt NRHybSur3dq8} and \nrsurqfifteen{}. In the first case, $q \in [1, 8]$,  $M \in [40, 140] \Msun$ 
	and dimensionless spins are $|\chi_i|<0.8$. In the second case $q\in[8,15]$,  $|\chi_{1}|\leq 0.5$ and $\chi_2=0$}
\end{figure}
The figure clearly indicates that the quasi-circular limit of \daliAN{} is acceptably consistent with the two NR surrogate models
and with the basic quasi-circular model.

\subsubsection{Direct comparison with GIOTTO}
\label{sec:comparison_giotto}
\begin{figure}[t]
	\center	
	\includegraphics[width=0.42\textwidth]{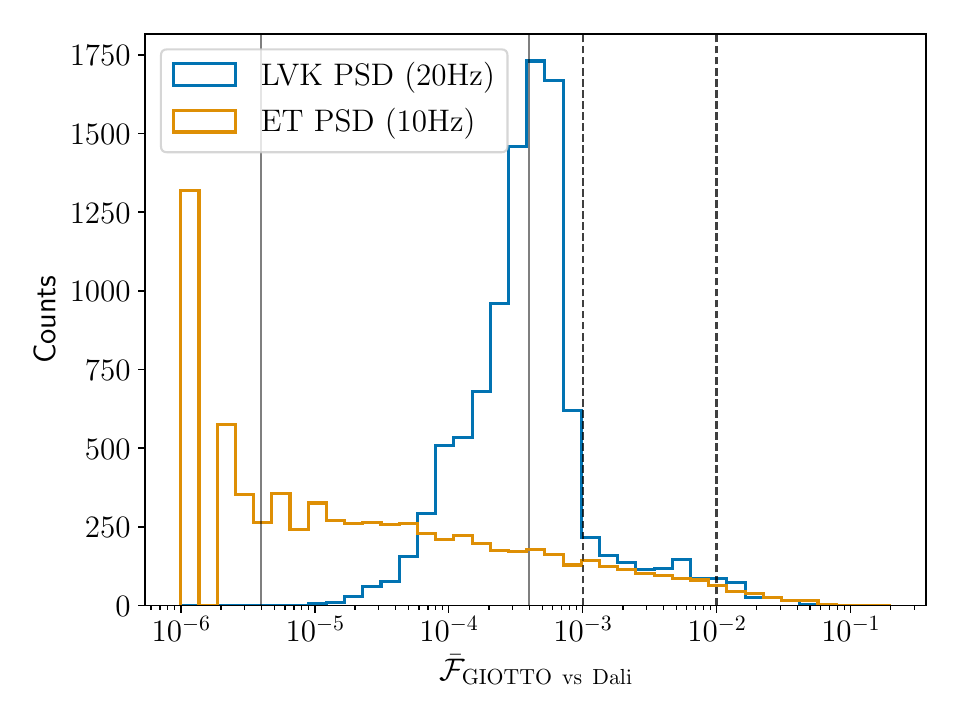}
	\caption{\label{fig:barF_dali_giotto}Direct comparison between \daliAN{} and {\tt TEOBResumS-GIOTTO}. 
	We compute mismatches for $10^4$ binaries with $q\in[1,15]$, $\chi_i \in [-0.9, 0.9]$. 
	We use two different noise curves for the computation: the Advanced LIGO PSD (blue) and the Einstein Telescope PSD (orange).
	For the former, mismatches are computed in the frequency range $[20, 2048]$~Hz, while for the latter, we use $[10, 2048]$~Hz.
	The two models are in very good agreement, in spite of their theoretical differences.}
\end{figure}

To further investigate differences between the two avatars of the model, we compute mismatches between 
\daliAN{} and {\tt TEOBResumS-GIOTTO} using the same settings employed for the mismatches above,
but extending the range of mass ratios and spins to $q\in[1,15]$, $\chi_i \in [-0.9, 0.9]$, and considering $10^4$ binaries.
The results are shown in Fig.~\ref{fig:barF_dali_giotto}. The two models are in very good agreement with one another, with mismatches
below the $1\%$ threshold for $98\%$ of the configurations and below $0.1\%$ for $88\%$ of them. 
The maximum mismatch is $\sim 4\%$, found for a $q \sim 13$, $\chi_1\sim 0.8$ configuration.

The agreement between the two models further improves when comparing them using a 
lower frequency cutoff of $10$~Hz for the mismatch computation and the Einstein Telescope PSD~\cite{LIGO-P1600143}.
While the higher mismatch tail remains similar to the one found with the Advanced LIGO PSD, both in terms
of value and of the corresponding configurations, the fraction of systems with mismatches below $0.1\%$ 
increases to $91\%$, with a considerable number of binaries having mismatches below $10^{-5}$. This result
indicates that the two models are in very good agreement with one another during the inspiral phase.

We remind the reader that this consistency is, a priori, not a trivial achievement because of the many
theoretical differences between the two models. We are now expecting that this new version of \daliAN{} will allow us 
to reduce (or eliminate) the systematics in parameter estimation that were found using previous versions of 
the model~\cite{Bonino:2022hkj}.

\begin{figure}[t]
  \includegraphics[width=0.45\textwidth]{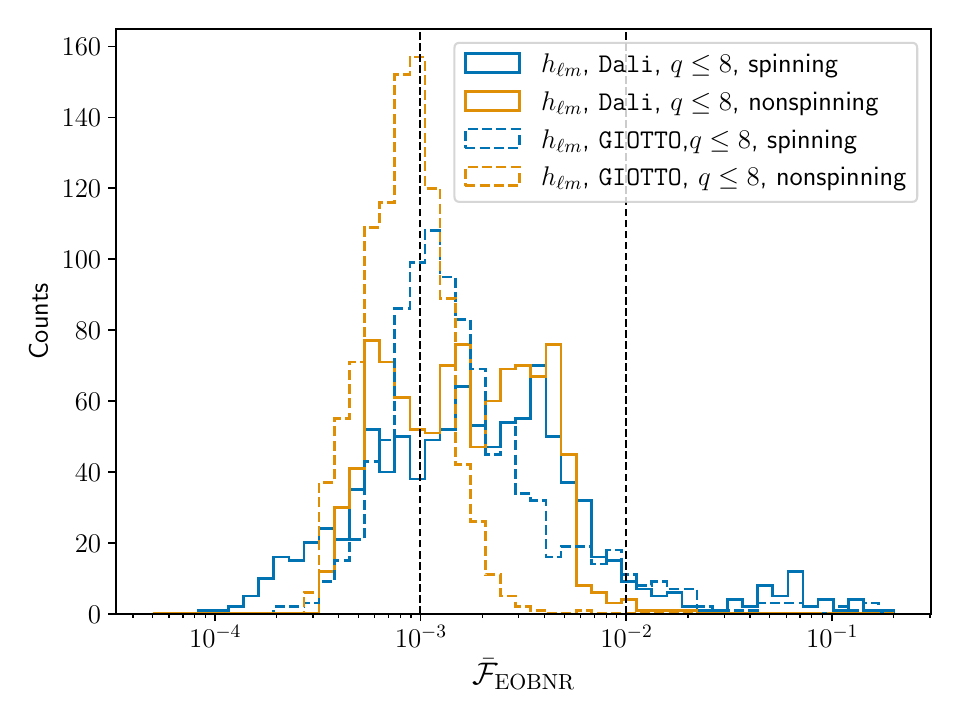}
  \caption{\label{fig:higher_modes}EOB/NR unfaithfulness for the higher modes, for the same configurations
  considered in Fig.~\ref{fig:barF_circ_sur}.}
\end{figure}

	\begin{table*}[t]
		\caption{\label{tab:SXS}SXS simulations with eccentricity analyzed in this work. From left to right: the 
			ID of the simulation from the SXS catalog; the mass ratio $q\equiv m_1/m_2\geq 1$ and the individual dimensionless 
			spins $(\chi_1,\chi_2)$; the time-domain NR phasing uncertainty at merger $\delta\phi^{\rm NR}_{\rm mrg}$ obtained comparing
			the highest and second highest resolution; the estimated NR eccentricity at first apastron $e_{\omega_a}^{\rm NR}$; 
			the NR frequency of first apastron $\omega_{a}^{\rm NR}$; 
			the initial EOB eccentricity $e^{\rm EOB}_{\omega_a}$ and apastron frequency $\omega_{a}^{\rm EOB}$ used to start the EOB evolution; 
			the maximal NR unfaithfulness uncertainty, $\bar{\F}^{\rm max}_{\rm NRNR}$ and the initial frequency $Mf_{\rm min}$ used in the EOB/NR unfaithfulness
			computation shown in Figs.~\ref{fig:barF_ecc} using the 4PN analytical information in the flux and in 
			Fig.~\ref{fig:barF_ecc_4PNtuned} using the corresponding NR-tuned one. The last two columns report the corresponding maximum values of $\bar{\F}_{\rm EOBNR}^{\rm max}$, the analytical, $\bar{\F}_{\rm EOBNR}^{\rm max,4PNan}$, and the NR-tuned one, $\bar{\F}_{\rm EOBNR}^{\rm max,4PNnr}$.}
		\begin{center}
			\begin{ruledtabular}
				\begin{tabular}{c| c  c c c c |l l| c|c cc} 
					$\#$ & id & $(q,\chi_1,\chi_2)$ & $\delta\phi^{\rm NR}_{\rm mrg}$[rad]& $e^{\rm NR}_{\omega_a}$ & $\omega_a^{\rm NR}$ &$e^{\rm EOB}_{\omega_a}$ & $\omega_{a}^{\rm EOB}$ & $\bar{\F}_{\rm NRNR}^{\rm max}[\%]$  & $Mf_{\rm min}$ &$\bar{\F}_{\rm EOBNR}^{\rm max,4PNan}$ &$\bar{\F}_{\rm EOBNR}^{\rm max,4PNnr}$ \\
					\hline
					\hline
					1 & BBH:1355 & $(1,0, 0)$ & $+0.92$ & 0.0620  & 0.03278728 & 0.0888    & 0.02805750  & 0.012 & 0.0055 & 0.173 &0.026 \\
					2 & BBH:1356 & $(1,0, 0)$& $+0.95$ & 0.1000 &  0.02482006  & 0.15038  & 0.019077  & 0.0077 & 0.0044 &  0.159& 0.052\\
					3 & BBH:1358 & $(1,0, 0)$& $+0.25$ & 0.1023 & 0.03108936 & 0.18082   & 0.021238 & 0.016 &  0.0061 & 0.328 & 0.065\\
					4 & BBH:1359 & $(1,0, 0)$& $+0.25$  & 0.1125 & 0.03708305 & 0.18240   & 0.021387 & 0.0024& 0.0065  & 0.441 & 0.327 \\
					5 & BBH:1357 & $(1,0, 0)$& $-0.44$  & 0.1096 & 0.03990101 & 0.19201   & 0.01960 & 0.028& 0.0061 &     0.198 & 0.101 \\
					6 & BBH:1361 & $(1,0, 0)$& +0.39    & 0.1634 & 0.03269520  & 0.23557   & 0.020991   & 0.057&0.0065  & 0.357 & 0.113\\
					7 & BBH:1360 & $(1,0, 0)$& $-0.22$ & 0.1604 & 0.03138220 & 0.2440  & 0.019508   &0.0094  & 0.0065 &  0.254 & 0.085\\
					8 & BBH:1362 & $(1,0, 0)$& $-0.09$ & 0.1999 & 0.05624375 & 0.3019     & 0.01914 & 0.0098 & 0.0065 & 0.244 &0.119 \\
					9 & BBH:1363 & $(1,0, 0)$& $+0.58$ & 0.2048 & 0.05778104 &  0.30479    & 0.01908 & 0.07 & 0.006 &  0.520 & 0.381\\
					10 & BBH:1364 & $(2,0, 0)$& $-0.91$ & 0.0518 &  0.03265995   & 0.0844    & 0.025231   & 0.049  & 0.062 & 0.089 & 0.054 \\
					11 & BBH:1365 & $(2,0, 0)$& $-0.90$ & 0.0650  &  0.03305974   & 0.110     & 0.023987 & 0.027&  0.062 &  0.109 & 0.073\\
					12 & BBH:1366 & $(2,0, 0)$& $-6\times 10^{-4}$ & 0.1109 & 0.03089493 & 0.14989   & 0.02577 &  0.017  & 0.0052 & 0.201 &0.148 \\
					13 & BBH:1367 & $(2,0, 0)$& $+0.60$ & 0.1102 & 0.02975257 & 0.15095    & 0.0260  & 0.0076  & 0.0055 & 0.108 & 0.095\\
					14 & BBH:1368 & $(2,0, 0)$& $-0.71$ & 0.1043 & 0.02930360 &  0.14951  & 0.02512    & 0.026 & 0.0065 & 0.169  & 0.201\\
					15 & BBH:1369 & $(2,0, 0)$& $-0.06$ & 0.2053 & 0.04263738 & 0.3134     & 0.0173386  & 0.011& 0.0041  & 0.559 & 0.560 \\
					16 & BBH:1370 & $(2,0, 0)$& $+0.12$ & 0.1854 &  0.02422231 &  0.31708  & 0.016779  & 0.07& 0.006 &  0.430 & 0.217\\
					17 & BBH:1371 & $(3,0, 0)$& $+0.92$ & 0.0628 & 0.03263026  & 0.0912     & 0.029058   & 0.12  & 0.006  & 0.179 & 0.115\\
					18 & BBH:1372 & $(3,0, 0)$& $+0.01$& 0.1035 & 0.03273944 & 0.14915      & 0.026070 & 0.06  & 0.006 & 0.105& 0.060\\
					19 & BBH:1373 & $(3,0, 0)$& $-0.41$ & 0.1028 & 0.03666911 & 0.15035    & 0.02529 & 0.0034 &  0.0061 & 0.749 & 0.705 \\
					20 & BBH:1374 & $(3,0, 0)$& $+0.98$ & 0.1956  & 0.02702594 & 0.314   & 0.016938   & 0.067 & 0.0059 & 0.473 & 0.385\\
					\hline
					21 & BBH:89   & $(1,-0.50, 0)$         &  $\dots$  & 0.0469  & 0.02516870 & 0.07194    & 0.01779  & $\dots$ & 0.0025 & 0.214 & 0.0749\\
					22 & BBH:1136 & $(1,-0.75,-0.75)$   &  $-1.90$ & 0.0777  &0.04288969 &0.1209      & 0.02728 & 0.074 & 0.0058 & 0.356 & 0.152\\
					23 & BBH:321  & $(1.22,+0.33,-0.44)$& $+1.47$ & 0.0527  & 0.03239001 &0.07621     & 0.02694 & 0.015  & 0.0045  &0.204 & 0.033\\
					24 & BBH:322  & $(1.22,+0.33,-0.44)$& $-2.02$  & 0.0658  &  0.03396319 &0.0984       & 0.026895 & 0.016 & 0.0061 & 0.203 & 0.0486\\
					25 & BBH:323  & $(1.22,+0.33,-0.44)$& $-1.41$ & 0.1033  & 0.03498377 &0.1438      & 0.02584 & 0.019 & 0.0058 &  0.131 & 0.0745\\
					26 & BBH:324  & $(1.22,+0.33,-0.44)$& $-0.04$ & 0.2018  & 0.02464165 &0.29425        & 0.01894 & 0.098 & 0.0058 & 1.209 & 0.671\\
					27 & BBH:1149 & $(3,+0.70,+0.60)$  &  $+3.00$  & 0.0371  &0.03535964 &$0.06237$   & $0.02664$ &0.025  & 0.005 & 0.660  & 1.166\\
					28 & BBH:1169 & $(3,-0.70,-0.60)$    &  $+3.01$ & 0.0364  &0.02759632 &$0.04895$     & $0.024285$ & 0.033 & 0.004& 0.178 &  0.129  %
					
				\end{tabular}
			\end{ruledtabular}
		\end{center}
	\end{table*}
\begin{figure}[t]
	\center
	\includegraphics[width=0.45\textwidth]{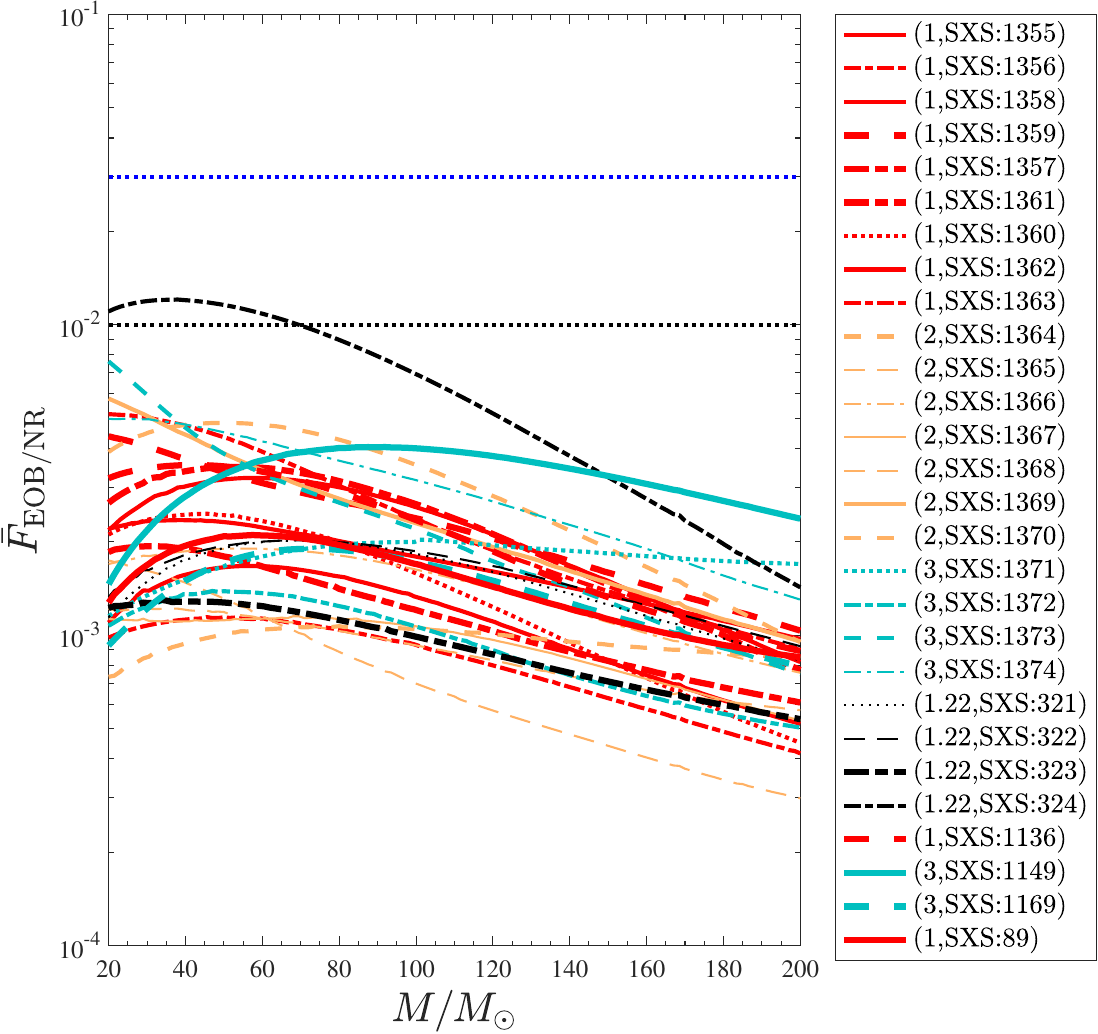}
	\caption{\label{fig:barF_ecc}EOB/NR unfaithfulness for the $\ell=m=2$ mode computed over the
	28 eccentric inspiral SXS simulations publicly available of Table~\ref{tab:SXS}. The horizontal lines mark the
	$0.03$ and $0.01$ values. $\bar{\F}^{\rm max}_{\rm EOBNR}$ is always below $0.01$, although
	the performance slightly degrades for large, positive, spins, consistently with the quasi-circular limit behavior.}
\end{figure}

\subsubsection{Higher waveform multipoles}
\label{sec:higher_modes}
Although the main focus of this work lies in the examination of the quadrupolar $(\ell, |m|) = (2,2)$ mode,
\TEOBResumS{} can be employed to generate waveforms encompassing higher modes as well. 
In the realm of inspiral-to-merger-only waveforms, the model can compute waveforms containing modes up to $\ell=m=8$, 
while the complete merger-ringdown (IMR) phase is accessible for $(2,|1|), (3,|3|),$ and $(4,|4|)$ modes, 
utilizing the fits described in~\cite{Pompili:2023tna, Nagar:2023zxh}. See in particular 
Ref.~\cite{Nagar:2023zxh} for a description of the modifications needed within the \TEOBResumS{} 
framework to use the NR-informed ringdown fits of Ref.~\cite{Pompili:2023tna}.
The careful reader will notice that the full content of IMR 
higher modes in this version of \TEOBResumS{} is comparatively lower than its quasi-circular 
counterpart, {\tt TEOBResumS-GIOTTO}~\cite{Nagar:2023zxh}. 
This discrepancy arises from the intricacies of modeling binaries on non-circularized orbits, 
where the conventional strategy employed for transitioning between inspiral-plunge and merger 
phases faces challenges.  EOB models designed for quasi-circular orbits incorporate 
next-to-quasi-circular NR-informed corrections (NQCs) in the waveform. These corrections account 
for non-circular effects during plunge, ensuring a seamless connection between pre- and post-merger waveforms.
In contrast, when the binary system is non-circularized from the outset, this conventional strategy must be 
reevaluated. Reference~\cite{Nagar:2021gss} introduced a sigmoid function to smoothly eliminate the 
non-circular Newtonian prefactor, that might become inaccurate close to merger, and progressively 
activate the NR-informed NQCs. While this method has proven simply effective for the $(2,2)$ mode,
for higher modes its interplay with the so-called NQC basis might generate inaccuracies around the
waveform peak in some regions of the parameter space. While an in-depth characterization of these
effects lies beyond the scope of this work here we present a preliminary, simple improvement to the $(2,1)$ 
mode that allows us to increase the NR faithfulness of the multipolar model.
Let us briefly review some basic information about NQC corrections within the present context. We address
the reader to Sec.~IIB of Ref.~\cite{Nagar:2021gss}.
The EOB multipolar NQC corrections to the waveform are formally given by
\be
\hat{h}_\lm^{\rm NQC}=\left(1+a_1^\lm n_1^\lm + a_2^\lm n_2^\lm \right) e^{i(b_1^\lm n_3^\lm + b_2^\lm n_4^\lm)} \ ,
\ee
where $\{a_{i}^\lm, b_{i}^\lm\}$ are coefficients determined following the procedure detailed in 
e.g. Ref.~\cite{Damour:2012ky}, and $\{ n_1^\lm, \dots, n_4^\lm \}$ are functions depending 
on the EOB dynamical variables. Similar to other EOB building blocks, there is some freedom in choosing
these functions, given their effective nature. So far, the choice of $n_i$'s implemented in the {\tt Dal\'i} model
differs slightly from those used in the {\tt GIOTTO} model and is the one detailed at the end of Sec.~IIB 
of Ref.~\cite{Nagar:2021gss}. Using these functions, we find that for some configurations, characterized
by large, positive spins, the NQC correction to the frequency evolution presents an unphysical repentine
increase during the late inspiral.
We find that this unwanted behavior is easily cured by using the following $n_4$ function
\be
n_4^{21} = r\Omega p_{r_*} \ ,
\ee
instead of $p_{r_*}/(r\Omega^{1/3})$ previously implemented. This simple modification allows us to obtain 
a more correct frequency evolution for the considered mode.

After performing this improvement, we compute the EOB/NR surrogate unfaithfulness considering higher 
modes for the same configurations considered in Fig.~\ref{fig:barF_circ_sur}. 
Following the same procedure as Ref.~\cite{Nagar:2023zxh}, we fix the inclination angle 
to $\iota=\pi/3$ and minimize the unfaithfulness over the sky-position of the binary.
Results are shown in Fig.~\ref{fig:higher_modes}. 
The EOB/NR unfaithfulness obtained with the generic-orbits model is overall consistent with the one computed with 
the quasi-circular model, though characterized by longer tails towards larger values of mismatch. 
Such tails can reach up to $\sim 10\%$ for binaries with large, positive spins, although the model is 
NR faithful to more than $1\%$ for a large portion of the parameter space.

\begin{table*}[t]
	\caption{\label{tab:chi_scattering}
		Comparison between EOB and NR scattering angle for nonspinning binaries using
	either $\rho_{22}^{\rm 3^{+2}PN}$ , the  resummed (analytical) $\rho_{22}^{\rm 4PN}$ or the NR-tuned one. The NR values 
	are taken from the nonspinning configurations from Ref.~\cite{Damour:2014afa} and Ref.~\cite{Hopper:2022rwo}.}
	  \begin{center}
		\begin{ruledtabular}
   \begin{tabular}{c c c c cc c| c c c| c  c c} 
   $\#$  & $r^{3^{+2}\rm PN}_{\rm min}$ & $r^{4\rm PNan}_{\rm min}$  & $r^{4\rm PNnr}_{\rm min}$& $E_{\rm in}^{\rm NR}/M$ & $J_{\rm in}^{\rm NR}/M^2$ & $\chi^{\rm NR}$ & 
   $\chi^{\rm EOB}_{3^{+2}\rm PN}$ & $\chi^{\rm EOB}_{\rm 4PN-Analytic}$ & $\chi^{\rm EOB}_{\rm 4PN-NRTuned}$& $[\%]$ 
   & $[\%]$ & $[\%]$\\
   \hline
   1 & 3.430 & 3.375   &  3.408   & 1.0225555(50) &  1.099652(36) & 305.8(2.6)  &  315.94 & 346.83  & 326.79&3.31 & 13.42 & 6.86\\
   2 & 3.760 & 3.738   &  3.751    &  1.0225722(50) &1.122598(37)  & 253.0(1.4)   &  258.54    & 265.87  & 261.06  & 2.19  &5.09& 3.18\\
   3 & 4.059 & 4.050    & 4.057   & 1.0225791(50)&1.145523(38) & 222.9(1.7)   &  225.25   &227.85 & 225.95 &1.05   &2.22 & 1.37\\
   4 & 4.862 & 4.862    &  4.863  & 1.0225870(50) &  1.214273(40)& 172.0(1.4)   & 171.62  & 171.77& 171.51 & 0.22&0.13 & 0.28\\
   5 & 5.352 & 5.353    &  5.353  &1.0225884(50)&1.260098(41) &152.0(1.3)   & 151.31     & 151.27 &  151.18 & 0.45&0.48 & 0.54 \\
   6 & 6.503 & 6.504     & 6.504   & 1.0225907(50)& 1.374658(45)&  120.7(1.5)  &  119.99  & 119.92 & 119.92& 0.58&0.64 & 0.64\\
   7 & 7.601  &7.602     &  7.602& 1.0225924(50)&  1.489217(48)& 101.6(1.7)  &  101.09 & 101.05 &101.05& 0.49 & 0.54 & 0.53\\
	8 & 8.675 & 8.675  & 8.675   & 1.0225931(50) &  1.603774(52)& 88.3(1.8)    &  87.98   & 87.95 & 87.96& 0.36 & 0.39 & 0.39\\
	9 &  9.735 & 9.735  & 9.735    & 1.0225938(50)&1.718331(55) &78.4(1.8)    &  78.18   &78.16 &78.16& 0.28 & 0.30 & 0.30\\
   10 & 10.788& 10.789   & 10.788 & 1.0225932(50)& 1.832883(58)&  70.7(1.9)    &  70.50      &70.49 &70.49&0.28 & 0.30 & 0.29\\
   \hline
   11 & 3.02  & $\dots$   & 2.97 & 1.035031(27)& 1.1515366(78)& 307.13(88)  &  338.0382  & {\it plunge} &  393.73 &10.06& $\dots$  & $28.2$ \\
   12 & 3.91 & 3.90  & 3.91& 1.024959(12) & 1.151845(12)&225.54(87) &  230.0844  &  234.04 & 231.37 & 2.01 & 3.77 & 2.58\\
   13 & 4.41 & 4.41  & 4.41& 1.0198847(82)& 1.151895(11)& 207.03(99) &  207.5565  &  208.43 &  207.6076 & 0.26& 0.68 &0.28\\
   14 & 4.99 & 4.99  & 4.99 & 1.0147923(76) & 1.151918(16)&195.9(1.3) &  194.6248  & 194.6735 & 194.4233 &  0.67 & 0.64& 0.77\\
   15 & 6.68 & 6.68   & 6.68  &  1.0045678(42) & 1.1520071(73) &201.9(4.8)   & 200.1620    & 199.9873 & 200.0012 & 0.87 &  0.95 & 0.94
   \end{tabular}
\end{ruledtabular}
\end{center}
\end{table*}
\begin{table*}[t]
	\caption{\label{tab:chi_scattering_spin}Comparison between EOB and the (average) NR scattering angle 
	for some of the equal-mass, spin-aligned, configurations of Ref.~\cite{Rettegno:2023ghr}. 
	All datasets share the same initial angular momentum $J_{\rm in}^{\rm NR}/M^2=1.14560$. The EOB angles
	are calculated either using the (resummed) analytical $\rho_{22}^{\rm 4PN}$ or the effective 4PN NR-tuned 
	one, with the corresponding values of $a_6^c$ and $c_3$. Note that for large values of the (anti)-aligned spins
	the EOB dynamics plunges instead of scattering.}
	  \begin{center}
		\begin{ruledtabular}
   \begin{tabular}{c c c c  c  c c |c c c c} 
   $\chi_1$  & $\chi_2$ & $\tilde{a}_0$ & $E_{\rm in}^{\rm NR}/M$ & $r^{4\rm PNan}_{\rm min}$  & $r^{4\rm PNnr}_{\rm min}$ & $\chi^{\rm NR}$ &  $\chi^{\rm EOB}_{\rm 4PN-Analytic}$ & $\chi^{\rm EOB}_{\rm 4PN-NRTuned}$& $\Delta\chi_{\rm 4PNan}^{\rm EOBNR}[\%]$  & $\Delta\chi_{\rm 4PNnr}^{\rm EOBNR}[\%]$ \\
   \hline
 $-0.3$     & $-0.30$    & $-0.30$    & 1.022690   &  $\dots$ & $\dots$ & $plunge$ & $plunge$   & $plunge$ & $\dots$ & $\dots$\\
 $-0.25  $ &  $-0.25$ & $-0.25$  &1.022680  &  $\dots$ & $\dots$&  367.55     &  $plunge$ & $plunge$   &    $\dots$ & $\dots$\\
$-0.23  $ &$-0.23$ & $-0.23$  &1.022670   & $\dots$  & $\dots$& 334.35     &  $plunge$    & $plunge$  &  $\dots$ & $\dots$\\
$-0.20 $ &$ -0.20$ & $-0.20$  &1.022660   & $3.46$  & 3.50 &303.88      & 386.9102  & 352.5517    &   27.32 & 16.02 \\
$-0.15 $ &$ -0.15$ & $-0.15$  &1.022650   & $3.65$  & 3.68 &272.60      & 305.6974  & 294.9987  &    12.14 &8.22\\
$-0.10 $ &$ -0.10$ & $-0.10$  &1.022650   & $3.80$   &3.82 &251.03      & 269.0546  &  263.6445 &    7.18 &5.03\\
$-0.05 $ &$ -0.05$ & $-0.05$  &1.022640   & $3.93$    &3.94 &234.57      & 245.3143  & 242.1832  &    4.58 &3.25\\
$+0.00 $ & $0.0$ & $0.0$  &1.022640 & $ 4.04$  &4.05 & 221.82    &   228.1024 & 226.1822  &     2.83 &1.97\\
$+0.10 $ &$ +0.10$ & $+0.10$  &1.022650  & $ 4.24$  &4.24 & 202.61    &   203.7849 & 203.0811  &     0.58 &0.23\\
$+0.20 $ &$ +0.20$ & $+0.20$  &1.022660  & $ 4.40$  &4.40 & 187.84    &   186.8409 & 186.7207  &     0.53&0.59\\
$+0.20 $ &$ -0.20$  &$0.0$  &1.022660  &  $4.04$   &4.05 &221.82     &  228.1338  & 226.2067 &     2.85 &1.98\\
$+0.30 $ &$ +0.30$ & $+0.30$  &1.022690  & $ 4.53$   &4.53 &176.59    &   174.0689  & 174.2778 &     1.43& 1.31\\
$+0.40 $ &$ +0.40$ & $+0.40$  &1.022740  &  $4.65$  &4.65  &167.54    &   163.9378  & 164.3545 &    2.15 &1.90\\
$+0.60 $ &$ +0.60$ & $+0.60$  &1.022880  &  $4.85$  &4.84  &154.14    &   148.6040  &149.3273 &     3.59&3.12\\
$+0.60 $ & $0.0$ & $+0.30$ &1.022760  &  $4.53$   &4.53 &177.63    &   174.2648  &  174.2686 &    1.89 & 1.89\\
$+0.70 $ &$ -0.30$ & $+0.20$  &1.022840  & $ 4.38$   &4.38 &190.41    &   187.2741  &  186.7755 &    1.65 &1.91 \\
$+0.80 $ &$ -0.80$ & $0.0$  &1.023090  &  $4.01$   &4.02 &221.68    &   229.5845  &  227.4832 &    3.57 & 2.62\\
$+0.80 $ &$ -0.50$ & $+0.15$  &1.022940  &  $4.30$   &4.30 &198.99    &   195.3268  & 194.4505  &    1.84 &2.28\\
$+0.80 $ &$ +0.20$ & $+0.50$ &1.022880  &  $4.75$   &4.75 &162.07    &   155.7832  & 156.1319  &    3.88&3.66\\
$+0.80$ &$ +0.50$ & $+0.65$  &1.022950  &  $4.89$   &4.88 &152.30    &   145.5650 & 146.3084   &     4.42&3.94\\
$+0.80$ &  $+0.80$ &$+0.80$ & 1.023090 &   $5.00$   &4.99 &145.36    &   137.3641 &  139.1984 &     5.50&4.24\\
   \end{tabular}
\end{ruledtabular}
\end{center}
\end{table*}

\subsection{Eccentric inspirals}
\label{sec:ecc}

Let us now consider the performance of the model for mildly eccentricy bound systems.
Figure~\ref{fig:barF_ecc} shows the EOB/NR unfaithfulness
versus $M$ computed with the 28 SXS simulations of eccentric inspirals currently publicly 
available~\cite{Hinder:2017sxy}. In spite of these datasets being rather old, to date these
remain the only SXS data available for non-circular orbits spanning a considerable number of orbits. 
Other eccentric NR waveforms do exist, e.g. from the RIT~\cite{Gayathri:2020coq} and MAYA catalogs~\cite{Ferguson:2023vta}, 
but are typically shorter.
The properties of the datasets considered are collected in Table~\ref{tab:SXS}.
Following previous works, when performing EOB/NR comparisons it is necessary to tune two
parameters -- the initial frequency at  apastron $\omega_0$ and initial nominal eccentricity $e_0$ -- to correctly match 
the EOB and NR inspirals. 
This is required, in our case, because for simplicity the EOB dynamics is always started at apastron, with zero initial radial momentum.
This choice is consistent with previous works of this lineage, from the very first development of an eccentric
model within the \TEOBResumS{} framework~\cite{Chiaramello:2020ehz}.
Notably, similar coverage of the parameter space can be obtained
by fixing the initial frequency, and allowing the initial (true or mean) anomaly\footnote{We remind the reader
that, for a given eccentricity and semi-latus rectum, anomalies uniquely identify the position of the bodies in the elliptic orbit. 
The inversion points (i.e., apastron and periastron) are characterized by zero initial radial momentum, while a generic point
on the orbit needs not follow this requirement, and may have nonzero initial radial momentum.} to vary. 
This is, for example, the choice made in Ref.~\cite{Ramos-Buades:2023yhy}. As also pointed out in this reference, (i) starting the eccentric inspiral 
at the apastron and varying on initial frequency and eccentricity is {\it equivalent} to (ii) starting the eccentric
inspiral at fixed initial frequency and varying on eccentricity and anomaly. Both choices entail a complete 
coverage of the parameter space, though the approach (ii) is intuitively closer to what usually done for 
quasi-circular binaries. In Appendix~\ref{sec:eccentric_ics} we discus the implementation of the anomaly and a description of 
initial data that is close, though different, to the one of Ref.~\cite{Ramos-Buades:2023yhy}. However, 
for consistency with previous work, we here keep giving initial data at apastron.
Figure~\ref{fig:barF_ecc} shows the EOB/NR unfaithfulness versus $M$. The results improve
with respect to previous work, with $\bar{\F}^{\rm EOBNR}\lesssim 1\%$ all over the dataset sample. The plot
is complemented by Table~\ref{tab:SXS}.

\subsection{Scattering configurations}
\label{sec:scattering}

We conclude this section by considering unbound configurations, and in particular BBH scatterings.
Rather than computing and comparing waveforms, a non-trivial feat from the NR side,
we directly gauge the goodness of the EOB dynamics by performing comparisons of the gauge-invariant EOB and NR scattering angles.
Following standard procedures already adopted
in previous work, we consider the nonspinning configurations of Refs.~\cite{Damour:2014afa,Hopper:2022rwo} (see Tab.~\ref{tab:chi_scattering}) 
and the spinning configurations of Ref.~\cite{Rettegno:2023ghr} (see in Tab.~\ref{tab:chi_scattering_spin}).
We do not perform detailed comparisons for the spinning simulations presented in Ref.~\cite{Hopper:2022rwo} because they
are limited to rather extreme cases, with large energies and spins, and do not present a systematic and 
detailed analysis of the numerical error, which was instead performed in Ref.~\cite{Rettegno:2023ghr}.
Collectively, these
result suggest that the analytical EOB description for spin-aligned binaries becomes less accurate 
in the part of the parameter space that is close to the capture threshold. This is evident for the 
data of Ref.~\cite{Rettegno:2023ghr} listed in Table~\ref{tab:chi_scattering_spin}, where we can
see the sequence as the effective spin $\tilde{a}_0$ is decreased, but the same phenomenology
is present also in the data of Ref.~\cite{Hopper:2022rwo}, although they were not such to 
systematically cover the transition.

Let us finally mention that, for the nonspinning configurations of Table~\ref{tab:chi_scattering}, 
we also list EOB calculations that use the model based on $3^{+2}$~PN Taylor-expanded $\rho_{22}$ function. 
For the most extreme configurations (first rows in Table~\ref{tab:chi_scattering}) the corresponding angles 
are closer to the NR ones than those obtained using the fully analytical 4PN $\rho_{22}$. The fact that a 
model that is {\it less} NR-faithful for quasi-circular configurations is actually {\it more} NR-faithful 
for scattering configurations highlights the difficulty in constructing a model capable of covering 
well all configurations, as well as the delicate interplay between dissipative and conservative effect
in the description of the dynamics. In the next section we will see that it is actually possible to do 
better by carefully improving the description of the radiation reaction using NR information.
The delicacy of the interplay between the various effects is evident by looking at dataset number 11 in
Table~\ref{tab:chi_scattering}: in the $3^{+2}$~PN case we have a scattering (that is, at least,
qualitatively consistent with the NR prediction) while in the 4PN case the system plunges.

\section{Noncircularized waveform model with NR-informed radiation reaction}
\label{sec:tuning_rho22}
\begin{figure}[t]
	\center	
	\includegraphics[width=0.45\textwidth]{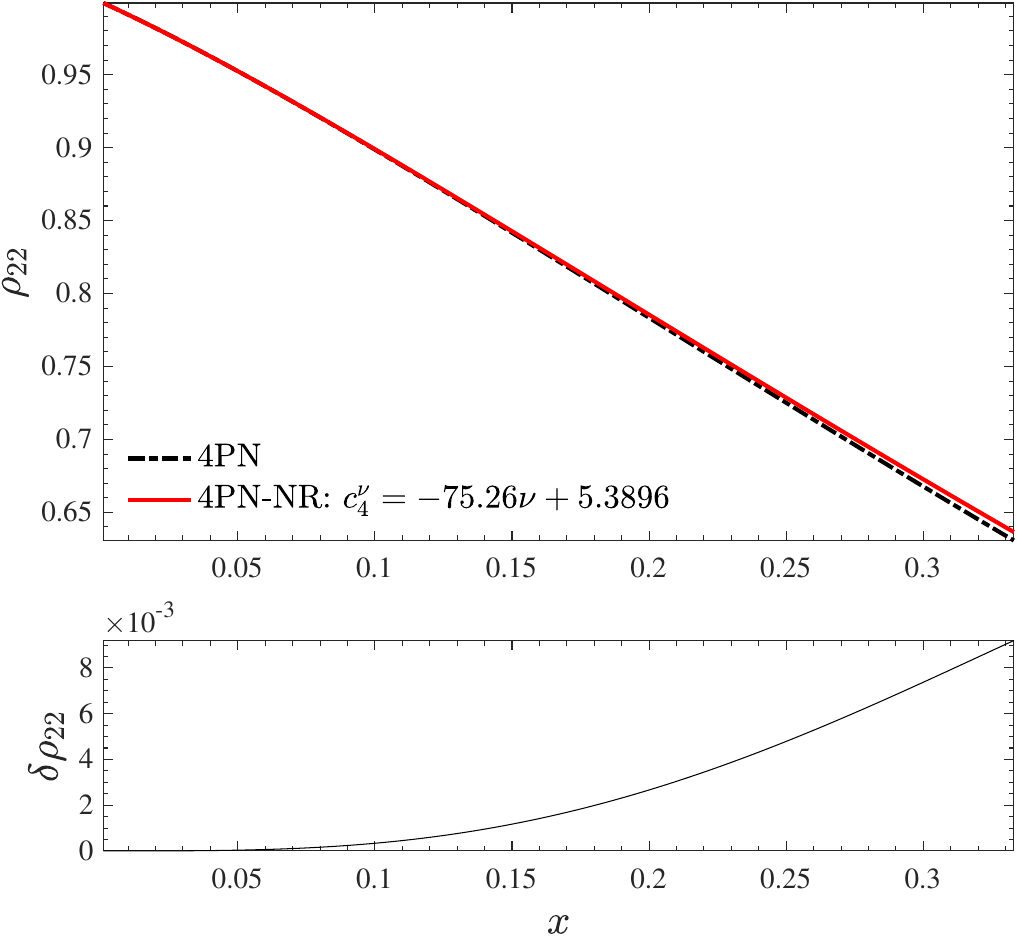}
	\caption{\label{fig:rho22_NR_q1}Informing an effective 4PN term $c_4^\nu$ in $\rho_{22}$ 
	using NR data, Eq.~\eqref{eq:c4_tuned} in the equal-mass case. The bottom panel shows the relative
	difference normalized to the 4PN analytic curve, $\delta \rho_{22}$, that  is ~$0.6\%$
	at $x_{\rm LSO}\sim 0.25$ (see Table~\ref{tab:LSO}). Such a (tiny) difference is actually 
	necessary to lower the EOB/NR unfaithfulness by approximately one order of magnitude, 
	up to~$\sim 10^{-4}$, see Fig.~\ref{fig:phasing_tuned}.}
\end{figure}

\begin{figure}[t]
	\center	
	\includegraphics[width=0.21\textwidth]{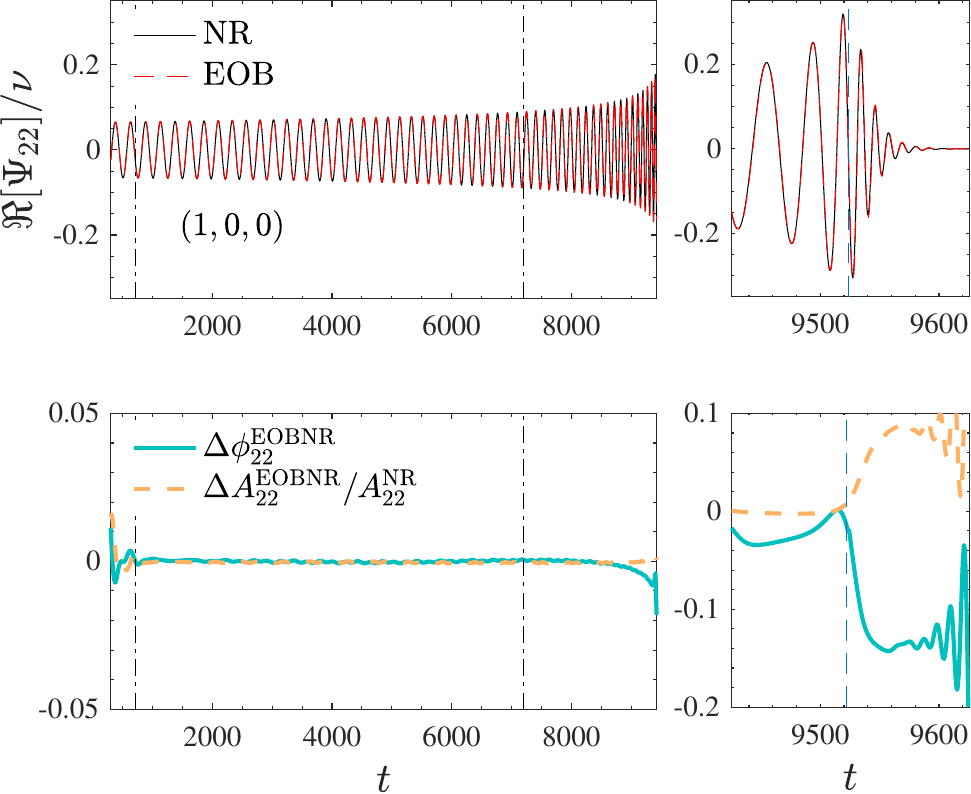}
	\qquad
	\includegraphics[width=0.21\textwidth]{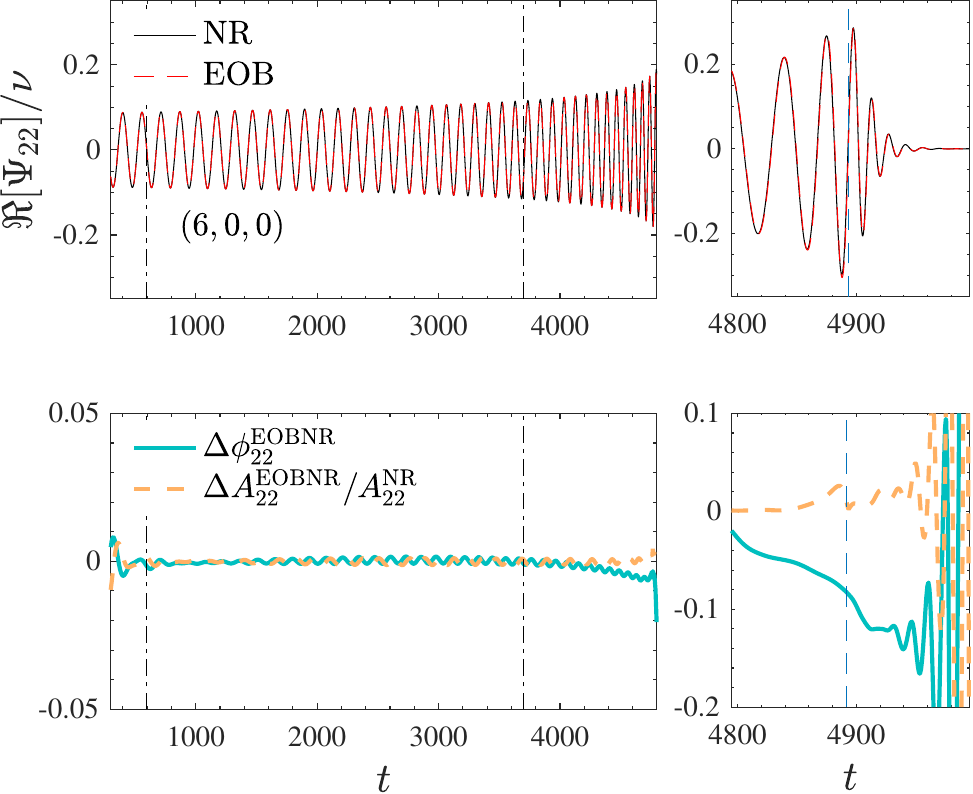}\\
	\vspace{5mm}
	\includegraphics[width=0.21\textwidth]{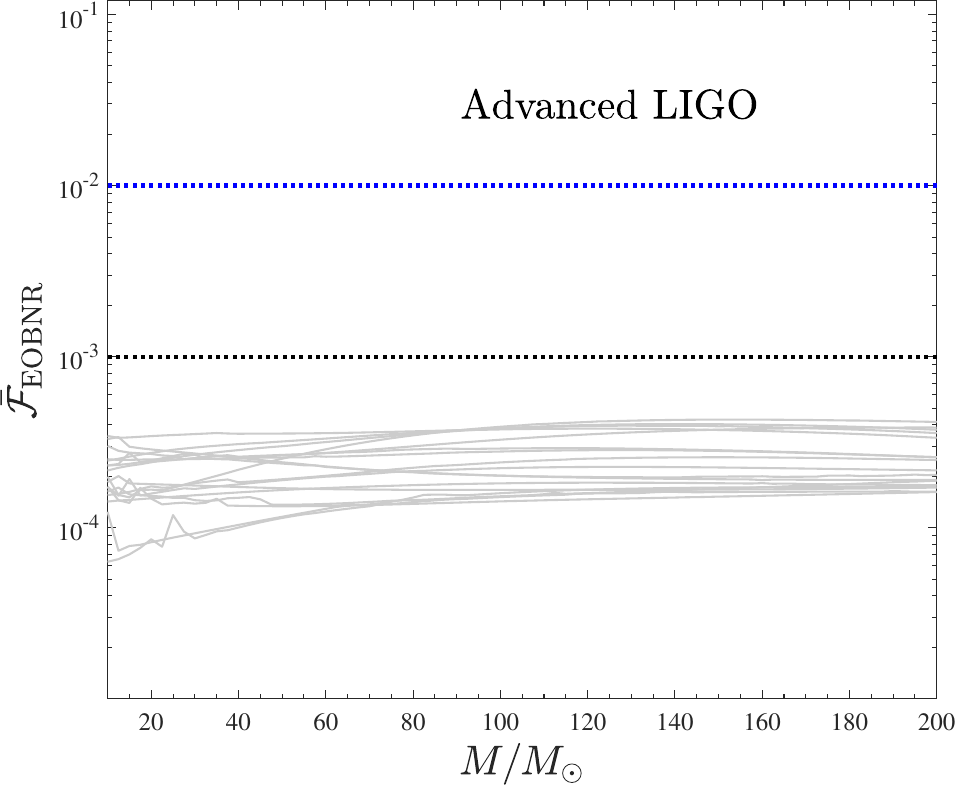}
	\qquad
	\includegraphics[width=0.21\textwidth]{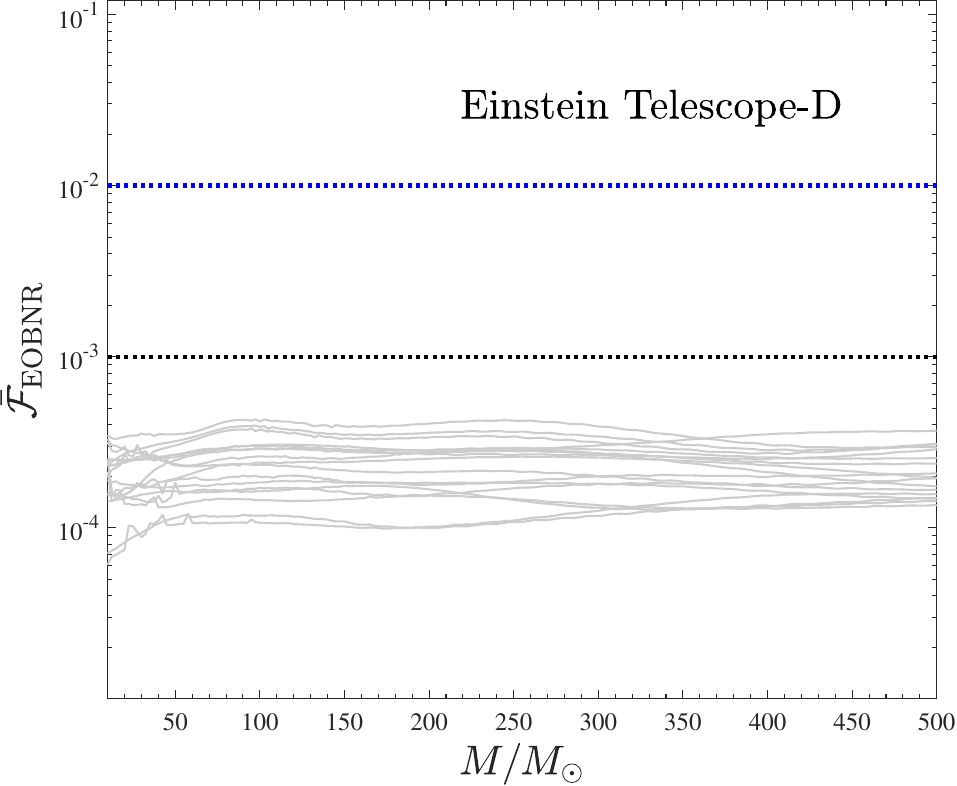}
	\caption{\label{fig:phasing_tuned}Nonspinning case: NR-informing at the same time the 
	EOB interaction potential (via $a_6^c$, Eq.~\eqref{eq:a6c_tuned}) and the resummed
	radiation reaction via $c_4^\nu$, Eq.~\eqref{eq:c4_tuned}. Top panels: two illustrative time-domain phasings.
	Bottom panels: EOB/NR unfaithfulness for all SXS nonspinning datasets available (up to $q=15$) 
	with both the Advanded-LIGO and ET-D~\cite{Hild:2008ng,Hild:2009ns, Hild:2010id} sensitivity designs. Phase differences accumulated 
	at merger are $\lesssim 0.1$~rad, that yield $\bar{\F}_{\rm EOBNR}^{\rm max}\sim 10^{-4}$.}
\end{figure}

\begin{figure*}[t]
	\center	
	\includegraphics[width=0.42\textwidth]{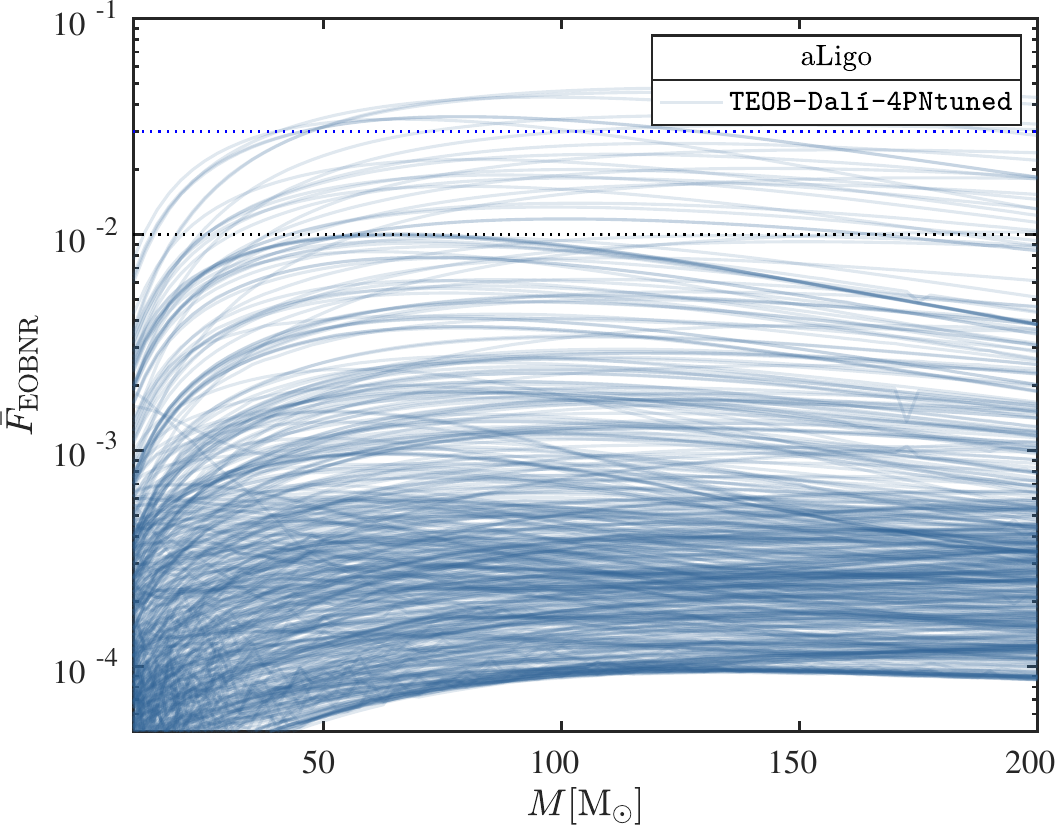}
	\includegraphics[width=0.45\textwidth]{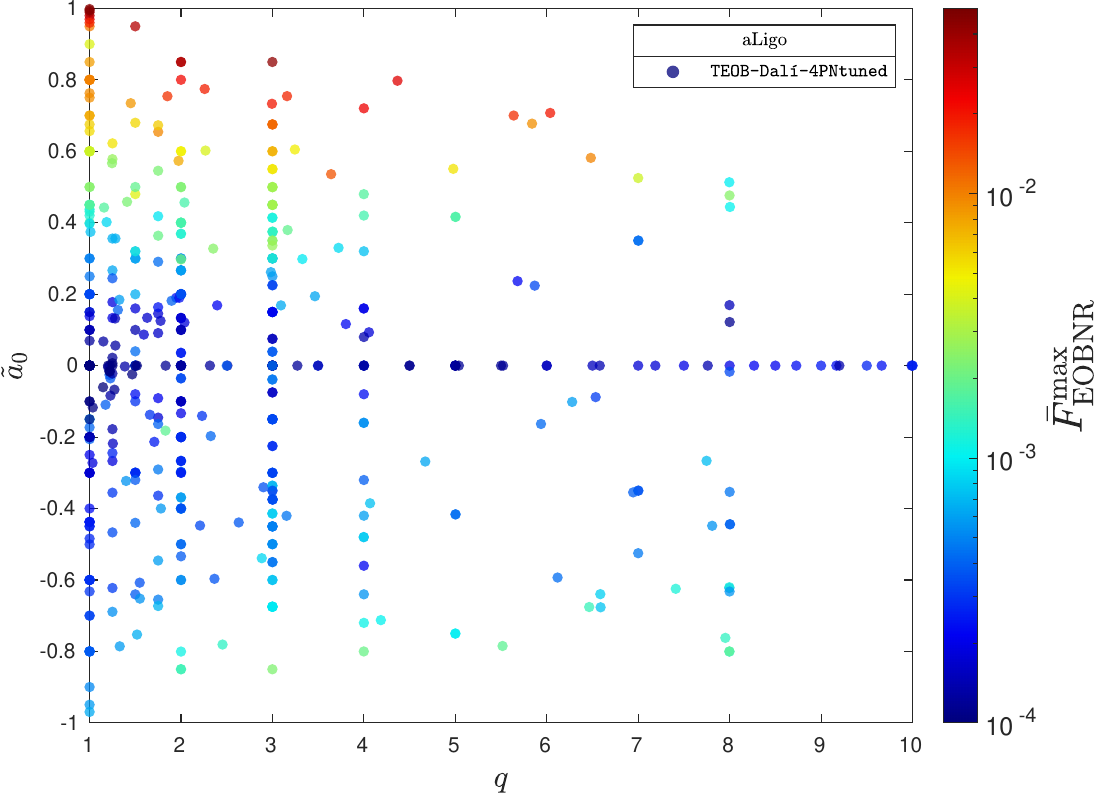}
	\caption{\label{fig:barF_circ_eobnr_4PNtuned}EOB/NR unfaithfulness for the $\ell=m=2$ mode obtained with
	the NR-informed effective 4PN term in $\rho_{22}^{\rm orb}$, the inverse-resummed spin-dependent
	radiation reaction and the consistently obtained fits given by Eqs.~\eqref{eq:a6c_tuned}, ~\eqref{eq:c4_tuned}
	and the second row of Table~\ref{tab:c3_coeff} for $c_3$. Comparing with Fig.~\ref{fig:barF_circ_eobnr} one sees the largely improved
	EOB/NR unfaithfulness for negative and mildly positive spins. By contrast, one finds a loss in accuracy for large, positive spins.
	This is understood as due to an overestimate of the action of the radiation reaction force. See text for discussion.}
\end{figure*}
\begin{figure}[t]
	\center	
	\includegraphics[width=0.42\textwidth]{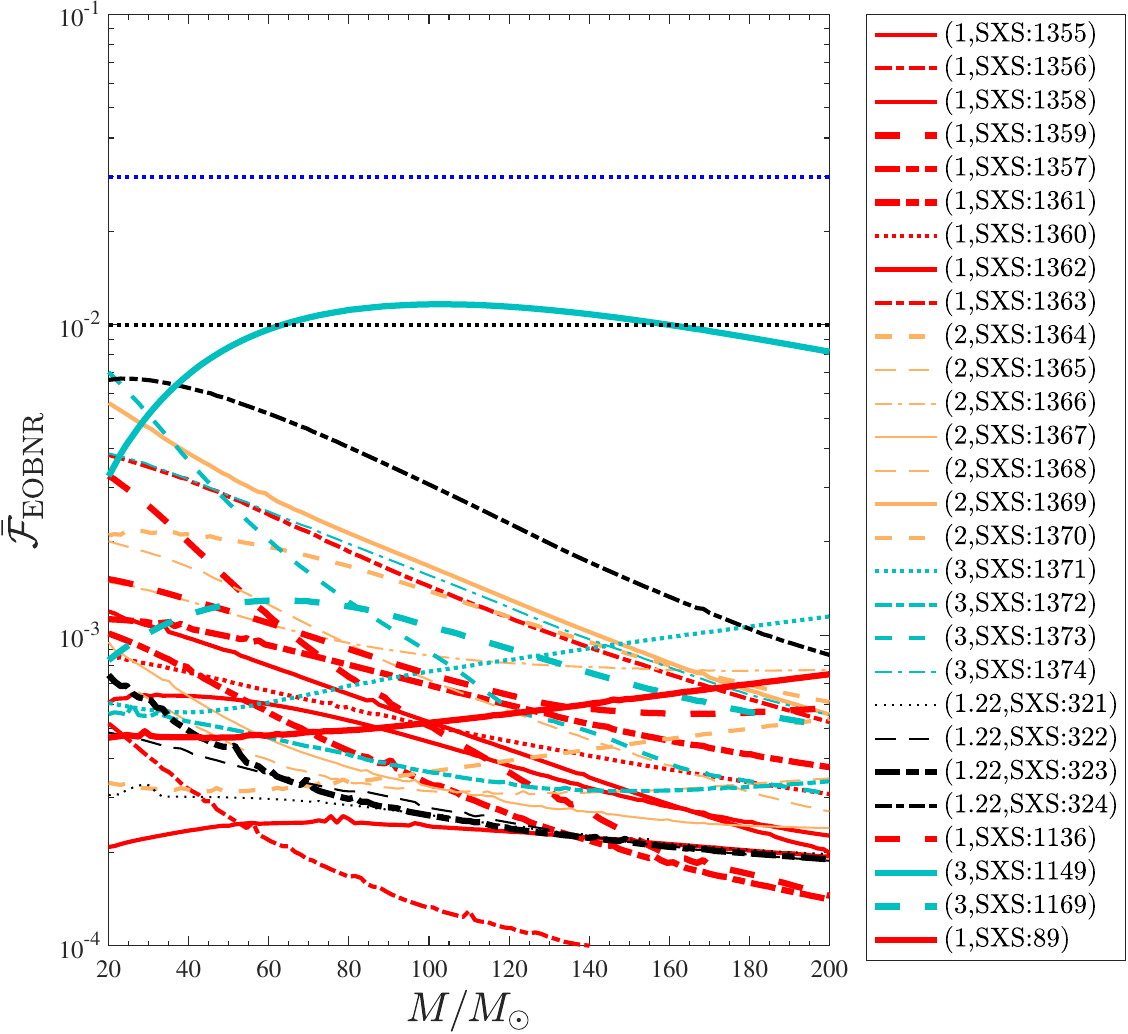}
	\caption{\label{fig:barF_ecc_4PNtuned}EOB/NR unfaithfulness for the $\ell=m=2$ mode computed over the
	eccentric SXS simulations publicly available using the ${\tt TEOB-Dal\'i_{4PN-NRTuned}}$ model, analogous
	of Fig.~\ref{fig:barF_ecc} above. The unfaithfulness is much smaller except for a single
	outlier $\sim 1\%$. This configuration has $\tilde{a}_0=0.675$ and $\bar{\F}_{\rm EOBNR}^{\rm max}$,
	is consistent with the quasi-circular value, $\sim 0.01$, at $(3,0.675)$, as illustrated by the 
	right panel of Fig.~\ref{fig:barF_circ_eobnr_4PNtuned}.}
\end{figure}
\begin{figure}[t]
	\center
	\includegraphics[width=0.23\textwidth]{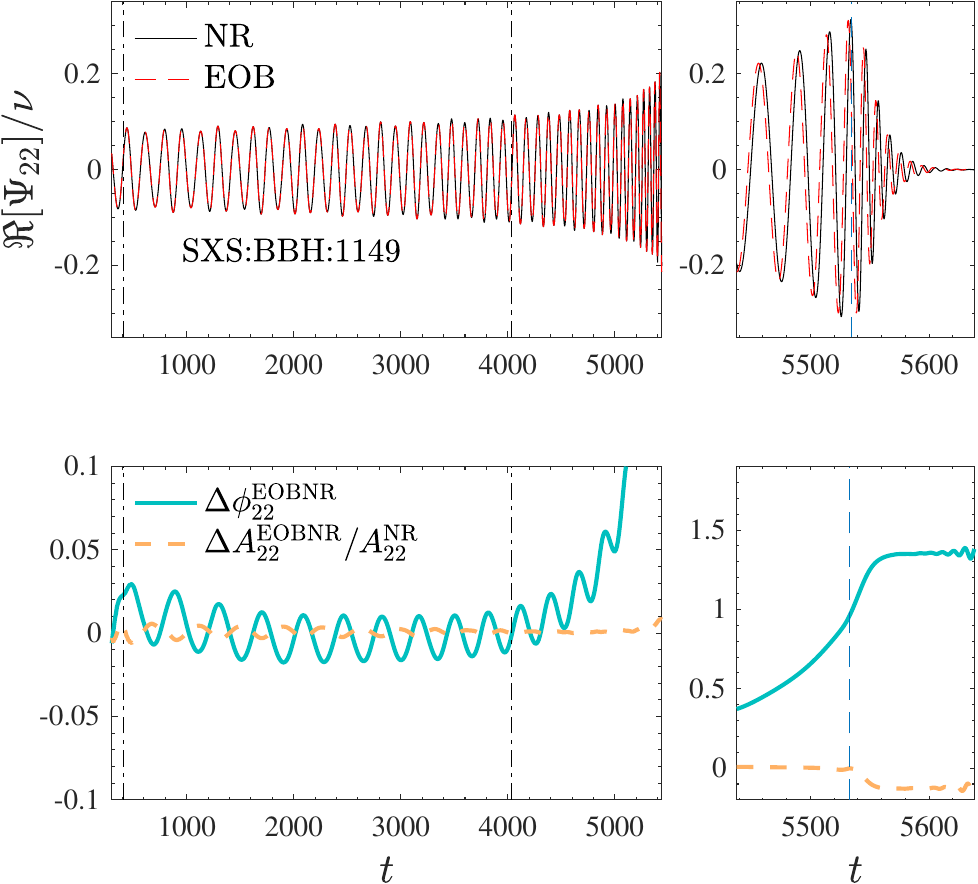}
	\includegraphics[width=0.245\textwidth]{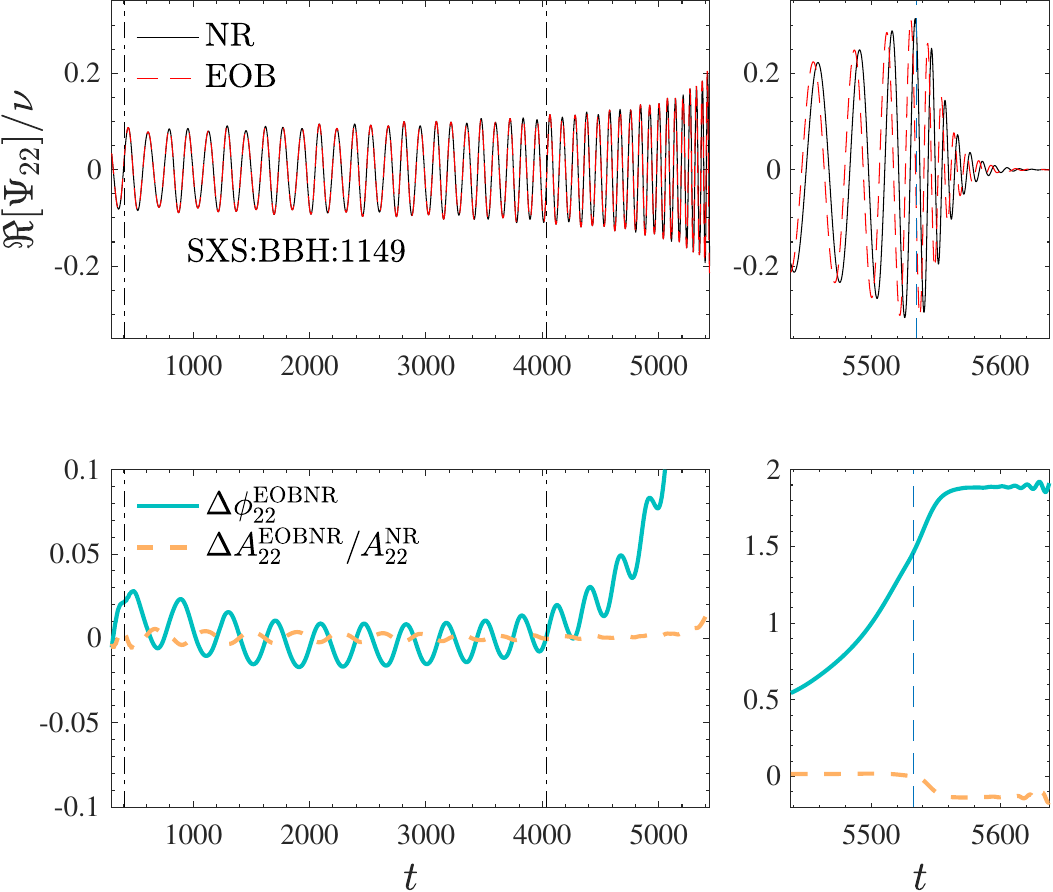}
        \caption{\label{fig:phasing_eccentric}Eccentric case: understanding the unfaithfulness of Figs.~\ref{fig:barF_ecc} 
        and~\ref{fig:barF_ecc_4PNtuned} in terms of time domain phasing for SXS:BBH:1149, i.e. $(3,+0.70,+0.60)$.
        The left panel corresponds to \daliAN{} and the right panel to \daliNR{}. The worsening of 
        the $\bar{\F}_{\rm EOBNR}^{\rm max}$ from $4\times 10^{-3}$ (in Fig.~\ref{fig:barF_ecc}) to $\sim 10^{-2}$ in Fig.~\ref{fig:barF_ecc_4PNtuned}
        is due to the (approximate) doubling of the phase difference at merger related to the improvable spin sector of  \daliNR{}. 
        See Sec.~\ref{sec:discussion} for discussion.}
\end{figure}
	
The analysis we have carried out so far has highlighted the importance of the analytical
choice made for the radiation reaction. In particular, it has shown that its effect cannot be
completely absorbed/corrected by NR tuning $a_6^c$. We have two models with different
performances versus NR waveforms. From the above discussion, it seems clear that the 
$3^{+2}$PN $\rho_{22}$ is too large and entails an incorrect phase acceleration, 
with a positive phase difference accumulated with respect to the NR waveform up to merger. 
On the other hand, the $P^2_2(\rho_{22}^{\rm 4PN})$ function, which is smaller, yields an accumulated 
phase difference up to merger is negative and nonnegligible.
On the basis of this analysis we thus expect that a function that is slightly larger than 
$P^2_2(\rho_{22}^{\rm 4PN})$ might succeed in improving the EOB/NR phasing agreement
up to the $0.01$~rad level during the latest orbits before the beginning of the plunge.
As a first attempt, we took $\rho_{22}$ at $4^{+1}$PN with an effective 5PN parameter
linear in $\nu$ that can be tuned. This is then resummed using either a $(2,3)$ or a $(3,2)$
Pad\'e approximant. Unfortunately we find that in both cases the Pad\'e approximant 
develops a spurious pole, which prevents us from following this route.
As an alternative, we can, instead, still work at 4PN accuracy, but replace the exact 4PN 
$\nu$-dependence with an {\it effective} term of the form $\nu c_4^\nu x^4$ where $c_4^\nu$ is a 
parameter to be determined via EOB/NR comparison. Schematically, the 4PN term in $\rho_{22}$
thus reads $(c_0 + \nu c_4^\nu)x^4$, where now $c_4^\nu$ is a parameter intended to replace
the analytical $\nu$ dependence of the function in Eq.~\eqref{eq:rho22}. For consistency with
our previous choice, we then take the $(2,2)$ Pad\'e approximant, that now depends on $c_4^\nu$.
It turns out that it is easy to tune $c_4$ to reduce the dephasing in the last part of the
inspiral; similarly, one can additionally tune $a_6^c$ so to adjust the phase difference 
through late plunge, merger and ringdown, so to to have it negative and monotonically 
decreasing\footnote{We remind the reader that the idea of NR-informing at the same time the 
conservative and nonconservative part of the dynamics is not new as it dates back to some
pioneering EOB/NR works~\cite{Damour:2007yf,Damour:2007vq,Damour:2008te}. 
In particular, note that Ref.~\cite{Taracchini:2012ig} already explored the possibility 
to NR-inform an effective 4PN corrections to the waveform and radiation reaction.}.
By iteratively tuning both $c_4^\nu$ and $a_6^c$ one eventually finds that the best 
values approximately lie on two straight lines and can be accurately fitted as follows
\begin{align}
\label{eq:a6c_tuned}
a_6^c &= -24.453 - 270.25\nu \ ,\\
\label{eq:c4_tuned}
c_4^\nu    &= 5.3896  -75.26\nu \ .
\end{align}
It is interesting to note that the fractional difference between the NR-tuned $c_4^\nu$ and the analytic one
is at most $\sim 0.6\%$ around the LSO crossing. In the flux, this means $\sim 1.2\%$ fractional difference
between the fluxes. 
Focusing first on the EOB/NR comparisons for nonspinning configurations, Fig.~\ref{fig:phasing_tuned} 
gives us a flavor of the EOB/NR performance that can be achieved this way. In the top panel we show the time-domain
phasing for two illustrative mass ratios, $q=1$ and $q=6$, while the bottom panels 
display $\bar{\F}_{\rm EOBNR}$
for 19 different values\footnote{We consider $1\leq q \leq 10$ with steps of $0.5$ plus the $q=15$ case} of $q$.
The plot shows that an accumulated phase difference $\simeq 0.1$ rad translates in 
$\bar{\F}_{\rm EOBNR}\sim 10^{-4}$, the level of accuracy that we may expect to be needed for 3G detectors.

When considering spinning system, we have to determine a new expression of $c_3$  by EOB/NR phasing 
comparison. In doing so, one quickly realizes that the current implementation of the spin-dependent waveform
terms yields an emission of gravitational radiation (and thus backreaction of the system) that exceeds the
NR prediction: the transition from inspiral to plunge occurs too fast. This points us towards the identification of
systematic inaccuracies {\it also} in this building block of the model, that thus should be modified accordingly. 
As a minimal attempt in this direction, for the $m={\rm even}$ modes up to $\ell=4$ we implement the
orbital-factorized (and resummed) amplitudes introduced in Refs.~\cite{Nagar:2016ayt,Messina:2018ghh}.
Analytical expressions constructed following this approach were found to agree well with the corresponding
numerical data in the test-mass limit, although their potentialities were not explored in full in the comparable
mass case. 
The $\rho_\lm$'s residual amplitudes are written in orbital-factorized form
\be
\rho_\lm = \rho_\lm^{\rm orb}\hat{\rho}_\lm^S \ ,
\ee
and then both functions are resummed. The $\rho^{\rm orb}_\lm$'s are the same considered in
the previous section, i.e. are resummed using Pad\'e approximants. The $\hat{\rho}_{\lm}^S$ 
are instead replaced by their inverse-Taylor resummed expressions, $\bar{\rho}_\lm^S$, 
that are defined as:
\be
\bar{\rho}_\lm^S = \left[T_n\left(\rho_\lm^{-1} \right) \right]^{-1} \ ,
\ee
where $T_n$ indicates the Taylor expansion of order $n$. The $\bar{\rho}_\lm^S$ are then functions 
that formally read:
\begin{align}
\label{eq:bar_rho22_S}
\bar{\rho}_\lm^S &= \bigg(1 + c^\lm_{3/2} x^{3/2} + c^\lm_2 x^2 \nonumber\\
                          &+ c^\lm_{5/2} x^{5/2} + c^\lm_{3} x^{3} + c^\lm_{7/2}x^{7/2}\dots\bigg)^{-1} \ ,
\end{align}
where $x$ is some (squared) velocity PN variables. Here integer powers correspond to terms even 
in the spins, while semi-integer powers to terms that are odd in the spin.
In particular, up to $\ell=4$, the $m=\text{even}$ functions that we consider are explicitly given by:
\begin{align}
\label{eq:bar_rho22_S}
&\bar{\rho}_{22}^S= \Bigg\{1+\left(\dfrac{\tilde{a}_0}{2}+\dfrac{1}{6}X_{12}\tilde{a}_{12}\right)x^{3/2}-\dfrac{\tilde{a}_0^2}{2}x^2\nonumber\\
                          & + \left[\tilde{a}_0\left(\dfrac{337}{252}-\dfrac{73}{252}\nu\right)+X_{12}\tilde{a}_{12}\left(\dfrac{27}{28}+\dfrac{11}{36}\nu\right)\right]x^{5/2} \nonumber\\
                          & + \left[\tilde{a}_0^2\left(\frac{11}{42}+\dfrac{31}{252}\nu\right)+\tilde{a}_1\tilde{a}_2\left(\dfrac{19}{63}-\dfrac{10}{9}\nu\right)-\dfrac{179}{252}X_{12}\tilde{a}_12\tilde{a}_0\right]x^3 \nonumber\\
                          &+\left[\dfrac{2083}{2646}\tilde{a}_0-\dfrac{13}{12}a_0^3 +X_{12}\tilde{a}_{12}\left(\dfrac{\tilde{a}_0^2}{12} + \dfrac{13367}{7938}\right)\right]x^{7/2}\Bigg\}^{-1} , \\
&\bar{\rho}_{32}^S  =\left[1+ \left(-\dfrac{\tilde{a}_0}{3-9\nu}+\dfrac{X_{12}\tilde{a}_{12}}{3-9\nu}\right)x^{1/2}\right]^{-1} \ , \\
&\bar{\rho}_{44}^S  = \left[1 +\left(\dfrac{19}{30}\tilde{a}_0 + \dfrac{1-21\nu}{30-90\nu}X_{12}\tilde{a}_{12}\right)x^{3/2}\right]^{-1} \ ,\\
&\bar{\rho}_{42}^S  = \left[1 +\left(\dfrac{\tilde{a}_0}{30} +\dfrac{19-39\nu}{30-90\nu}X_{12}\tilde{a}_{12}\right)x^{3/2}\right]^{-1} \ .
\end{align}

With this analytic choice, we proceed determining a new expression for $c_3$, with the same functional form discussed
above. The corresponding fitting coefficients are listed in the second row of Table~\ref{tab:c3_coeff}. The model, now dubbed \daliNR{},
is then validated computing the unfaithfulness (using Advanced Ligo sensitivity) with all SXS quasi-circular NR simulations.
The result is reported in Fig.~\ref{fig:barF_circ_eobnr_4PNtuned}. The left panel of the figure shows $\bar{\F}_{\rm EOBNR}$
versus the total mass $M$, while the right panel gives $\bar{\F}^{\rm max}_{\rm EOBNR}$ versus $(\tilde{a}_0,q)$.
This analysis indicates that the tuning of the {\it nonspinning} flux eventually yields an improved EOB/NR agreement
for negative and mild spins, with a global shift of all values towards the $10^{-4}$ goal.
The performance for eccentric configurations is reported in Fig.~\ref{fig:barF_ecc_4PNtuned}. Not surprisingly,
the NR-tuning of the nonspinning radiation reaction allows for a general reduction of the EOB/NR unfaithfulness
even for eccentric bound systems.
We similarly recompute the scattering angle for all configurations previously considered. The corresponding 
values are listed in Tables~\ref{tab:chi_scattering}-\ref{tab:chi_scattering_spin}. Also in this case one sees that
the NR-tuning of the (nonspinning) flux eventually yields an improve agreement between the NR and EOB
scattering angles.

\begin{figure}[t]
	\center	
	\includegraphics[width=0.24\textwidth]{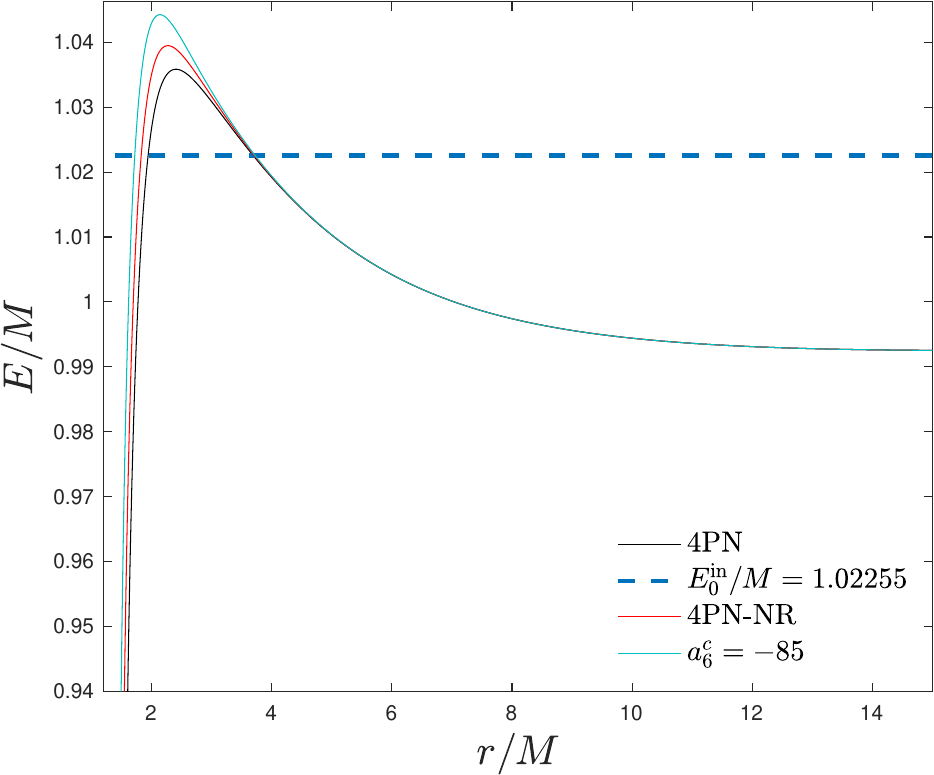}
	\includegraphics[width=0.215\textwidth]{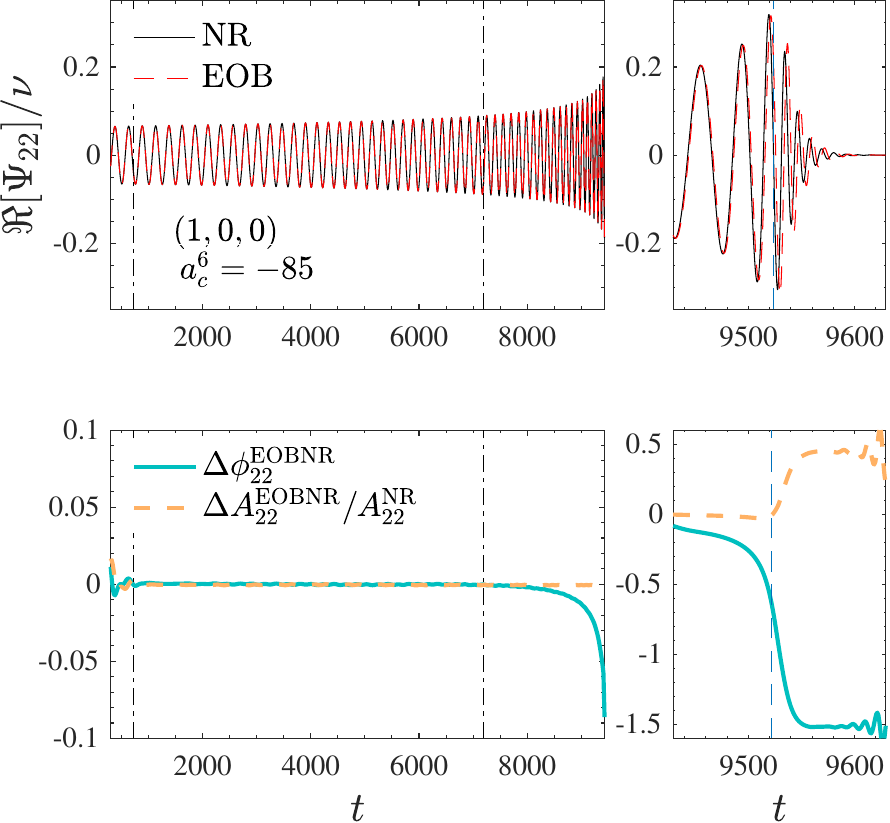}
	\caption{\label{fig:chi_tuned}Configuration $\#1$ in Table~\ref{tab:chi_scattering}, comparing
	different potential energies that yield different values of the scattering angle.}
\end{figure}

To better understand the impact of these changes on the EOB dynamics and put these numbers into perspective, 
it is instructive to observe how the changes in the model reflect on the potential energy. 
The left panel of Fig.~\ref{fig:chi_tuned} shows
$E_{\rm circ}/M=\sqrt{1+2\nu(\hat{E}_{\rm eff}^{\rm circ}-1)}$ where $\hat{E}^{\rm circ}_{\rm eff}=A(1+p_\varphi^2 u^2)$
for configuration $\#1$ in Table~\ref{tab:chi_scattering} for various choices of the potential $A$.
The black line corresponds to $\chi^{\rm EOB}_{\rm 4PN-Analytic}=346.83$, while  the red curve
to $\chi_{\rm 4PN-NRTuned}^{\rm EOB}=326.79$. The smaller value of the scattering angle is due to
the fact that the peak of the potential energy, corresponding to the unstable orbit, is higher. 
By keeping the NR-informed 4PN-like radiation reaction, we find that fixing $a_6^c=-85$, instead
of the value $a_6^c\simeq -92$ coming from Eq.~\eqref{eq:a6c_tuned}, result in an increase of the peak of
the potential energy such to yield for the scattering angle  $\chi_{\rm EOB}=308.76$, i.e.
with approximately $1\%$ fractional difference with the NR prediction $\chi_{\rm NR}= 305.8$.
This shows that it is actually possible to match the NR values consistently with their nominal
error bars by just a fine tuning of the $A$ function (improved EOB/NR agreement is evidently
found also for the other configurations).

\begin{figure}[t]
	\center	
	\includegraphics[width=0.24\textwidth]{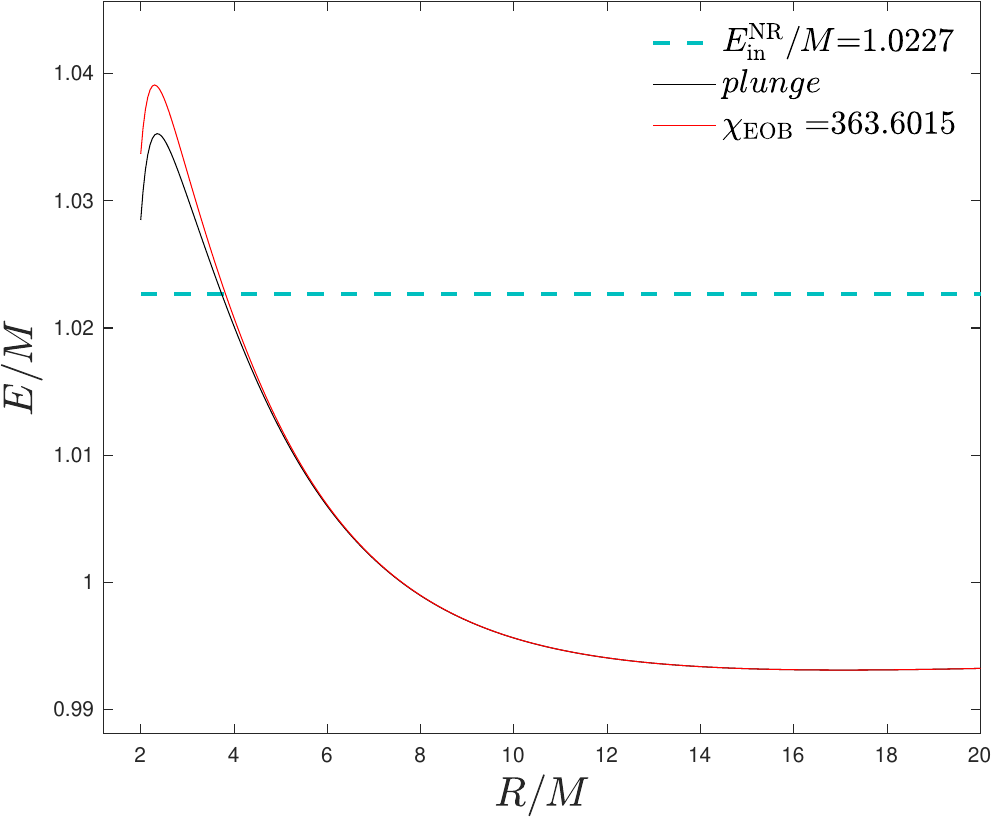  }
	\includegraphics[width=0.225\textwidth]{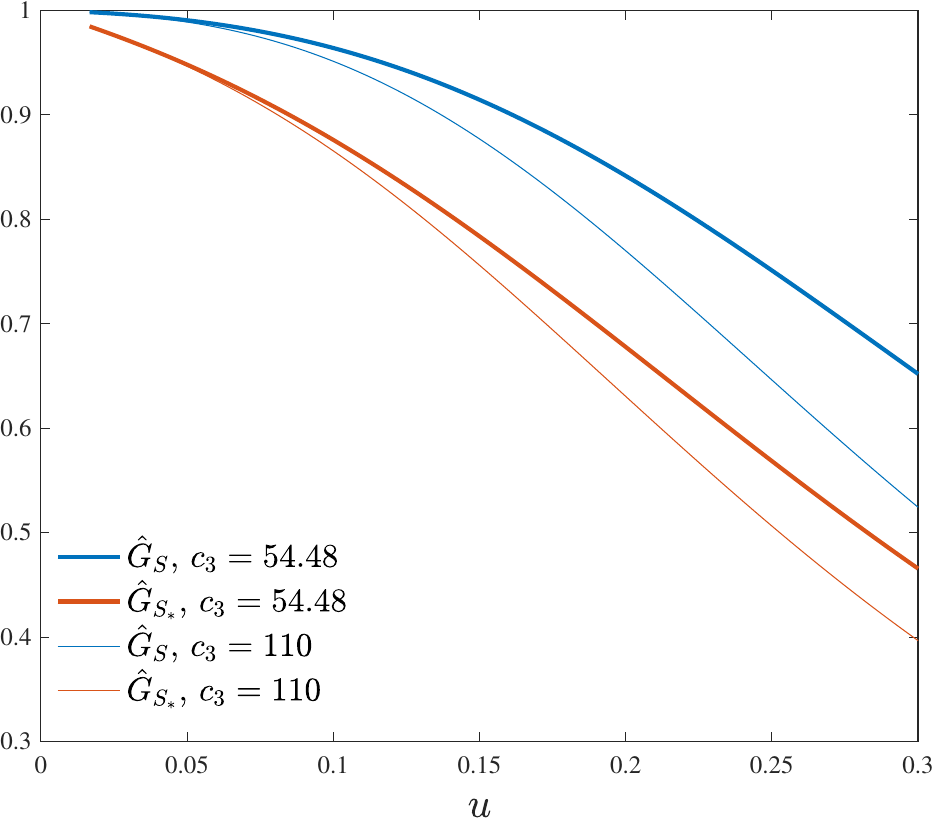}
	\caption{\label{fig:chi_spin}Configuration $\#1$ in Table~\ref{tab:chi_scattering_spin} $(1,-0.25,-0.25)$.
	Left panel: the potential energy. Right panel, the gyro-gravitomagnetic ratios. The value $c_3=110$ corresponds
	to the red curve in the left panel that yields a value of the angle compatible with the NR one. To do so, the
	magnitude of the spin-orbit coupling has to be reduced with respect to standard case 
	(thick line versus thin lines in the right panel).}
\end{figure}

An analogous explanation  holds in the spinning case, as highlighted in Fig.~\ref{fig:chi_spin}.
The figure refers to the second configuration of Table~\ref{tab:chi_scattering_spin}, 
$(1,-0.25,-0.25)$, with the NR-informed 4PN-effective radiation-reaction term. In this case,
the EOB model predicts a plunge, while NR gives $\chi_{\rm NR}=367.55$~deg. Since the 
EOB and NR values in the nonspinning case are rather consistent, with a fractional difference 
$\sim 2\%$, we argue that the spin-sector, though NR-informed by quasi-circular simulations,
might need to be further modified to properly match the NR scattering angle. In principle the effects are
expected to be shared between both the conservative and nonconservative part of the spin-sector
of the model. As a first exploratory step, we only decide to modify the Hamiltonian, looking for a 
value of $c_3$ such to yield an acceptable EOB/NR consistency. This is obtained by fixing $c_3=110$,
that  determines a rise in the peak of the potential such to yield $\chi_{\rm EOB}\simeq 363$. 
This corresponds to a large modification to the normalized gyro-gravitomagnetic 
functions $(\hat{G}_S,\hat{G}_{S_*})$ shown in the right panel of Fig.~\ref{fig:chi_spin}.
Clearly, this value of $c_3$ will not yield an accurate phasing in the quasi-circular case.
This simple analysis thus highlights the complication of finding full consistency between 
the quasi-circular case and configurations that are close to direct plunge. By contrast, the
flexibility (and robustness) of the model is such that each case can be matched accurately
with the tuning of one single parameter. Note that these effects were already pointed 
out in Ref.~\cite{Hopper:2022rwo}, using however configurations with higher values 
of the (negative) spins.
Finally, is worth stressing that the current analysis should be seen as essentially illustrative
and qualitative. A reduction of the EOB/NR disagreement between scattering angles close to
the threshold of capture might be also obtained by modifying other sectors of the model, 
like the radiation reaction or the noncircular part of the conservative dynamics, e.g. 
the $D$ function (see e.g. an exploratory analysis along these lines in Ref.~\cite{Nagar:2020xsk}).
Our findings are just supposed to highlight the delicate interplay of various effects in the
subtle regime around the threshold of immediate merger and will deserve more
dedicated studies in the future.

%

\subsection{Discussion: understanding the results}
\label{sec:discussion}
\begin{figure}[t]
	\center	
	\includegraphics[width=0.42\textwidth]{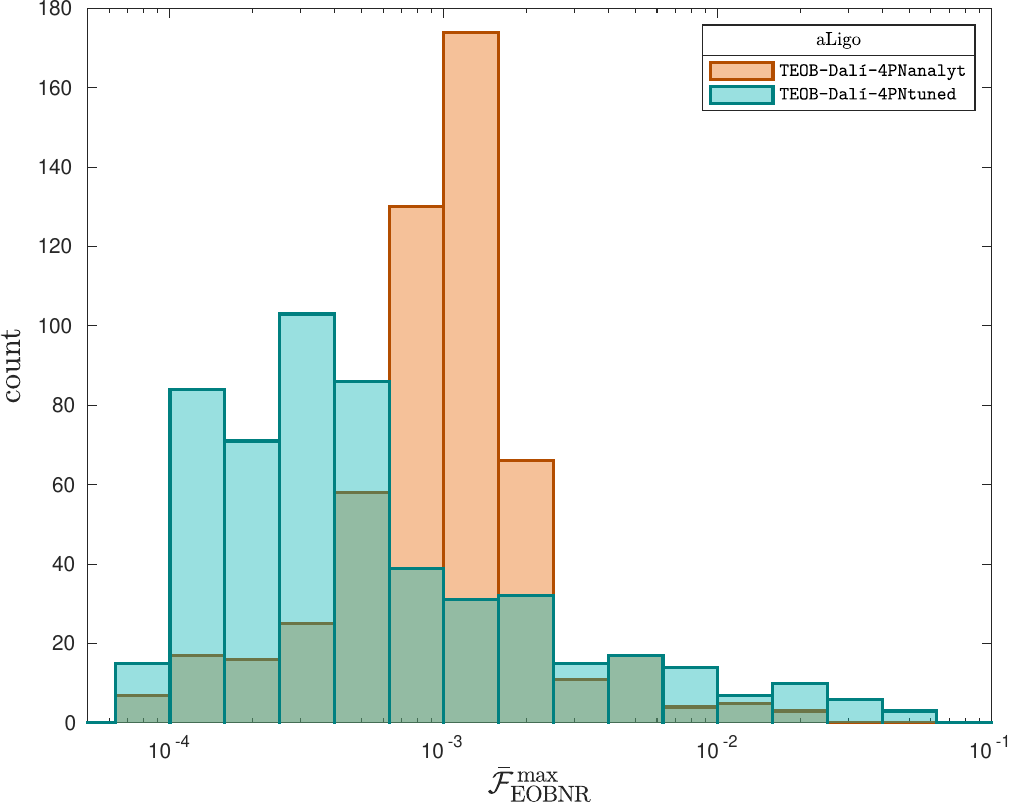}
	\caption{\label{fig:hist_4pn_4pntuned}Quasi-circular limit: comparing $\bar{\F}_{\rm EOBNR}^{\rm max}$ 
	for \daliAN{} and \daliNR{}. Despite the tail towards values of $\bar{\F}_{\rm EOBNR}^{\rm max}\sim 0.1$ 
	(corresponding to large, positive, spins), thanks to the NR-tuning of the (nonspinning) flux, 
	\daliNR{} performs globally better all over the SXS catalog of spin-aligned
	waveforms, with median $\sim 3.92\times 10^{-4}$. The corresponding value for \daliAN is instead $1.06\times 10^{-3}$,
	although $\bar{\F}_{\rm EOBNR}^{\rm max}$ is at most $\sim 0.01$. The performance of \daliNR{}  suggests 
	that a careful NR-tuning of the dissipative part of the  dynamics might be eventually needed to construct a 
	highly faithful (say $\simeq 10^{-4}$) model all over the BBH parameter space.}
\end{figure}
\begin{figure*}[t]
	\center	
	\includegraphics[width=0.23\textwidth]{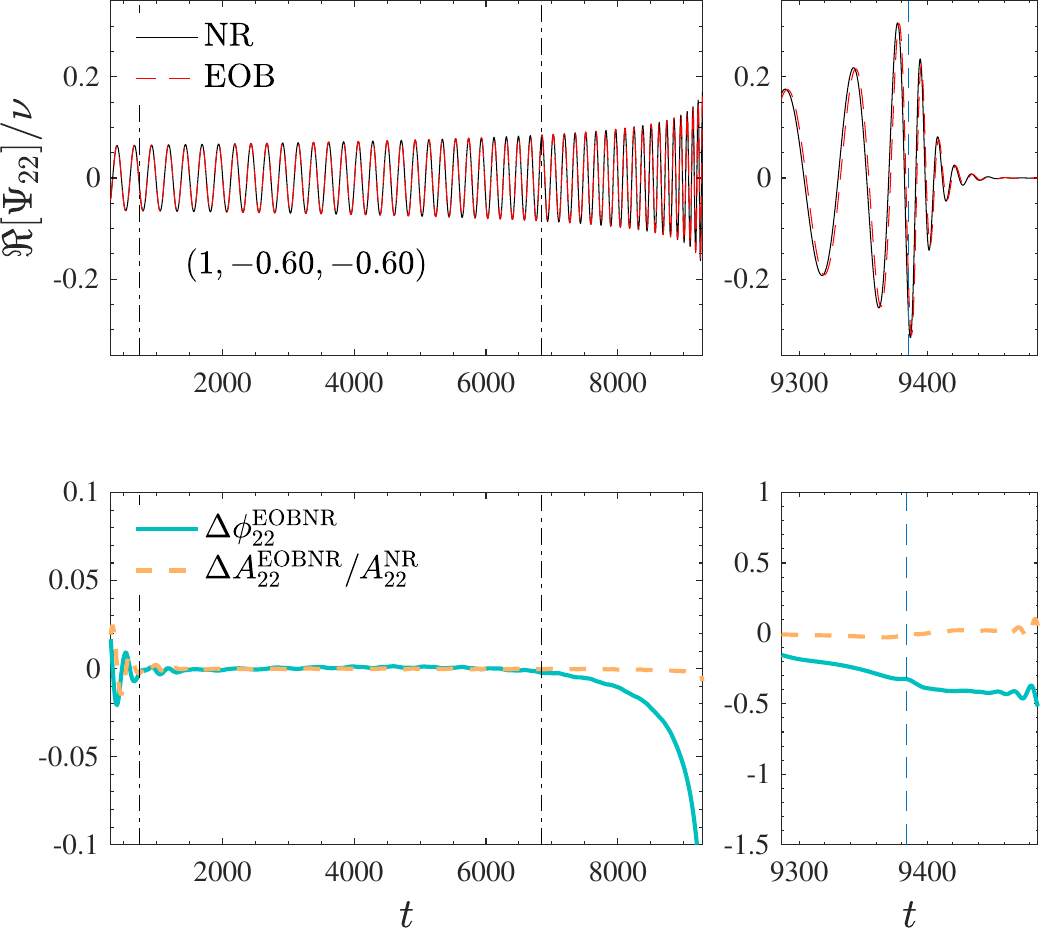}
	\includegraphics[width=0.23\textwidth]{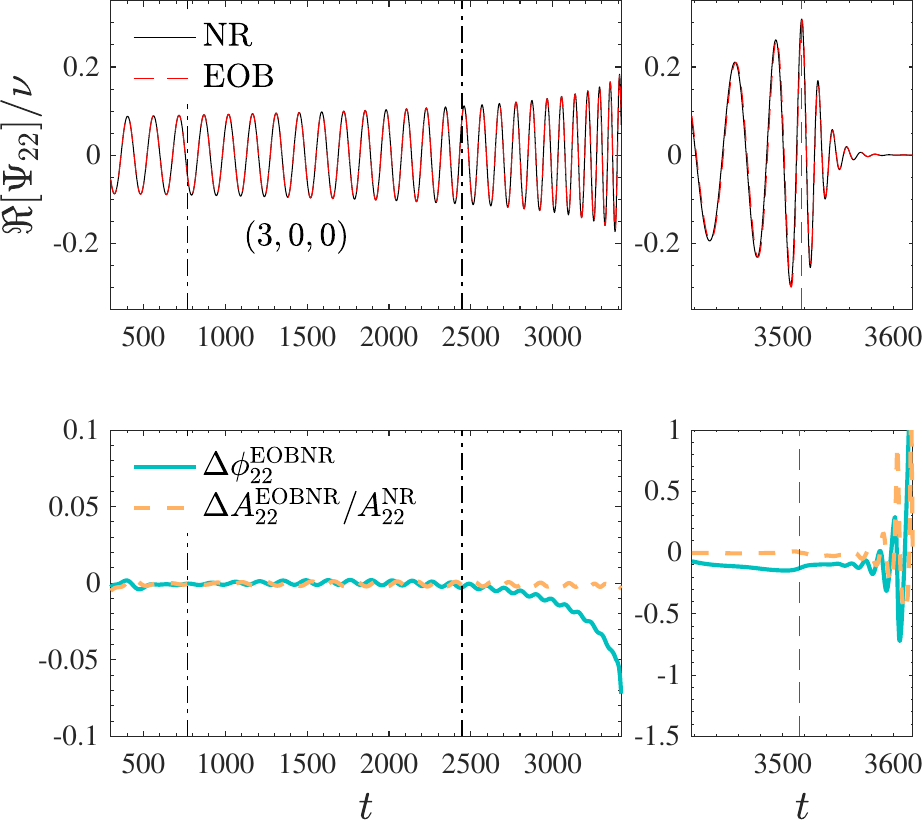}
	\includegraphics[width=0.23\textwidth]{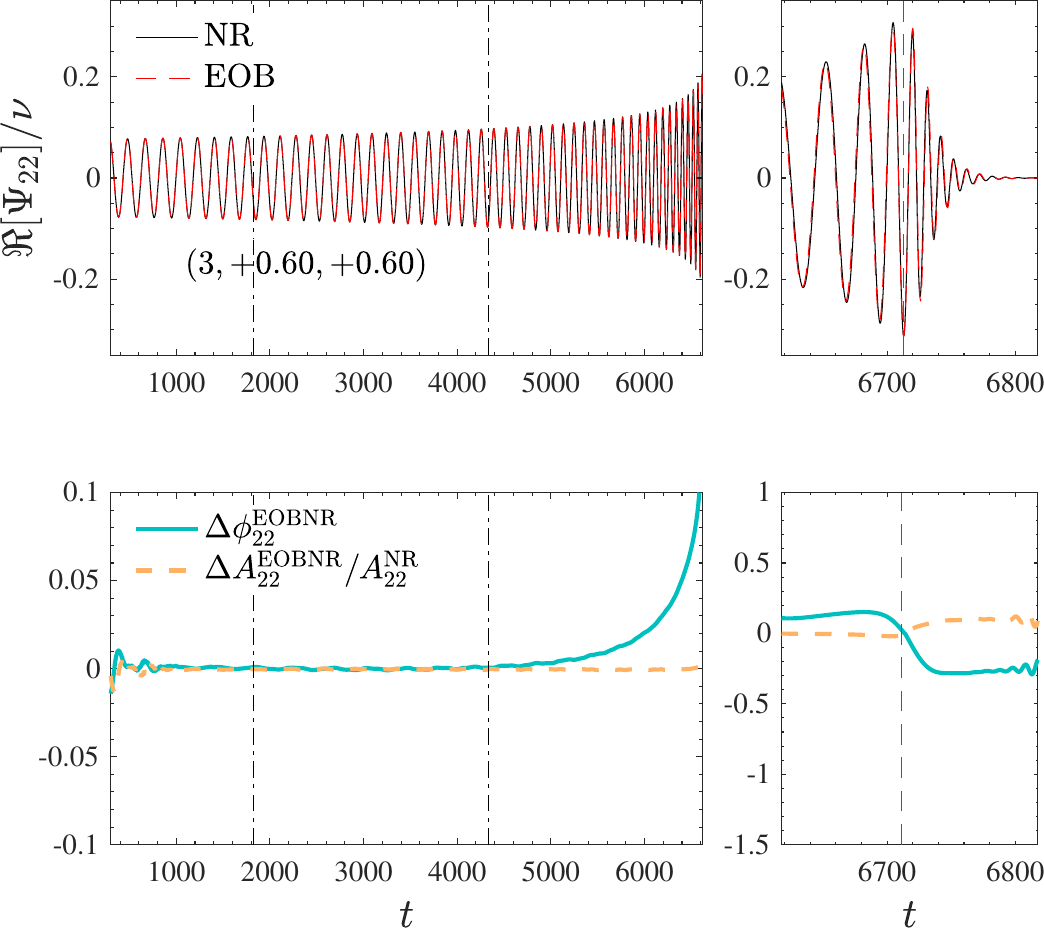}
	\includegraphics[width=0.23\textwidth]{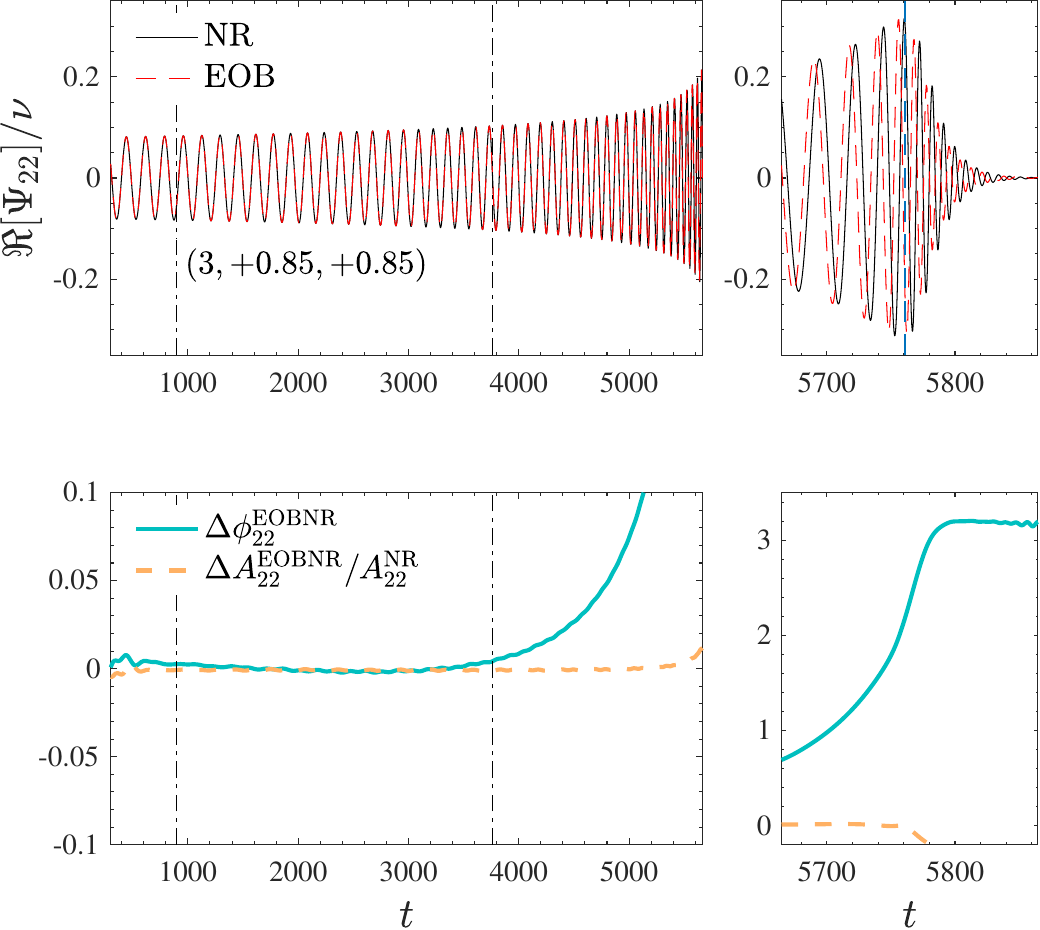}\\
	\vspace{5mm}
	\includegraphics[width=0.23\textwidth]{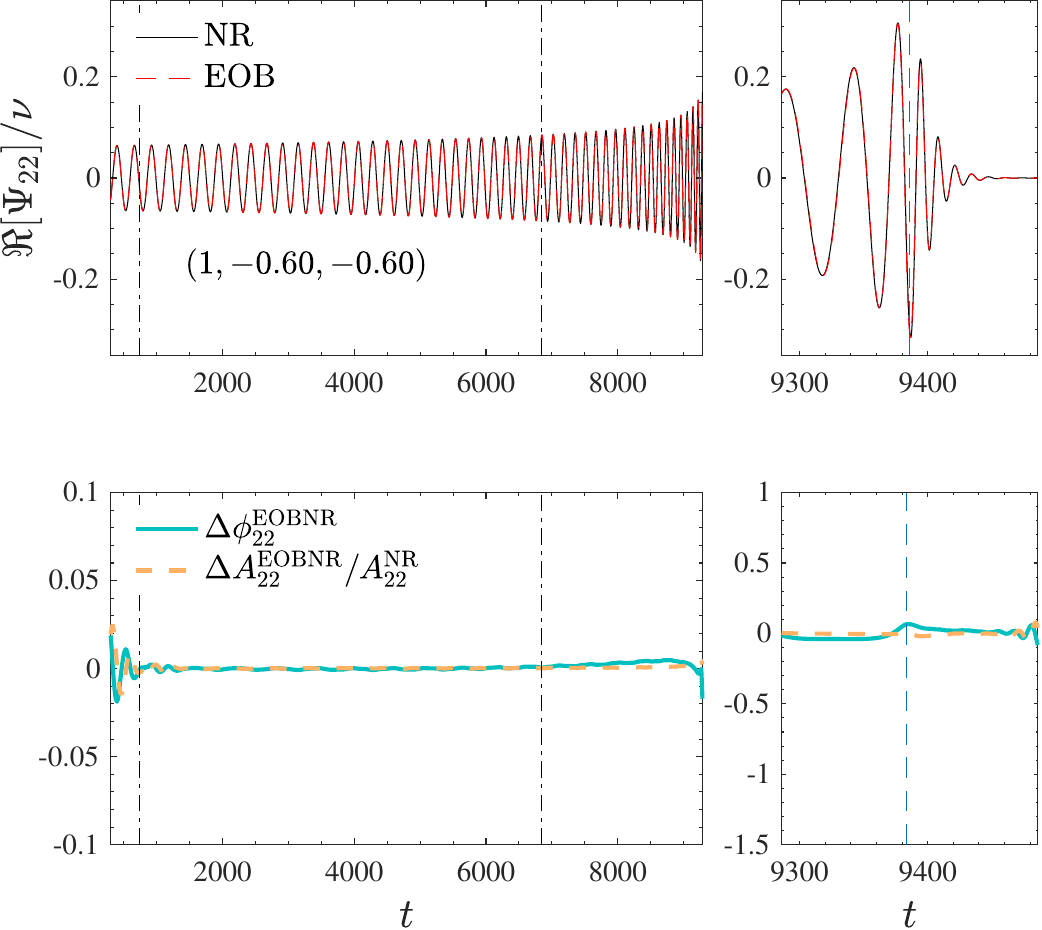}
	\includegraphics[width=0.25\textwidth]{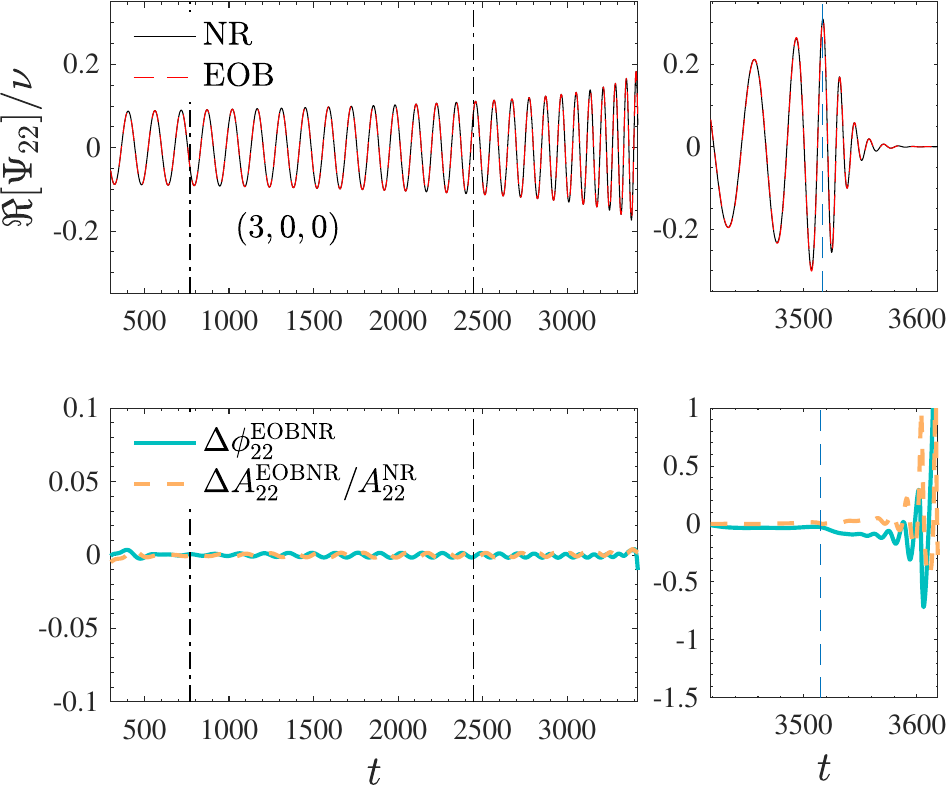}
	\includegraphics[width=0.23\textwidth]{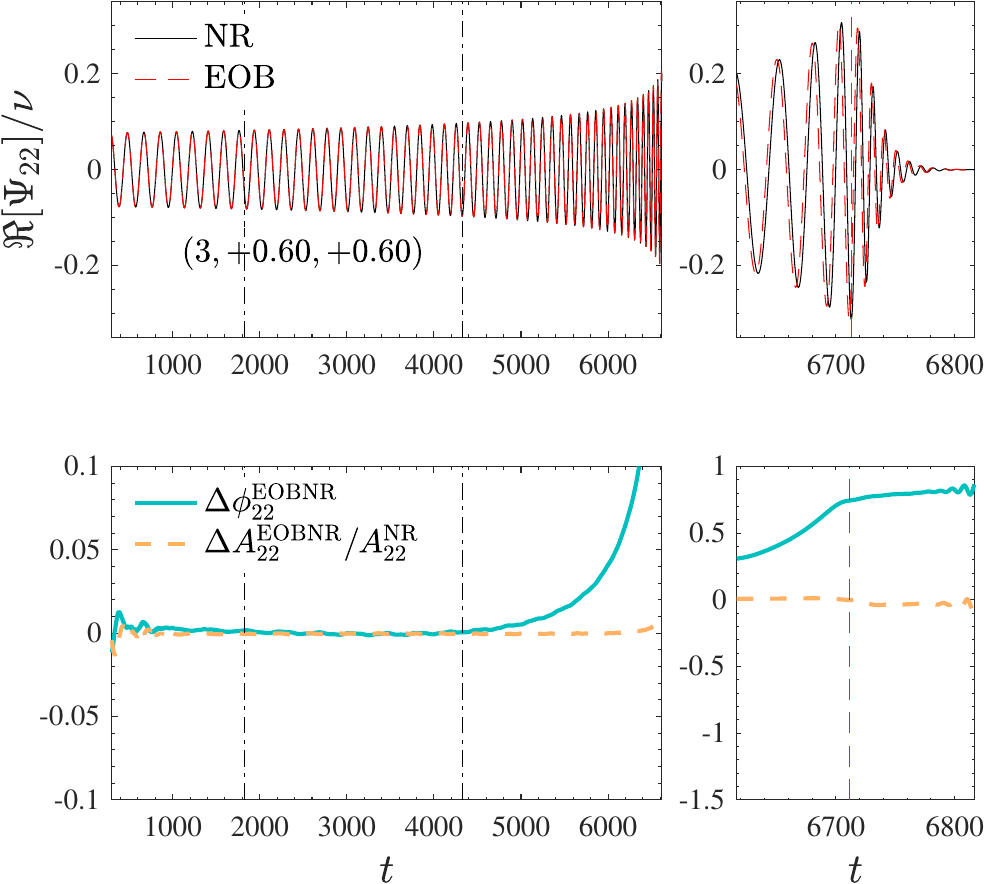}
	\includegraphics[width=0.23\textwidth]{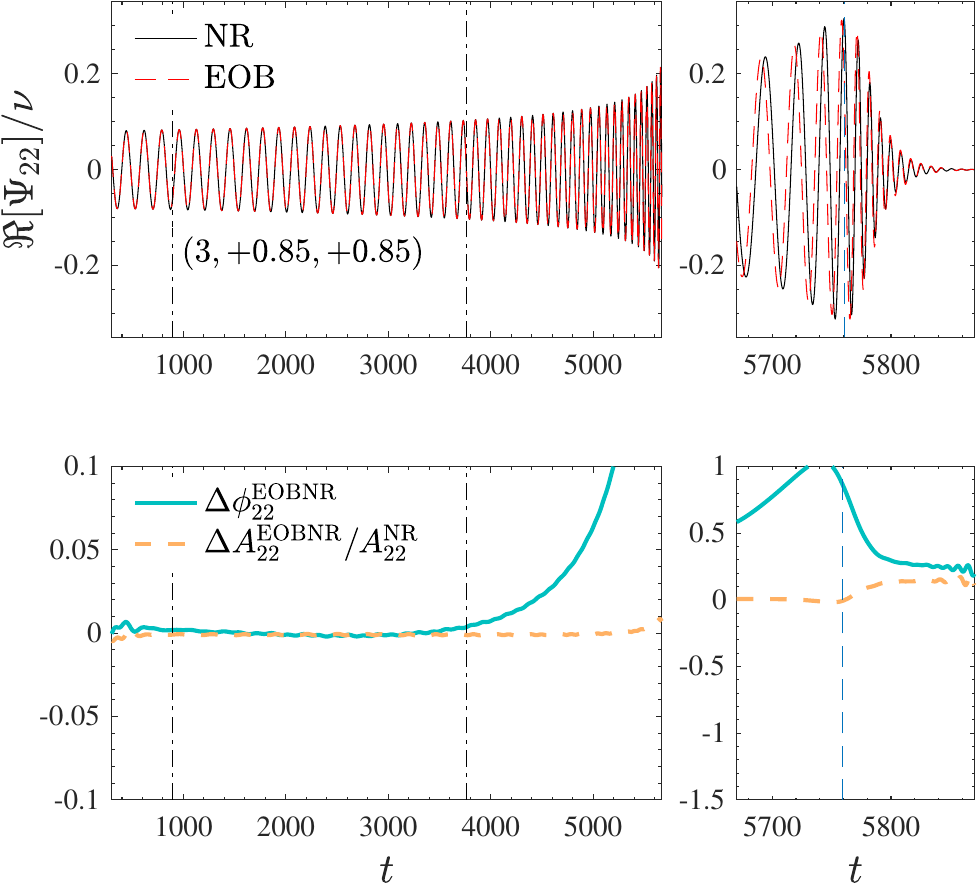}
	\caption{\label{fig:phasing_circular}Quasi-circular case: understanding the unfaithfulness of 
	Figs.~\ref{fig:barF_circ_eobnr} and~\ref{fig:barF_circ_eobnr_4PNtuned} 
	with selected time-domain phasing analysis. Top panels: model \daliAN{}. Bottom panels: model \daliNR{} (that also differs for the spin part of radiation reaction).
	The bottom-left panels show the excellent EOB/NR phasing agreement  brought by the NR-tuning of the flux. By contrast, progressively large dephasings 
	as the spins are increased are found because of inaccuracies in the spin-dependent part of the flux. This well explains the behavior of 
	$\bar{\F}_{\rm EOBNR}^{\rm max}$ shown in the right panel of Fig.~\ref{fig:barF_circ_eobnr_4PNtuned}. Note in particular that in the bottom-right panel
	the phase difference is nonmonotonic in time around merger, which eventually yields large values of $\bar{\F}_{\rm EOBNR}$.}
\end{figure*}
%
\begin{figure*}[t]
	\center
	\includegraphics[width=0.30\textwidth]{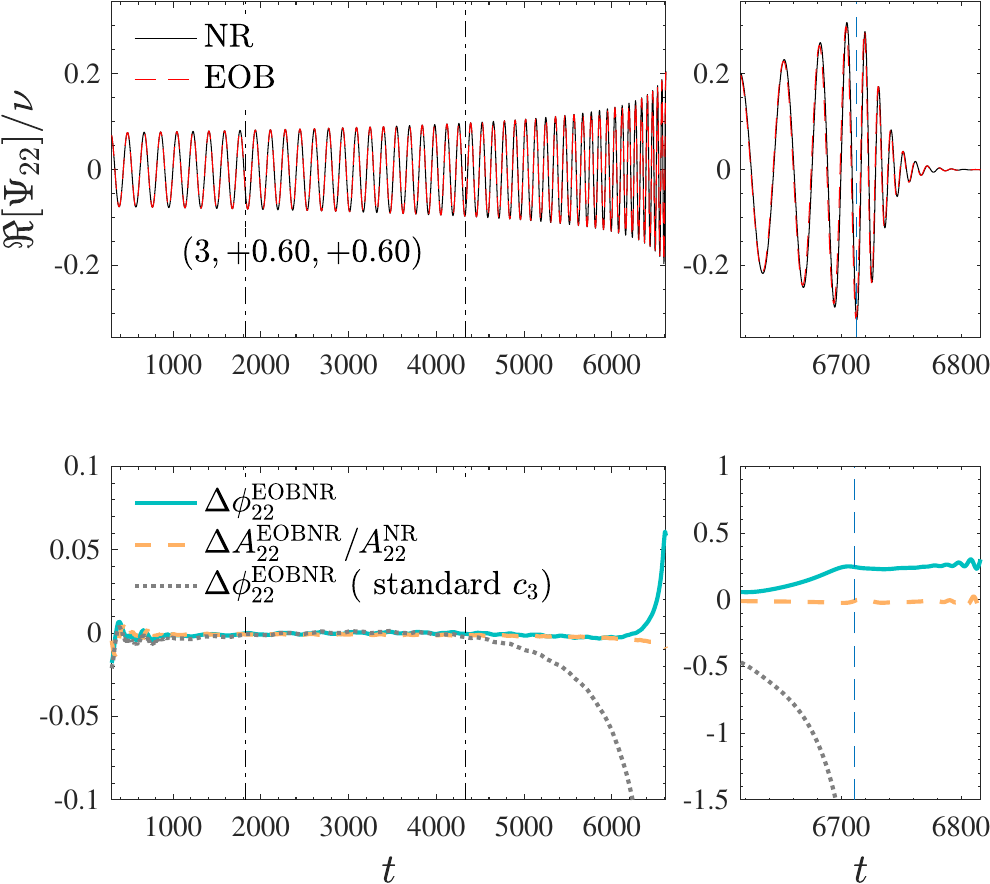}
	\includegraphics[width=0.33\textwidth]{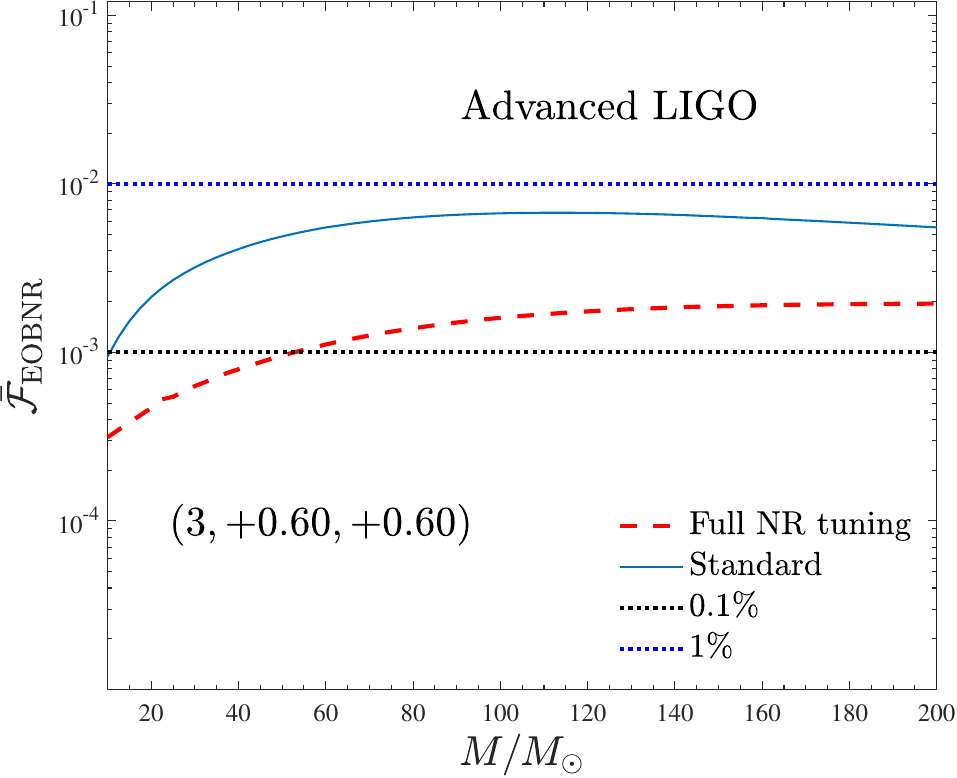} 
	\includegraphics[width=0.345\textwidth]{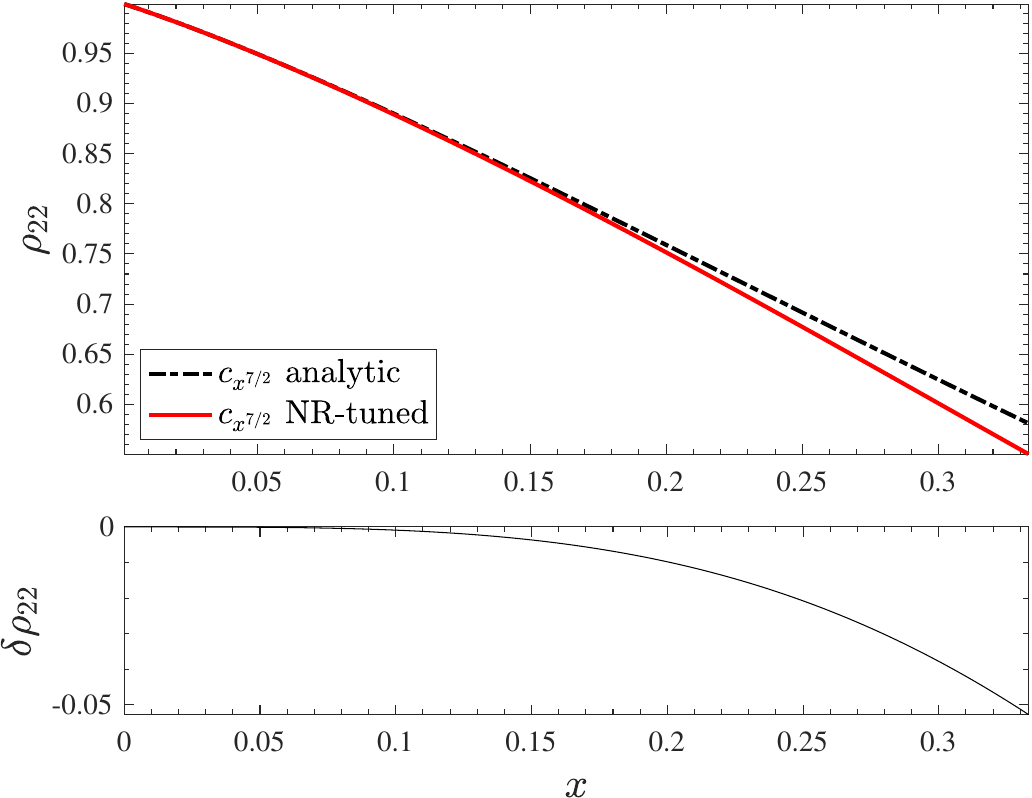}
        \caption{\label{fig:AllTuned}Proof of principle: effect of tuning, at the same time, the spin-dependent part of the 
        waveform (and radiation reaction), with an effective 3.5PN coefficient, and the N$^3$LO effective spin-orbit parameter $c_3$. 
        The phase difference during the early inspiral is now flat (cf. the corresponding panel in the bottom row of Fig.~\ref{fig:phasing_circular}) 
        and approximately zero, increasing then monotonically up to only $\sim 0.2$~rad around merger (left panel). 
        Consistently, the EOB/NR unfaithfulness (middle panel)  is at most $\sim 1.9\times 10^{-3}$ for large masses.
        The right panel shows the reduction of the NR-tuned $\rho_{22}$ with respect to the analytical one. This entails a fractional 
        reduction of $\sim 10\%$ in the radiation reaction force around $x\sim 0.2$, that eventually yields the dotted-line phase difference
        in the leftmost panel of the figure. The further tuning of  $c_3$ yields the $\sim 0.2$~rad at merger displayed and the rather
        low $\bar{\F}_{\rm EOBNR}\sim 10^{-3}$ all over the mass range.}
\end{figure*}
So far we have explored two, different, NR-informed routes to obtain an eccentric waveform model 
that is consistent with the quasi-circular, spin-aligned, SXS waveform data as well as with the 
NR surrogates \nrsurqeight{} and \nrsurqfifteen.
In one case, we use 4PN-resummed analytical radiation reaction and we find a satisfactory model
with $\bar{\F}_{\rm EOBNR}^{\rm max}\simeq 1\%$ all over the parameter space of spin-aligned quasi-circular
configurations. In the other case, we additionally NR-tune the spin-independent part of the radiation reaction force 
and change the analytic description of the $m=even$ waveform (and radiation) modes up to $\ell=4$: this gives
rather low EOB/NR unfaithfulness values for negative and mildly positive values of the effective spin
($10^{-4}\lesssim {\bar{\F}}_{\rm EOBNR}^{\rm max}\lesssim 10^{-3}$), though they can be as large as a few 
parts in $10^{-2}$ for large, positive spins. 
The performance of both models is summarized in Fig.~\ref{fig:hist_4pn_4pntuned}, that shows together the
two distributions of $\bar{\F}_{\rm EOBNR}^{\rm max}$. Despite the tail towards values of 
$\bar{\F}_{\rm EOBNR}^{\rm max}\sim 0.1$ (corresponding to large, positive, spins), the model with the NR-informed, 
effective, 4PN radiation reaction  (and waveform) performs globally better all over 
the SXS catalog, with median $\sim 3.92\times 10^{-4}$, approximately three times smaller than the $1.06\times 10^{-3}$
corresponding to the 4PN analytical model. This suggests that a careful NR-tuning (or at least analytical improvement) of 
the dissipative part of the (spin-dependent) dynamics might be eventually needed to construct a highly faithful 
(say $\simeq 10^{-4}$) model all over the BBH parameter space.
Although this task is beyond the scope of the present work, it is pedagogically useful to connect 
some selected values of $\bar{\F}^{\rm max}_{\rm EOBNR}$ to the time-domain phasing, so to 
get a sense of their actual meaning. This is done in Fig.~\ref{fig:phasing_circular} for four selected configurations. 
The top row of the figure is obtained with the \daliAN{} model, while the bottom panel with the \daliNR{} model. 
The leftmost panels, $(1,-0.60,-0.60)$ and $(3,0,0)$, connect the values $\F^{\rm max}_{\rm EOBNR}\simeq 10^{-4}$
with phase differences around merger $\lesssim 0.05$~rad.
Similarly, the increase of the values of $\bar{\F}^{\rm max}_{\rm EOBNR}$ for larger spins is mirroring either 
a larger value of $\Delta\phi^{\rm EOBNR}_{22}$ at merger, or the fact that the phase difference is not monotonic 
through late plunge, merger and ringdown. As throughly discussed in Ref.~\cite{Nagar:2023zxh} this is one of
the features of the phase difference that is mirrored into large values of $\bar{\F}_{\rm EOBNR}^{\rm max}$.  
The fact that the phasing inaccuracies increase with the value of the effective spin $\tilde{a}_0$ is explained
as follows. The figure shows that,  for {\it both} models, and in the presence of positive spins, the EOB dynamics 
predicts a transition from inspiral to plunge and merger that is less adiabatic (i.e., faster) than the NR one, 
with a (positive) phase difference that accumulates progressively during the late inspiral. This phase difference 
cannot be reduced only by the tuning of the dynamic parameter $c_3$, as it happens at spatial separations (or frequencies)
where its tuning is practically ineffective. The reason for this is that $c_3$ parametrizes spin-orbit corrections that are
proportional to $r^{-3}$, and thus that become important only when $r$ is small enough.
In any case, the fact that the phasing predicted by \daliNR{} is highly NR-faithful for $(3,0,0)$, with a dephasing 
of approximately $-0.02$~rad at merger, while it is not for larger spins indicates that the spin sector of the model 
should be improved in some way\footnote{Note in this respect that \daliNR{} already uses the 
factorized expression of  the $\rho_\lm$'s with $m=\text{even}$ instead of the additive ones, that, we verified, give
even larger differences.}. Improving the spin sector means controlling the subtle interplay between conservative
and nonconservative effects, similarly to what we discussed already in the nonspinning case. In particular, 
the fact that the phase difference is positive and grows during the late inspiral already suggests that the radiation 
reaction force is inaccurate as the two objects get close and should be modified in some way.
We thus explore, as a proof of principle, whether the EOB/NR agreement can be improved by NR-tuning, at the same
time, the spin-dependent part of the radiation reaction and consistently the spin-orbit Hamiltonian via $c_3$.
Focusing on $\bar{\rho}_{22}^S$, we recall that it is given in inverse resummed form at NNLO (i.e., 3.5PN accuracy). 
As a first attempt, we explored the effect of adding higher-order terms (i.e., beyond 3.5PN) to Eq.~\eqref{eq:bar_rho22_S}. 
Such terms, notably $(c_4^{22}x^4,c_{9/2}x^{9/2},c_5 x^5)$, were obtained by 
extrapolating to the comparable mass case the corresponding terms in the test-particle limit, following the procedure 
introduced in Sec.~VB of Ref.~\cite{Messina:2018ghh}. Not surprisingly, we found no effect on the late-inspiral behavior.
We decided then to tune an effective 3.5PN term, i.e. replacing the analytical $c_{7/2}^{22}$ coefficient with an effective one. 
Since our aim is only to understand the origin of the physical effect, we consider the single configuration $(3,+0.60,+0.60)$. 
Analogously to the nonspinning case discussed above, we realized that it is possible to tune,
iteratively, both $(c_{7/2}^{22},c_3)$, so to reduce the EOB/NR phase difference through the late 
inspiral, merger and ringdown and have it growing monotonically with time.
Figure~\ref{fig:AllTuned} reports our final result: it is obtained with $c^{22}_{7/2}=3.1$ and $c_3=33.8$.
Analytically, for this configuration (from Eq.~\eqref{eq:bar_rho22_S}) one has $c^{22}_{7/2}=0.49$
and the previously NR-tuned value of $c_3$ was $c_3=21.11$
 The meaning of these numbers is as follows. One needs to reduce the action of the analytical radiation reaction 
(and thus the amplitude $\bar{\rho}_{22}^{\rm S}$) so to slow the rate of inspiral down. The value $c^{22}_{7/2}=3.1$
corresponds to the red line in the rightmost panel of Fig.~\ref{fig:AllTuned}, that lies below the analytical curve.
One has a fractional difference of $\sim 2\%$ at $x\sim 0.20$, that corresponds to a (fractional) reduction of the flux 
of $\sim 10\%$ at the same value of $x$. The effect of this reduction on the EOB/NR phase difference is illustrated 
by the dotted, gray, line in the leftmost panel of Fig.~\ref{fig:AllTuned}, that however still retains $c_3=21.11$,
as obtained from the second row of Table~\ref{tab:c3_coeff}. At this stage, it is additionally possible to modify $c_3$,
and thus reduce the magnitude of the spin-orbit interaction (i.e., {\it shortening} the EOB waveform), 
until one obtains the $\Delta\phi_{22}^{\rm EOBNR}$ curve depicted in light blue in the leftmost panel of the figure.
This corresponding to $c_3=33.8$. As mentioned above, this result was obtained tuning iteratively 
the two parameters whose action is, partly, degenerate.
Although it is certainly possible to increase both parameters to further reduce the phase difference at merger, we
here content ourselves to show that this is feasible and that it is necessary to NR-tune the radiation reaction force
so to obtain an inspiral waveform that is more NR-faithful. Although in this case we reached this goal by tuning 
one additional parameter,  it might be possible that other analytical representation of the resummed waveform 
(and radiation reaction) exist such to eventually yield a similar result.
The important take away message is that an improved analytical representation of the (spin-dependent) part 
of the flux might be important in order to get to the $10^{-4}$ unfaithfulness level {\it also} for large-positive spins.

\section{Conclusions}
\label{sec:conclusions}
In this work we present an updated model for spin-aligned, coalescing black hole binaries for generic 
(i.e., noncircularized) planar orbits, from eccentric inspirals to scattering configurations. 
This model builds upon, improves and replaces previous work in the \TEOBResumS{} 
lineage~\cite{Chiaramello:2020ehz,Nagar:2020xsk,Nagar:2021gss,Nagar:2021xnh,Albanesi:2022xge,Bonino:2022hkj,Nagar:2023zxh}, 
notably Refs.~\cite{Nagar:2021xnh,Nagar:2023zxh}. The most important feature of the new eccentric model is
that its quasi-circular limit shows an excellent consistency with the latest avatar
of the quasi-circular model {\tt TEOBResumS-GIOTTO}~\cite{Nagar:2023zxh}.
The new physical understanding of this paper is a fresh look at the importance of the radiation reaction force 
in correctly modeling the late-inspiral dynamics and waveform. In particular, we explored the  influence of various version 
of the azimuthal component, ${\cal F}_\varphi$, that drives the backreaction on the orbital motion due to the loss of angular momentum 
through gravitational waves. We thus analyze the class of analytic {\it waveform systematics} related to the dissipative
part of the dynamics, complementing similar studies reported in Refs.~\cite{Nagar:2021gss,Nagar:2023zxh}
that were focused only on systematics related to changes to the conservative part of the dynamics.
In doing this exploration, we ended up with two different, though consistent, prescriptions for building an 
improved waveform model for eccentric binaries. These two main results can be summarized as follows.
\begin{itemize}
\item[(i)]We took advantage of the recently computed 4PN waveform terms in the $\ell=m=2$ mode~\cite{Blanchet:2023bwj,Blanchet:2023sbv,Blanchet:2023soy} 
and updated the model with this new analytical information. We argued that the use of a resummed 4PN residual amplitude 
is important and carefully compared (in the nonspinning case) the performance of the $\rho_{22}^{\rm 4PN}$ with the $\rho_{22}$ 
at $3^{+2}$PN accuracy used in all implementations of \TEOBResumS{} since 2009~\cite{Damour:2009kr}. 
We clarified that in one case the actual flux seems to be overestimated (and thus the transition from inspiral to plunge occurs faster 
than the NR prediction) while in the other case it is slightly underestimated (and thus the transition is slower), although in this second 
case the performance of the model is generally better. Therefore, we conclude that the 4PN-resummed $\rho_{22}$ function looks like 
the current best analytical choice to build EOB radiation reaction and waveform. The model is then informed by quasi-circular NR-data 
so as to determine the usual coefficients $(a_6^c,c_3)$, respectively modeling effective 5PN correction in the orbital interaction potential and effective
4.5PN (or N$^3$LO) spin-orbit effects~\cite{Damour:2014sva}.
The model performance is then evaluated all over the parameter space currently covered by public NR simulations
or data, in particular: (i) in the quasi-circular limit, it is compared with the full SXS catalog~\cite{SXS:catalog} 
of public NR simulations (up to $q=15$) as well as with the quasi-circular NR surrogates {\tt NRHybSur3dq8} and {\tt NRHybSur2dq15}; 
(ii) for eccentric inspiral, it is compared with the 28 public SXS simulations; (iii) scattering angles.
Figure~\ref{fig:barF_circ_sur} shows the excellent consistency between {\tt TEOBREsumS-GIOTTO}~\cite{Nagar:2023zxh} 
and the 4PN-based \TEOBd{} model for quasi-circular configurations. 
For the considered eccentric configurations, $\bar{\F}^{\rm max}_{\rm EOBNR}$ is always well below $1\%$ 
(except a single outlier, that also corresponds to a rather noisy dataset). 
Furthermore, the scattering-angle comparisons (see Table~\ref{tab:chi_scattering})
are satisfactory and consistent with previous literature.

\item[(ii)]From the understanding that $\rho_{22}^{\rm 4PN}$ underestimates the effect of the actual radiation reaction,
while $\rho_{22}^{3^{+2}}$ overestimates it, we decided to attempt charting an unexplored territory by NR-informing, 
{\it  at the same time} both the conservative and nonconservative part of the EOB dynamics. This is done NR-tuning
both $a_6^c$ {\it and} an effective 4PN term entering the Pad\'e resummed $\rho_{22}^{\rm 4PN}$ that replaces the
analytical 4PN information of Ref.~\cite{Blanchet:2023bwj}. In the nonspinning case, one finds that just a small modification 
to the analytically known $P^2_2(\rho_{22}^{\rm 4PN})$ (together with a new $a_6^c$) is by itself sufficient to bring 
the EOB/NR phase difference  at merger below $\sim 0.1$~rad, a value that is consistent with the expected NR uncertainty. 
This results in $\bar{\F}_{\rm EOBNR}^{\rm max}\sim 10^{-4}$ for all available nonspinning datasets up to mass ratio $q=15$.
In the presence of spins, we, again, clearly highlighted the importance of the spin-dependent part of the radiation reaction and
evaluated the influence of different analytical prescriptions for the resummed EOB waveform that were discussed in the literature.
For example, we concluded that the additive expression $\rho_{22}^{\rm orb}+\rho_{22}^{\rm S}$ implemented in any version of \TEOBResumS{} 
is overestimating the flux for positive spins and that a better (though certainly improvable) representation of the residual
amplitude corrections is obtained by the factorized and inverse-resummed prescription discussed in 
Refs.~\cite{Nagar:2016ayt,Messina:2018ghh}.
With this choice, and a new expression of the NR-informed $c_3$, we may eventually end up having a model, 
dubbed {\tt TEOBResumS-Dali-4PNTuned}, that is globally more NR faithful than the current  {\tt TEOBResumS-Dali}.
A new look at the analytical representation of the EOB-resummed radiation reaction is postponed to future work.
\end{itemize}

In conclusion, we have now at hand two waveform models for non-circularized binaries that differ because of 
(i) the analytic content and 
(ii) the amount of NR-information included. 
Although in the quasi-circular limit {\it none} of these two model is as NR-faithful as {\tt TEOBResumS-GIOTTO}, they will
hopefully allow us to give a very precise quantitative meaning to the actual impact of waveform systematics 
on current and future GW detectors~\cite{Hild:2009ns,Evans:2021gyd}.

\begin{acknowledgments}
V.~F. is supported by the ERC-SyG project ``Recursive and Exact New Quantum Theory'' (ReNewQuantum), 
which received funding from the European Research Council (ERC) within the 
European Union's Horizon 2020 research and innovation program under Grant No. 810573.
S.~B. knowledges support by the EU Horizon under ERC Consolidator Grant, no. InspiReM-101043372.
P. ~R. and S.~B. thank the hospitality and the stimulating environment 
of the IHES, where part of this work was carried out. We thank M.~Panzeri for cross-checking
some results presented in Appendix~\ref{sec:alter_resum} and G.~Carullo for comments and
a careful reading of the manuscript. The present research was also partly 
supported by the ``\textit{2021 Balzan Prize for Gravitation: Physical and Astrophysical Aspects}'', 
awarded to Thibault Damour.
\end{acknowledgments}

\appendix

\section{Alternative Pad\'e resummation of $\rho_{22}$ and implications.}
\label{sec:alter_resum}
In Sec.~\ref{sec:4PN} the $\rho_{22}^{\rm 4PN}$ function was resummed by taking a global 
Pad\'e approximant obtained by replacing the $\log(x)$ terms with some formal constants and 
then reinserting them back. Historically this has been a standard approach within the EOB 
model (see e.g. Ref.~\cite{Damour:2009kr} and references therein), that, however, was never 
carefully tested with alternatives. It should also be noted that in the test-mass limit this approach 
was extensively used in Refs.~\cite{Nagar:2016ayt,Messina:2018ghh} and found sufficiently 
satisfactory at the time. In this Appendix we point out that this method introduces some
analytic systematics that were overlooked so far and that might be important at the level of 
accuracy we are currently pushing our models. Despite this, the results discussed in the main 
text are expected to stand even against these systematics.
To start with, let us review in detail our resummation  procedure so to highlight its drawbacks. 
The 4PN-accurate function of Eq.~\eqref{eq:rho22} schematically reads
\be
\label{eq:rho}
\rho = 1 + c_1 x + c_2 x^2 + x^3[c_3+c_3^{\log}\log(x)] + x^4[c_4+c_4^{\log}\log(x)] \ ,
\ee
For pedagogical purpose, let us first assume all the coefficients equal to one. 
Then, one poses $\log(x)=c$ and takes the $(2,2)$ Pad\'e approximant. This reads
\be
\label{eq:P22rho}
\rho'\equiv P_2^2(\rho)=\dfrac{1+x-c x^2}{1-(1+c)x^2} \ .
\ee 
There are two sorts of inconsistencies. First, by expanding this expression in powers of $c$, we find 
\be
\rho'=\frac{1}{1-x}+\frac{x^3}{(1+x)(1-x)^2} c + O(c^2) \ ,
\ee
and we should compare it with the original function $\rho$, Eq.~\eqref{eq:rho}, after 
replacing $c$ with $\log(x)$. In particular, we notice that even though 
\be
\frac{x^3}{(1+x)(1-x)^2}-(x^3+x^4)= O(x^5)\ ,
\ee
which is consistent with the error of Pad\'e approximation, the rational function $\frac{x^3}{(1+x)(1-x)^2}$ 
has a degree $3$ denominator which is an unreasonable approximation of $x^3(1+x)$. 

Second, by expanding $\rho'$ in Eq.~\eqref{eq:P22rho} at higher order, e.g. 5PN, we have 
\begin{align}
\rho'&\sim1+x+x^2 + (1+c)x^3 + (1+c)x^4 \nonumber\\
   & + (1+c)^2x^5 + O(x^6) \ .
\end{align}
When the constant $c$ is replaced by the $\log(x)$, one immediately sees that a 5PN
term of the form $\log^2(x) x^5$ appears. In the general case, where the coefficients
are not equal to one, the 5PN term guessed by the Pad\'e approximant has the 
structure
\be
c_{\rm 5PN}^{\rm guess}= \dfrac{n_0 + n_1 c + n_2 c^2}{d1 + d_2 c} \ ,
\ee
\begin{table}[t]
	\caption{\label{tab:comparison_LSO}Fractional differences 
	$\delta \rho_{22}=(\rho_{22}^X-\rho_{22}^{\rm Exact})/\rho_{22}^{\rm Exact}$ at $x_{\rm LSO}=1/6$ obtained 
	with different analytical approximations. The function where the $\log(x)$ are treated separately is generally 
	closer to the exact data at each PN order.}
	\begin{center}
  \begin{ruledtabular}
	\begin{tabular}{lccc}
	      & Pad\'e $\log(x)$ separated & Pad\'e $\log(x)=c$ & Taylor\\
	      \hline
	  4PN & $-0.000950$ & $-0.004563$  & $0.001804$ \\
	  5PN & $-0.001234$ & $-0.001358$  & $-0.001621$ \\
	  6PN & $-0.000317$ & $-0.000654$  & $-0.000425$\\
	  7PN & $-0.000187$ &   $-0.000393$  &   $+0.000001$ \\
	  8PN & $-0.000168$ & $-0.000143$  & $-0.000106$
 \end{tabular}
  \end{ruledtabular}
  \end{center}
  \end{table}
%
\begin{figure*}[t]
	\center	
	\includegraphics[width=0.45\textwidth]{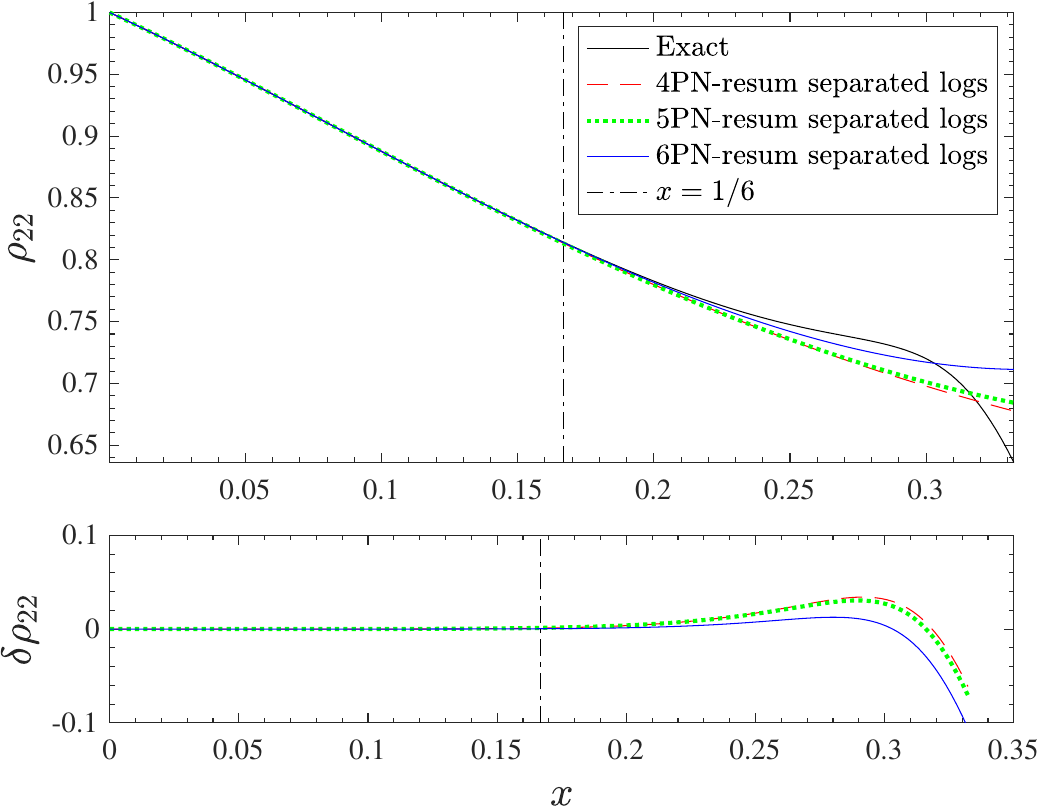}
        \includegraphics[width=0.45\textwidth]{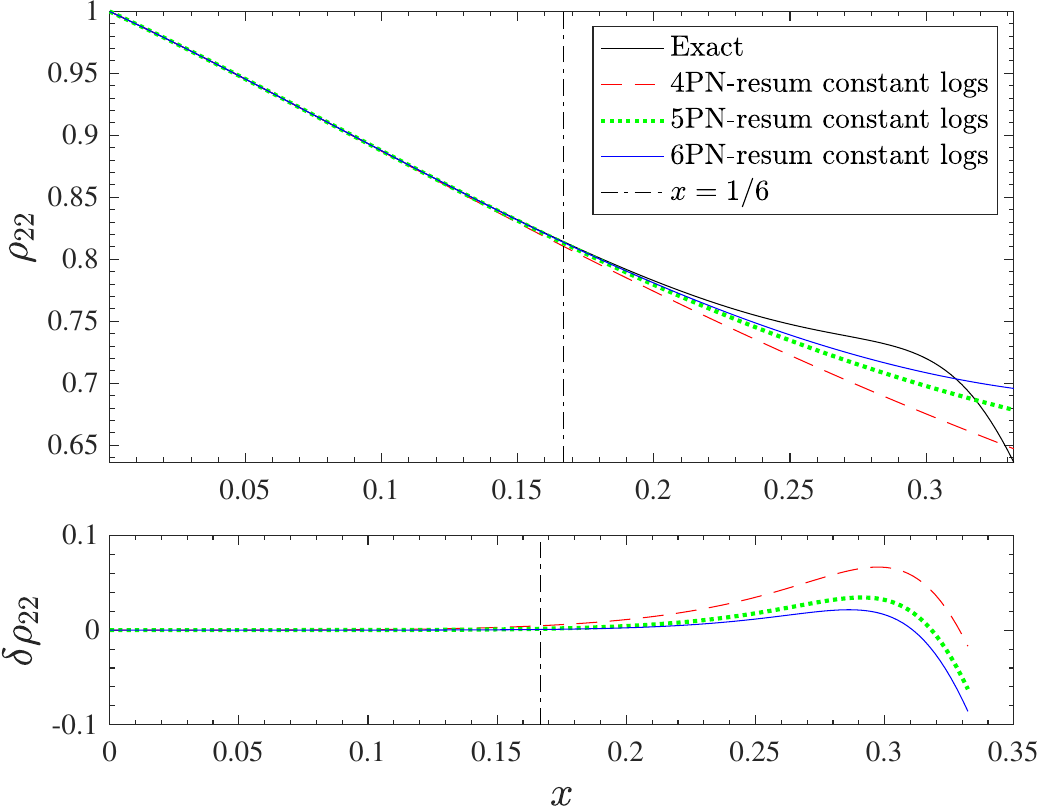}	
	\caption{\label{fig:rho22_old_new}Test-mass limit ($\nu=0$). Resummed $\rho_{22}$ at 4PN, 5PN and 6PN using 
	Pad\'e approximants $(2,2)$, $(3,2)$ and $(4,2)$, accordingly to the PN order. Left panel: factorize the logarithmic terms and separately 
	resum the polynomial coefficients. Right panel: set $c=\log(x)$, resum the polynomial with parameter $c$, and finally 
	substitute back $c=\log(x)$. The bottom panels show the relative differences with the exact curve obtained numerically, 
	$\delta\rho_{22}\equiv (\rho_{22}^{X}-\rho_{22}^{\rm Exact})/\rho_{22}^{\rm Exact}$.}
\end{figure*}
where, again, the $c=\log(x)$. So, in the general case, the PN expansion of the Pad\'e
where the logarithms are considered as constant introduces an even more intricate
logarithmic structure. Unfortunately, this is qualitatively inconsistent with the PN expansion 
of $\rho_{22}$, where the $\log^2(x)$ terms are know to only appear at 6PN order, as first
shown in Ref.~\cite{Fujita:2011zk}, Eq.~(7) therein. The same inconsistency pointed out here at
4PN is present also in the 6PN-based resummed amplitude of Refs.~\cite{Messina:2018ghh},
where a $(4,2)$ Pad\'e approximant (with constant logs) was used for most multipoles.
This affects, quantitatively, the (test-mass) radiation-reaction driven dynamics of
Refs.~\cite{Albanesi:2023bgi,Nagar:2022fep,Albanesi:2021rby} as well as the comparable-mass
dynamics of several works that were incorporating the approach of Ref.~\cite{Messina:2018ghh}
where different versions of \TEOBResumS{} were developed, e.g. Refs.~\cite{Nagar:2020pcj,Nagar:2023zxh}.
Nonetheless,  it should be noted that, since the EOB dynamics is additionally informed by 
NR simulations, this inconsistency is not expected to have a dramatic influence on well established results.
In this respect, in the main text we showed that a NR-informed effective 4PN term, within
the same Pad\'e resummed structure, may eventually yield an improved waveform
model, with unfaithfulness $\simeq 10^{-4}$. This, together with the inconsistency in
our resummation strategy, calls for an alternative approach to resumming 
$\rho_{22}^{\rm 4PN}$ such that the trascendental structure of the function is preserved. 
A very simple procedure consists in resumming separately the polynomial part 
and the terms that are proportional to the $\log(x)$. In this way, the trascendental order 
of the function is guaranteed not to be changed by the resummation procedure 
and, a priori, we may expect results more consistent with the exact function.
Schematically, $\rho_{22}^{\rm 4PN}$ can be recasted as
\be
\rho_{22}^{\rm 4PN}(x) = p_0^{(4)}(x) +  p_{\log}^{(4)}(x) \log(x)\ ,
\ee
where $p_{0}^{(4)}, p_{\log}^{(4)}$ are polynomials of the form
\begin{align}
\label{eq:p0}
p_0^{(4)}(x)     &= 1 + c_1 x + c_2 x^2 + c_3 x^3 + c_4 x^4 \ , \\
\label{eq:p1}
p_{\log}^{(4)}(x) & = c_3^{\log} x^3 + c_4^{\log} x^4 \ .
\end{align}
Then, we resum $p_0^{(4)}$ and $p_{\log}^{(4)}$ separately. For $p_0^{(4)}$ we use a $(2,2)$
Pad\'e approximant. For $p_0^{(4)}$, we factorize the $x^3$ term in front and the 
rest is resummed taking at $(0,1)$ Pad\'e approximant. When evaluated in the 
test-mass limit, the resulting analytical function is found to be closer to the
exact one, obtained numerically (see e.g.~\cite{Messina:2018ghh} and references therein),
than the our standard choice discussed in the main text. In particular the
fractional difference at $x_{\rm LSO}=1/6$ is $\sim -0.000969$ versus the 
value $-0.00456$ obtained with the Pad\'e approximant with constant logs. 
We will come back to the impact of this case on the comparable-mass case below.
Before this, since most of the established test-mass results mentioned above 
are based on 6PN-accurate $\rho_\lm$'s (see Table~I in Ref.~\cite{Messina:2018ghh}),
we also briefly investigate the effect of the new resummation at 6PN. 
A more comprehensive analysis of all multipoles will be reported elsewhere~\cite{Panzeri:2024prep}.
Schematically, $\rho_{22}^{\rm 6PN}$ can be recasted as
\be
\rho_{22}^{\rm 6PN}(x) = p_0^{(6)}(x) +  p_{\log}^{(6)}(x) \log(x) + p_{\log^2}^{(6)}(x) \log^2(x)\ ,
\ee
where $p_{0}^{(6)}, p_{\log}^{(6)}, p_{\log^2}^{(6)}$ are polynomials of the form
\begin{align}
\label{eq:p06}
p_0^{(6)}(x)  &= 1 + c_1 x + c_2 x^2 + c_3 x^3 + c_4 x^4 + c_5 x^5 + c_6 x^6 \ , \\
\label{eq:p16}
p_{\log}^{(6)}(x) & = c_3^{\log} x^3 + c_4^{\log} x^4 + c_5^{\log} x^5+ c_6^{\log} x^6 \ , \\
\label{eq:p26}
p_{\log^2}^{(6)}(x) & = c_6^{\log^2} x^6 \ .
\end{align}
Then we observe that resumming only $p_0^{(6)}$ (and taking the Taylor expansion of $p_{\log}^{(6)}$ and $p_{\log^2}^{(6)}$) gives a better approximation than resumming both $p_0^{(6)}$ and $p_{\log}^{(6)}$. For $p_0^{(6)}$ we use a $(4,2)$
Pad\'e approximant. As already noticed at 4PN, when evaluated in the 
test-mass limit, the resulting analytic function is found to be closer to the
exact one, obtained numerically than our standard choice discussed. 
In particular the fractional difference at $x_{\rm LSO}=1/6$ is $\sim -0.000317$ versus the value $-0.000654$ obtained using a $(4,2)$ Pad\'e approximant 
with constant logarithms.\footnote{When we resum both $p_0^{(6)}$ and 
$p_{\log}^{(6)}$ (using repsectively a $(4,2)$ and $(3,3)$ Pad\'e approximant), 
the fractional difference at $x_{\rm LSO}=1/6$ is $\sim -0.000525$.} This suggests 
that at 6PN the logarithmic terms $p_{\log}^{(6)}$ and $p_{\log^2}^{(6)}$ are 
well approximated by the Taylor expansion, while the polynomial part $p_0^{(6)}$ 
needs to be resummed. The same reasoning applies at 7PN and 8PN, where 
we observe that resumming only the polynomial part gives a better approximation 
than resumming separately the polynomials and the logarithic terms~\footnote{We resum $p_0^{(7)}$ with Pad\'e approximant $(5,2)$, and $p_0^{(8)}$ with Pad\'e approximant $(6,2)$. In both cases, the poles are complex conjugate.}. 
We collect the fractional differences at $x_{\rm LSO}=1/6$ in Table~\ref{tab:comparison_LSO}. 
Notably, comparing the fractional differences at $x_{\rm LSO}=1/6$, the resummation procedure 
described above gives a better approximation than the Taylor expansion (an exception, just by chance, is given by the 7PN, while at 8PN they are essentially equivalent). 
In addition, looking at the fractional differences, we see that resumming only the polynomial part is stable from 6PN to 8PN. 
The stability of this resummation at higher PN orders will be investigated 
elsewhere~\cite{Panzeri:2024prep}. A priori we expect the scheme to remain robust up to 10PN, but things might become
more subtle at higher orders, since fractional powers of $x$ appear.
Whereas, going from 4PN to 6PN, we saw that different resummation methods of $\rho_{22}$ are 
effective as shown in Fig.~\ref{fig:rho22_old_new}; in particular, the logarithmic terms were 
resummed with a Pad\'e approximant only at 4PN and 5PN, and they were not resummed at 6PN. 
Furthermore, at 6PN we also have $\log^2(x)$ terms which give a better approximation of the singular 
behavior of $\rho_{22}$. Hence, the different summation procedures at 4PN and 6PN can be justified 
as follows: Pad\'e approximants of $p_0^{(k)}$ for $k=4,5,6$ capture well the singular 
behavior of $p_0(x)$ which seems ``dominant'' also at lower PN orders. Conversely, from 6PN (and at least up to 8PN) the singular behavior 
of $p_{\log}(x)+\log(x)p_{\log^2}(x)$ is better captured by the presence of $\log^2(x)$, thus the Taylor expansion of $p_{\log}$ gives a good approximation. 
Figure~\ref{fig:rho22_old_new} summarizes our results at 4PN, 5PN and 6PN comparing the old resummation 
strategy (left panel) with the new one (right panel). It is remarkable the improvement found already at 4PN.
\begin{figure}[t]
	\center	
	\includegraphics[width=0.42\textwidth]{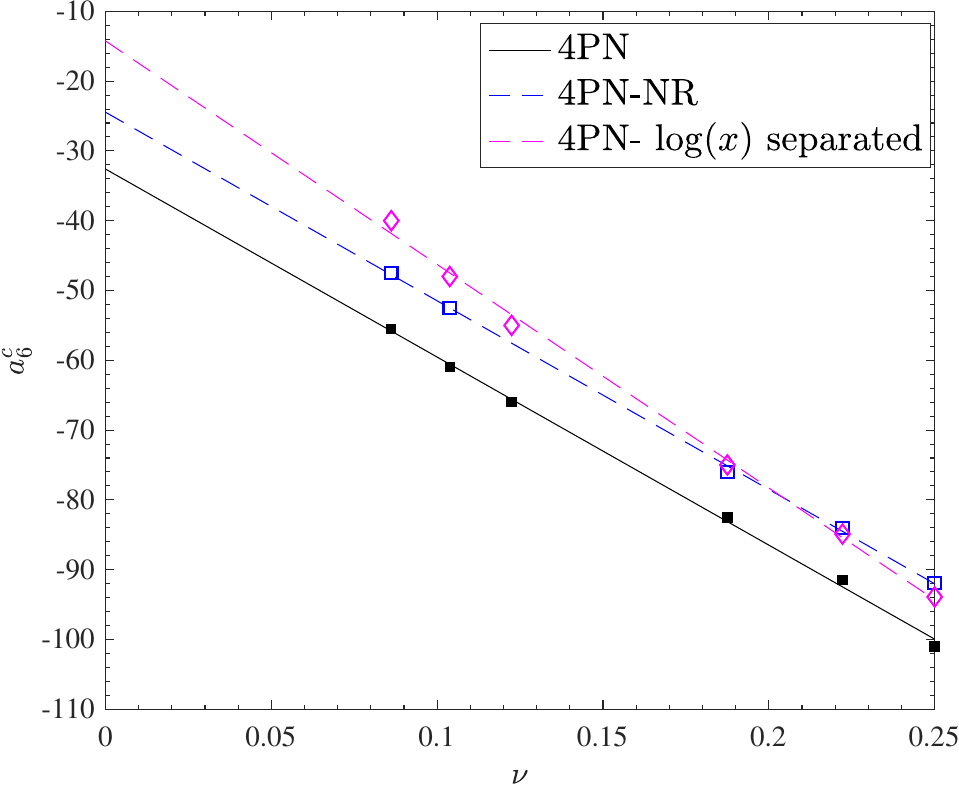}
	\caption{\label{fig:a6_log_sep}New values of the NR-informed $a_6^c$ obtained using
	the resumed $\rho_{22}^{\rm 4PN}$ based on the $\log$-separation of Eqs.~\eqref{eq:p0}-\eqref{eq:p1}.
	Note that the values of $a_6^c$ are close to the values corresponding to the NR-informed 
	$c_4^\nu$ coefficient, but progressively differ as $\nu$ decreases. This has implications on the phasing
	and the global EOB/NR performance of the model, that turns out to get improved with respect to the versions
	discussed in the main text.}
\end{figure}
\begin{figure}[t]
	\center	
	\includegraphics[width=0.22\textwidth]{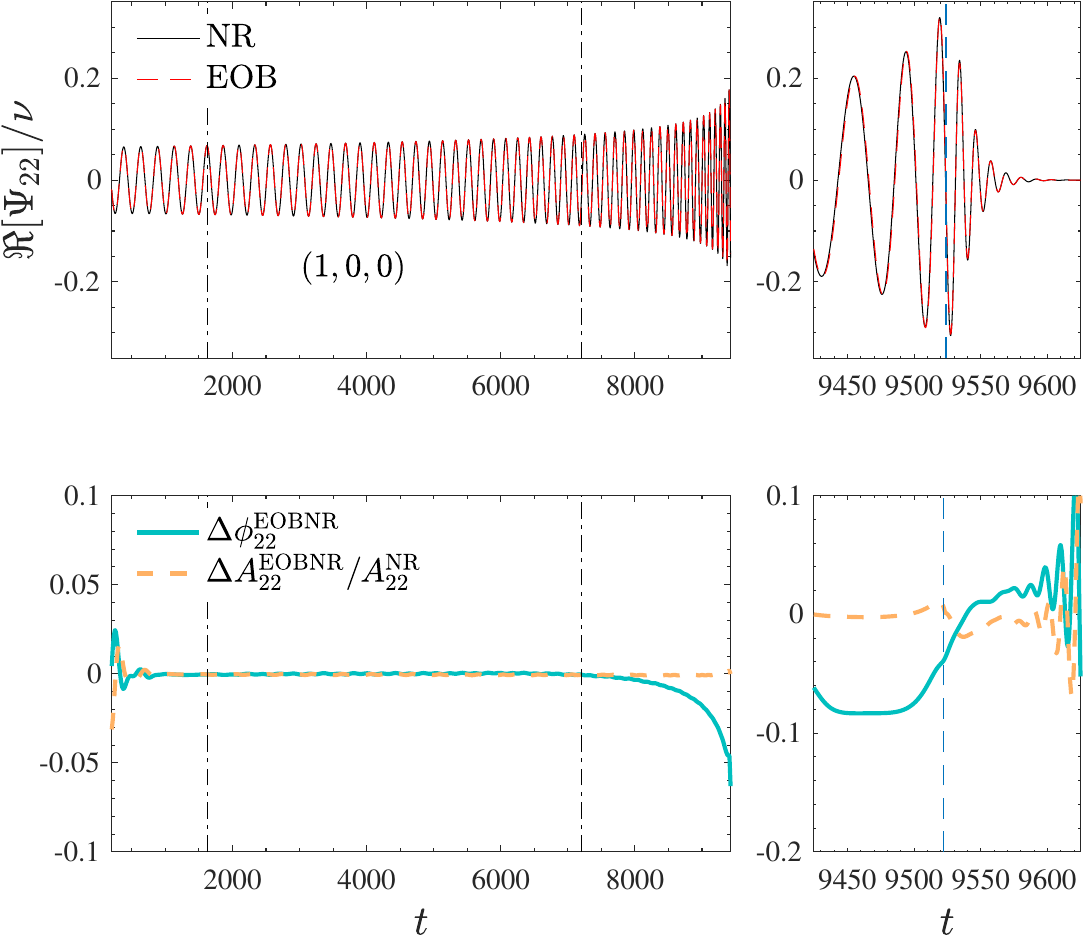}
        \includegraphics[width=0.22\textwidth]{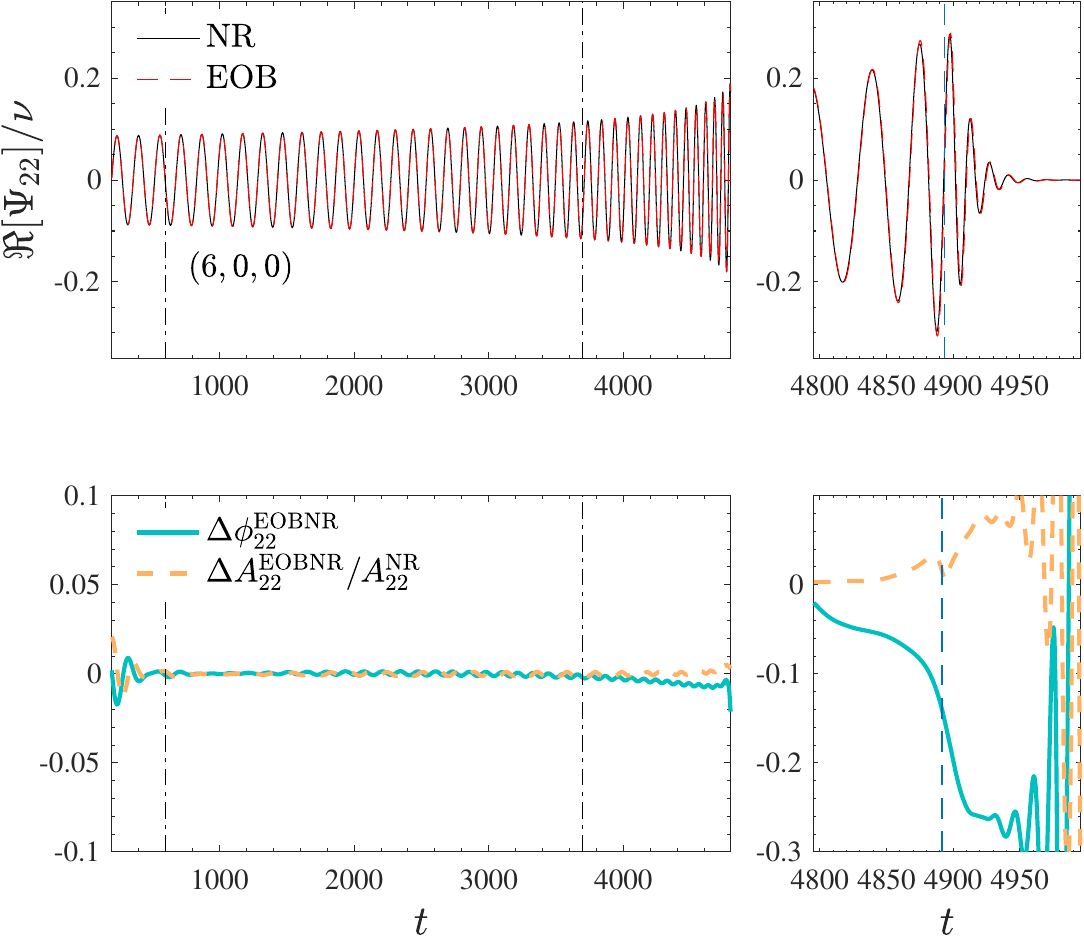}	
	\caption{\label{fig:phasings_newlog}EOB/NR time-domain phasings for $q=1$ and $q=6$.
	The EOB performance is comparable to the case with the NR-informed flux discussed 
	in the main text, see Fig.~\ref{fig:phasing_tuned}.}
\end{figure}

\begin{figure}[t]
	\center	
	\includegraphics[width=0.42\textwidth]{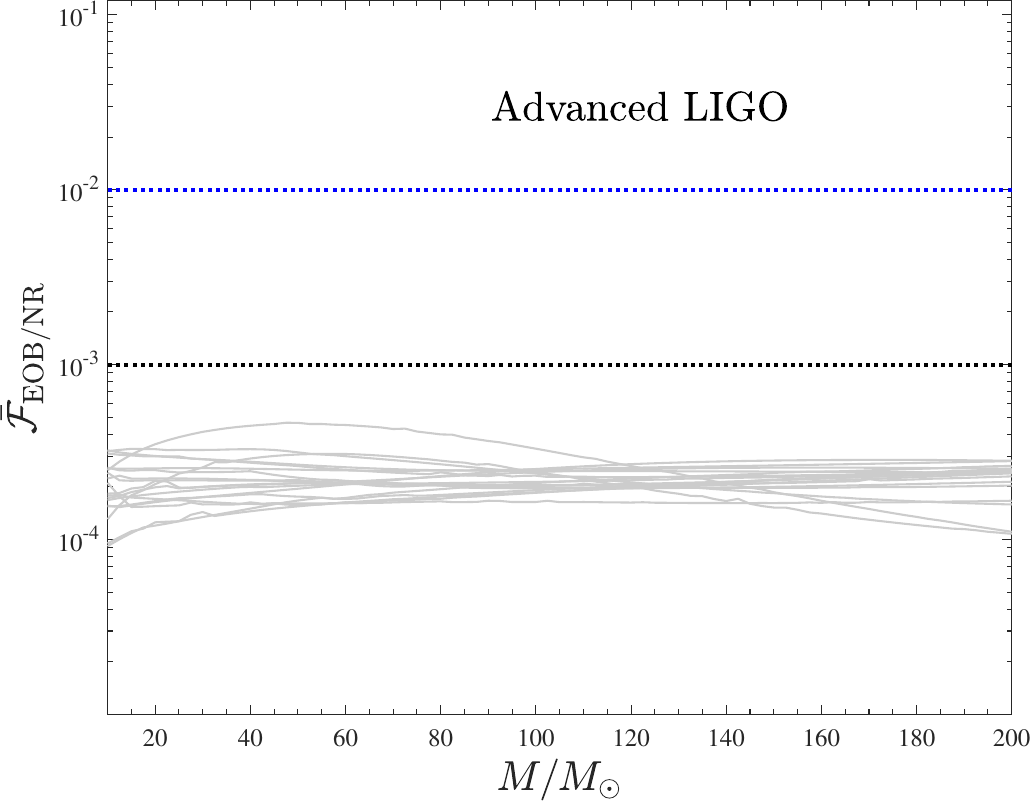}
	\caption{\label{fig:barF_log_sep}EOB/NR unfaithfulness for nonspinning configurations up to $q=15$.
	It is remarkable that the global EOB performance is substantially comparable to the left panel of 
	Fig.~\ref{fig:phasing_tuned}, where the radiation reaction was also tuned to NR simulations.}
\end{figure}

\begin{figure*}[t]
	\center
	\includegraphics[width=0.31\textwidth]{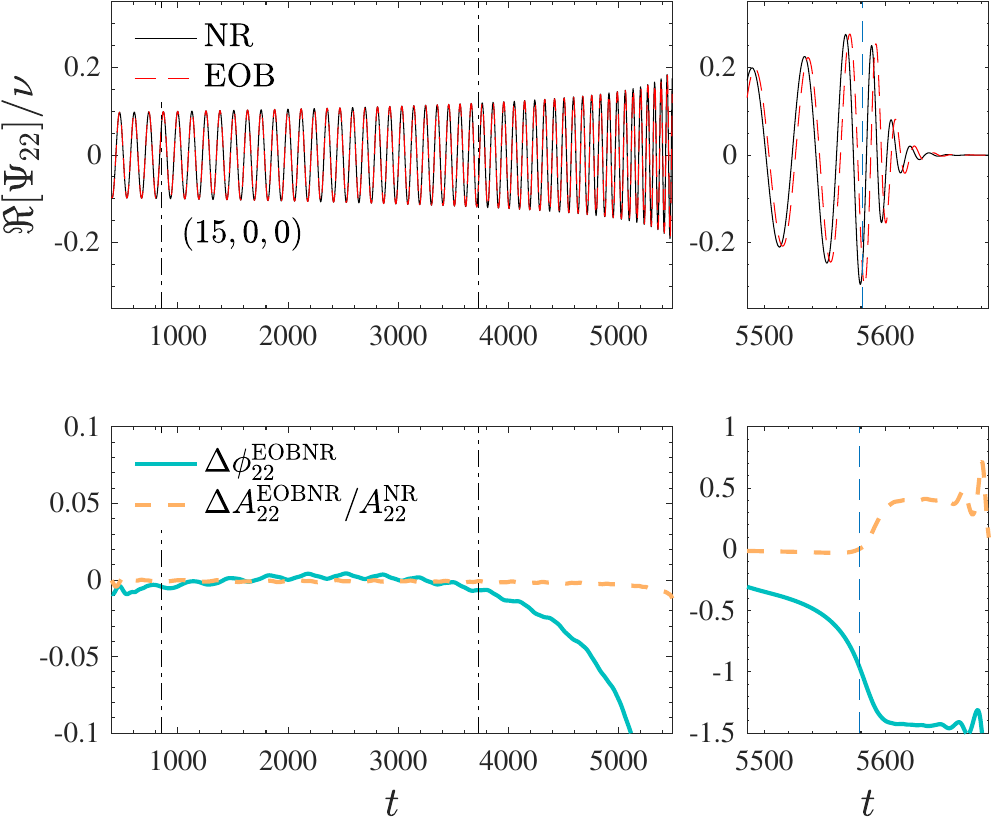}
	\includegraphics[width=0.31\textwidth]{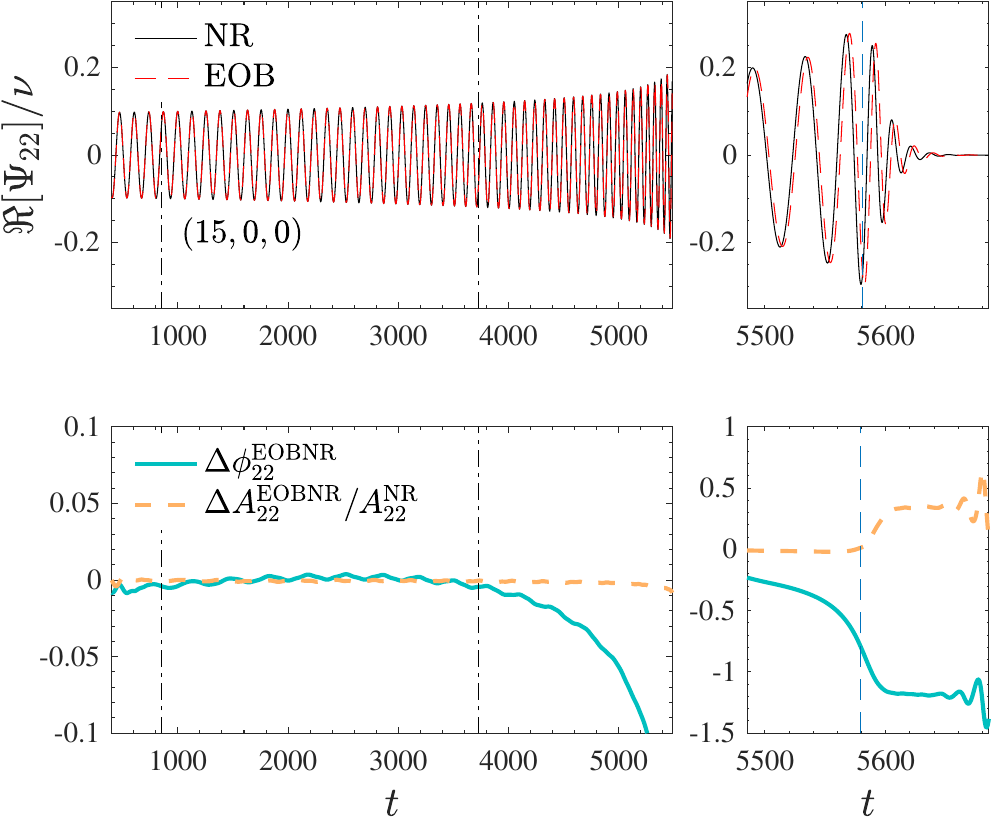} 
	\includegraphics[width=0.31\textwidth]{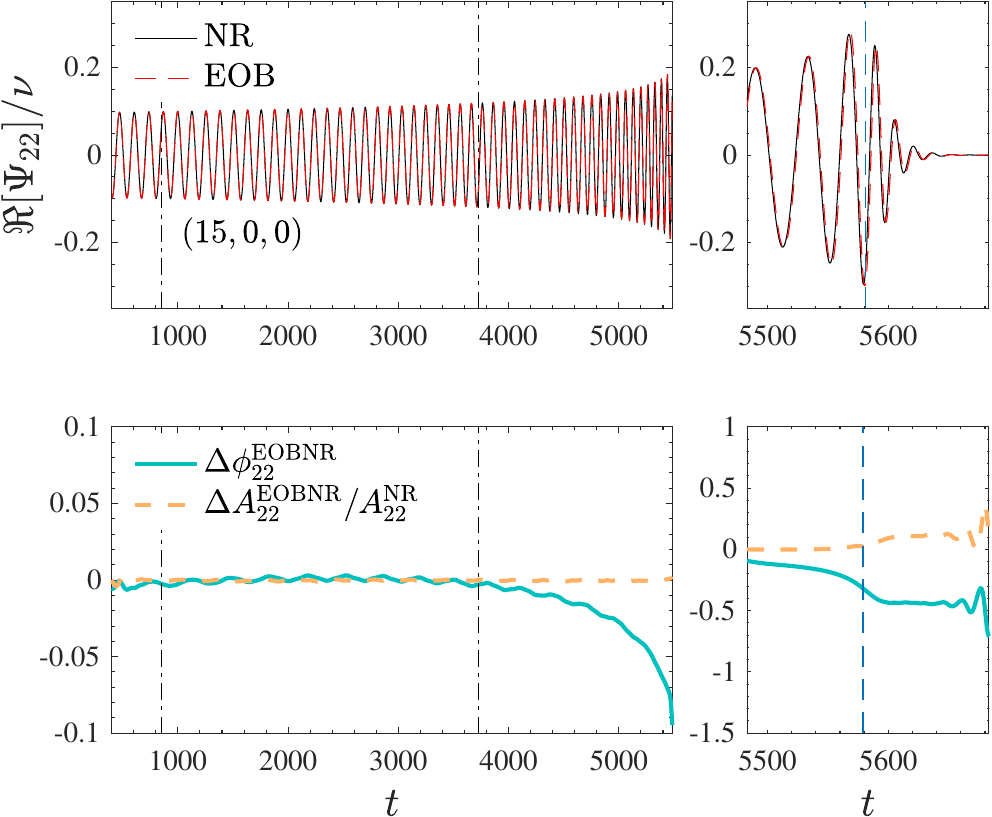}
        \caption{\label{fig:q15_test}EOB/NR phasing for $q=15$ comparing three different representations
        of $\rho_{22}$ (at formal 4PN accuracy) with the corresponding values of $a_6^c$. Left panel: analytical 4PN with
        Pad\'e $(2,2)$ resummation taking the $\log(x)$ as constant. Middle panel: same function but with the NR-tuned
        effective $c^\nu_4$ term of Eq.~\eqref{eq:c4_tuned}. Right panel: new resummation of $\rho_{22}^{\rm 4PN}$ separating the $\log(x)$'s from
        the rational part as described in Appendix~\ref{sec:alter_resum}. The reduction of the EOB/NR dephasing at merger in
        this case is evident.}
\end{figure*}

Now that we have a better understanding of the test-mass case, let us move to considering
comparable mass binaries. We work then with $\rho_{22}^{\rm 4PN}$ resummed as described above
(evidently, including the $\nu$-dependent terms) that thus yields a new waveform amplitude 
and radiation reaction. We then NR-inform a new function \hbox{$a_6^c=-14.24 - 320.26\nu$},
whose behavior is shown in Fig.~\ref{fig:a6_log_sep}. It is interesting to note
that for $\nu\sim 0.25$ the values are compatible with those obtained with the NR-informed value of $c_4^\nu$.
By contrast, the slope of the straight line is different than before. The model performance is then evaluated by computing
either phasings in the time domain or the EOB/NR unfaithfulness with the SXS datasets available. We remind
the reader that we consider mass ratios $1\leq q \leq 10$ spaced by 0.5 as well as the $q=15$ dataset of
Ref.~\cite{Yoo:2022erv}. Figure~\ref{fig:phasings_newlog} displays the time-domain phasing for $q=1$ and $q=6$
obtained with the new treatment of the logarithmic terms. It is quantitatively consistent with Fig.~\ref{fig:phasing_tuned} 
of the main text, though without NR-tuning of radiation reaction. 
Figure~\ref{fig:barF_log_sep} shows that, on the $\bar{\F}_{\rm EOBNR}$ quantity,
the model performance gives, on average, $\bar{\F}_{\rm EOBNR}^{\rm max}\sim 10^{-4}$ and it
substantially equivalent to the same analysis done with the NR-tuned $c_4^\nu$, see Fig.~\ref{fig:phasing_tuned}.
This remarkable fact suggests the following two considerations. On the one hand it is an example that,
by a (simple) improvement on the analytical side, one can obtain an excellent waveform model {\it reducing}
the amount of NR-tuning . This is an important conceptual lesson that should be kept in mind for future studies (see below).
On the other hand one has here an example of the extreme robustness of our EOB framework: even
when an {\it analytic systematic} is present, it can be corrected by careful NR-tuning of some parameters
and the actual performance {\it without} this systematic can be (substantially) obtained.
It must be noted, however, that the model with the new $a_6^c$ and resummation of $\rho_{22}$ actually
performs {\it better} than the totally NR-tuned one. This is apparent for the $q=15$ case. Figure~\ref{fig:q15_test}
shows the EOB/NR time-domain phasing for the three models considered in the paper. From left to right: (i) Pad\'e resummation
of $\rho_{22}^{\rm 4PN}$ with the $\log(x)$ taken as constant when doing the Pad\'e, NR-tuning of $a_6^c$ only; (ii) same
Pad\'e approximant but tuning both $(a_6^c,c_4^\nu)$; (iii) the model discussed in this Appendix.
It is remarkable that the dephasing at merger in this case is $\sim 0.3$~rad, with more than a factor two 
gained with respect to the standard approach. We may argue that additional improvement should be brought
once that a similar treatment of the $\log(x)$-dependent term is applied also to the subdominant modes,
that are more relevant in this case than for $q=1$.
Let us finally mention that the same  problem with the Pad\'e resummations performed under the assumption 
$\log(U)=c$, where $u \equiv GM/(R c^2)$, is present also in the EOB conservative dynamics, through the functions $A$ and $\bar{D}$ that 
are similarly resummed as discussed in Ref.~\cite{Nagar:2021xnh}. As a preliminary investigation, we considered
the 5PN accurate Taylor-expanded $A$ (with $a_6^c$ undetermined parameter) and resummed it using
a $(3,3)$ Pad\'e approximant for the polynomial part and a $(0,1)$ one for that proportional to the $\log(u)$'s,
once that the term $u^5\log(u)$ is factored out. In the adiabatic limit, one finds that the new resummed 
function (and in particular the effective photon potential $u^2A$) is sufficiently flexible to match the one 
obtained with the current model once a suitable value of $a_6^c$ is chosen, that is found again to be linear in $\nu$.
In conclusion, we state that the results discussed in the main text are expected to stand (and possibly improve)
even with the correct treatment of the $\log$-dependent terms in the potentials. This analysis was recently 
completed and is detailed in Ref.~\cite{Nagar:2024oyk}.

\section{{\tt TEOBResumS-GIOTTO} with 4PN flux}
\label{sec:giotto4PN}
In the main text we discussed the use of the 4PN-accurate resummed waveform (and flux) only to
improve the \TEOBd{} model valid for generic orbits for spin aligned binaries, while briefly mentioning that the 
same strategy would have not been equally successful for the simple quasi-circular model \TEOBg{}.
The aim of this Appendix is to support this statement by explicitly considering a version of the \TEOBg{} 
model where $\rho_{22}^{\rm 3^{+2} PN}$ is replaced by $\rho_{22}^{\rm 4PN}$, though in resummed form
and either treating the logarithm as constant within the Pad\'e resummetion (dubbed as {\tt 4PN-oldlogs}) or factoring
them out (dubbed as {\tt 4PN-newlogs}). The other elements of the model precisely coincide with those described in 
Ref.~\cite{Nagar:2023zxh}, except evidently for $a_6^c$ that needs to be redetermined for each version 
of radiation reaction considered. We consider only the nonspinning case, since this is sufficient to justify the 
choices we made in the main text. We recall that, differently from the case of \TEOBd{}, with \TEOBg{} the 
NQC corrections are also included in the radiation reaction, so that 3 iterations are needed
to get the NQC parameters converged (see Ref.~\cite{Riemenschneider:2021ppj} for details).
\begin{table}[t]
	\caption{\label{tab:a6c_giotto}First-guess values of $a_6^c$ for the \TEOBg{} model with 4PN (resummed)
	information. These numbers are then accurately fitted with the functional forms of Eqs.~\eqref{eq:a6c_oldlogs} and~\eqref{eq:a6c_newlogs} 
	respectively.}
	\begin{center}
\begin{ruledtabular}
	\begin{tabular}{ccc|cc}
	  $\#$ & ID & $q$ & $a^c_{\rm 6, {\tt 4PN-oldlogs}}$ & $a^c_{\rm 6, {\tt 4PN-newlogs}}$\\
	  \hline
1 & SXS:BBH:0180 & $1$ & $-118$ & $-100$\\ 
2 & SXS:BBH:0169 & $2$ & $-103.5$ & $-89$\\ 
3 & SXS:BBH:0168 & $3$ & $-91$ & $-76$\\ 
4 & SXS:BBH:0166 & $6$ & $-70$ & $-51$\\ 
5 & SXS:BBH:0298 & $7$ & $-65$ & $-46$ \\
5 & SXS:BBH:0299 & $7.5$ & $-62$ & $-43.5$\\ 
6 & SXS:BBH:0302& $9.5$ & $-54$ & $-35.5$ 
\end{tabular}
\end{ruledtabular}
\end{center}
\end{table}
\begin{figure*}[t]
	\center	
	\includegraphics[width=0.32\textwidth]{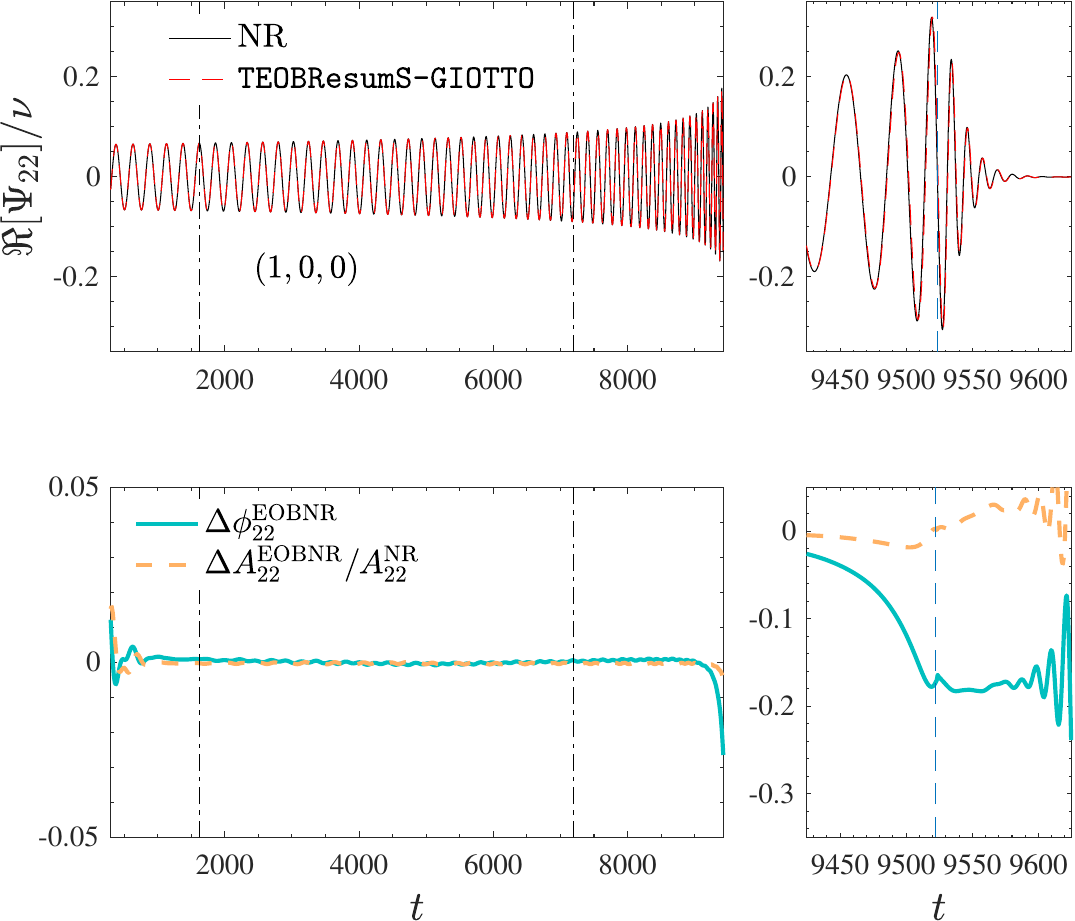}
	\includegraphics[width=0.32\textwidth]{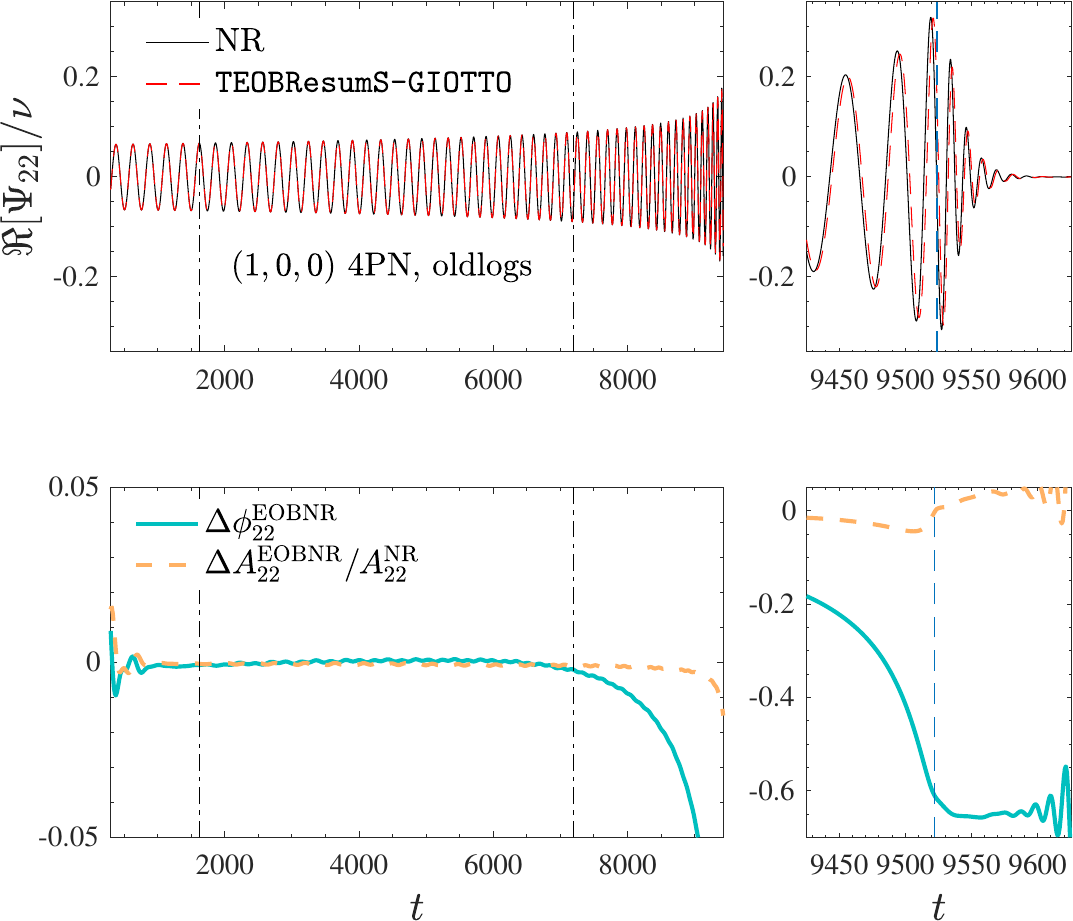}
	\includegraphics[width=0.32\textwidth]{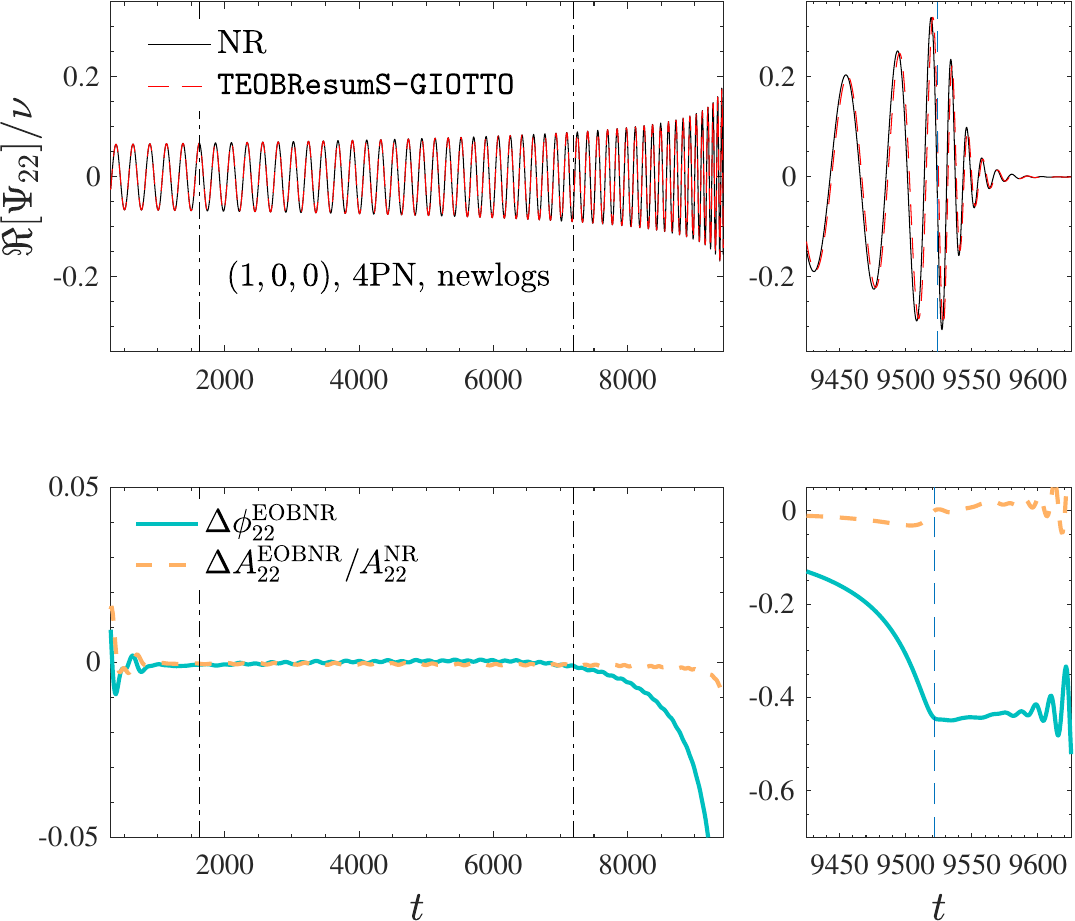}
	\caption{\label{fig:q1giotto}Phasing performance for (various flavors of) the quasi-circular \TEOBd{} model with $q=1$. 
	The standard $\rho_{22}^{\rm 3^{+2} PN}$ (leftmost panel) is replaced by two  different $\rho_{22}^{\rm 4PN}$ in resummed 
	form: one with the logarithms assumed to be constant when computing the Pad\'e approximant (oldlogs)
	and the other with the logarithms factored out and the coefficient Pad\'e resummed (newlogs). 
	Each choice of radiationa reaction yields a different determination of $a_6^c$, see Eqs.~\eqref{eq:a6c_oldlogs} 
	and~\eqref{eq:a6c_newlogs} respectively. It turns out that the use of 4PN resummed information (in any form) effectively
	reduces the flexibility of the model and the  tuning of $a_6^c$ is unable to match the phasing performance of the standard 
	model with the $\rho_{22}^{\rm 3^{+2} PN}$ (nonresummed) function.}
\end{figure*}
For both choices of the resummed $\rho_{22}^{\rm 4PN}$ we determined $a_6^c$ using
the same procedure discussed in the main text. Table~\ref{tab:a6c_giotto} reports the chosen
values of $a_6^c$ (with the corresponding SXS datasets) in the {\tt 4PN-oldlogs} 
and {\tt 4PN-newlogs} case. The points in Table~\ref{tab:a6c_giotto} 
are accurately fitted as
\be
\label{eq:a6c_oldlogs}
a^c_{\rm 6,oldlogs}=-13583\nu^3 + 6670.8\nu^2 - 1390.9\nu + 25.238 \ ,
\ee
in one case, and by
\be
\label{eq:a6c_newlogs}
a^c_{\rm 6,newlogs}=132.98\nu^2-432.68\nu+0.19937 \ 
\ee
in the other. The left panel of Fig.~\ref{fig:q1giotto} shows the EOB/NR phasing comparison for the standard
version of \TEOBg{}, i.e. the model of Ref.~\cite{Nagar:2023zxh}. In fact, this plot is the current
version of the top-left panel of Fig.~1 of~\cite{Nagar:2023zxh} (where the model was actually 
dubbed {\tt D3Q3\_NQC}) and the slightly smaller dephasing accumulated during merger and
ringdown is due to the fact that the fit used to determine the NQC corrections now is different
from the one used for Fig.~1 of~\cite{Nagar:2023zxh}, as explained in the same paper.
The middle panel of  Fig.~\ref{fig:q1giotto} shows the performance obtained using the oldlogs
resummation (and Eq.~\eqref{eq:a6c_oldlogs} for $a_6^c$) and the rightmost panel
using the newlogs resummation (and Eq.~\eqref{eq:a6c_newlogs}) for $a_6^c$. One sees 
that for the current choice of $a_6^c$ the use of 4PN (resummed) information does not allow
to reduce further the phase difference around merger. Note that, as explained in Ref.~\cite{Nagar:2023zxh},
$a_6^c$ is determined requiring that $\Delta\phi^{\rm EOBNR}_{22}$ decreases monotonically
and its derivative does not change sign, since this would eventually determine a worsening of 
the unfaithfulness. The global performance of the {\tt TEOBResumS-GIOTTO-4PN} models
for all mass ratios covered by (public) SXS simulations in explored in Fig.~\ref{fig:barF_giotto},
that displays the EOB/NR unfaithfulness $\bar{\F}_{\rm EOBNR}$ versus the total mass of
the binary $M$. For the reader's ease, in the leftmost panel of the figure we reported $\bar{\cal F}_{\rm EOBNR}(M)$
obtained with \TEOBg{}, which is equivalent to the quantities displayed in top panel of Fig.~2 of Ref.~\cite{Nagar:2023zxh}.
Quantitatively, for $q=1$ \TEOBg{} gives $\bar{\F}_{\rm EOBNR}^{\rm max}=0.052\%$,
which worsens to $0.3\%$ when using {\tt 4PN-oldlogs} or to $0.21\%$ when using {\tt 4PN-newlogs}.
This simple analysis, together with similar investigation done starting from the \TEOBd{} model and reported in the main 
text, eventually convinced us that it was not worth to attempt to improve the \TEOBg{} any further with 4PN information, 
and just focus on \TEOBd{} in this paper.
\begin{figure*}[t]
	\center	
	\includegraphics[width=0.32\textwidth]{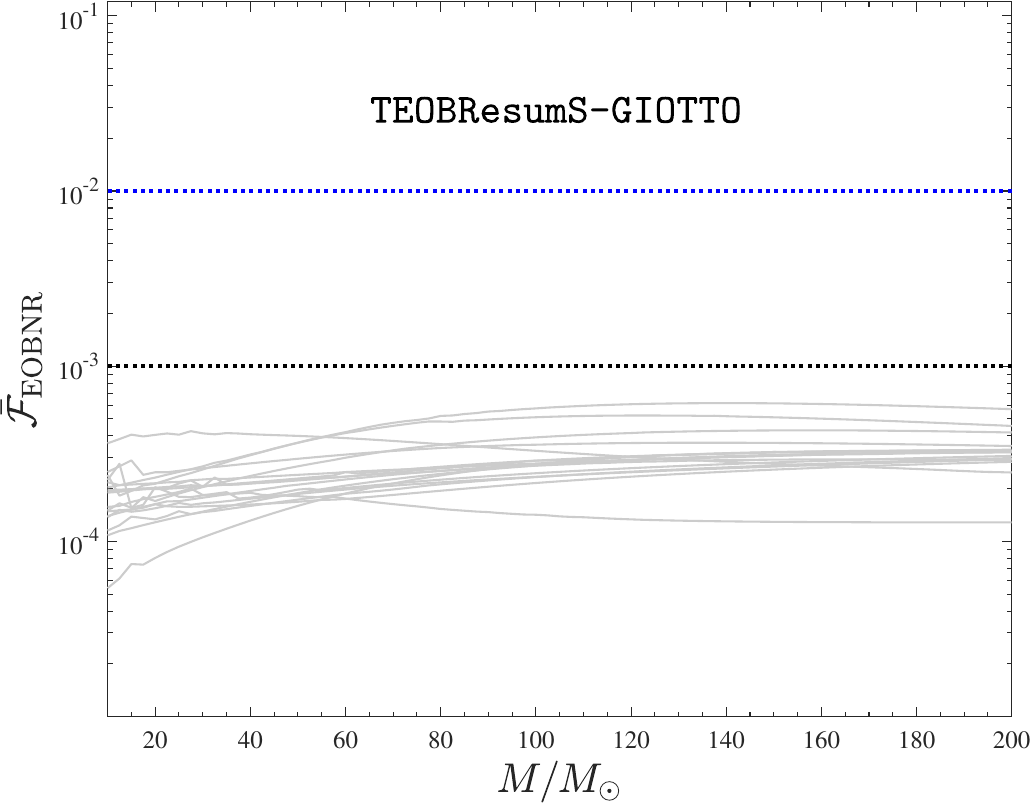}
	\includegraphics[width=0.32\textwidth]{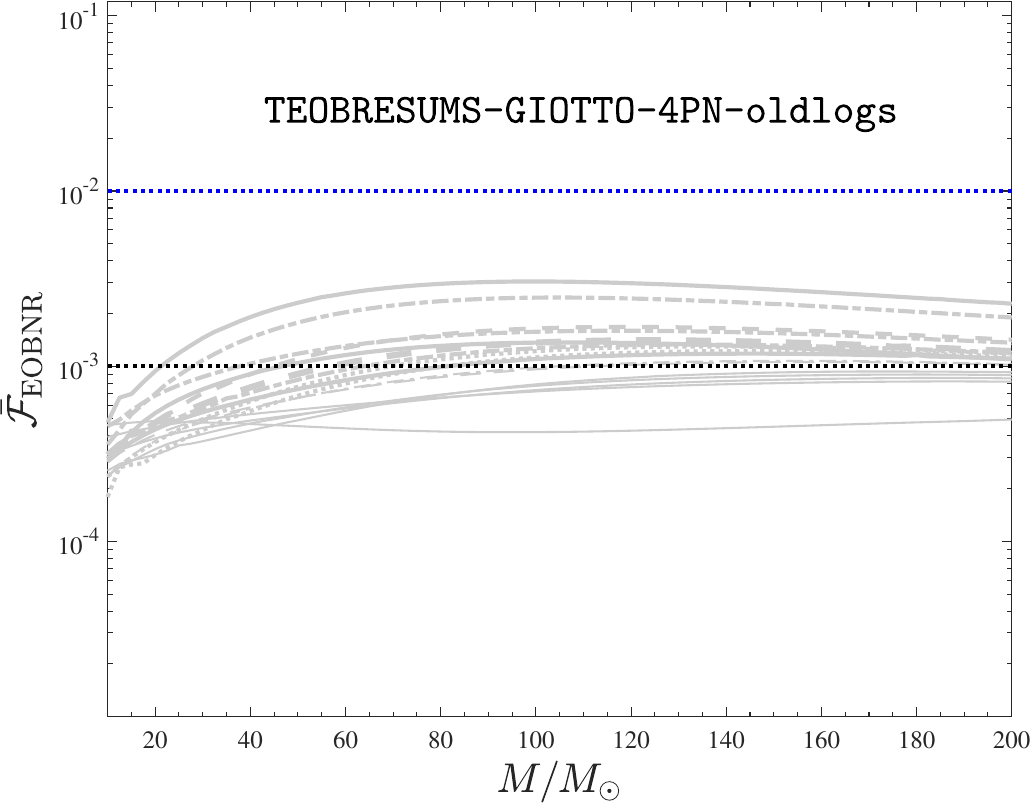}
	\includegraphics[width=0.32\textwidth]{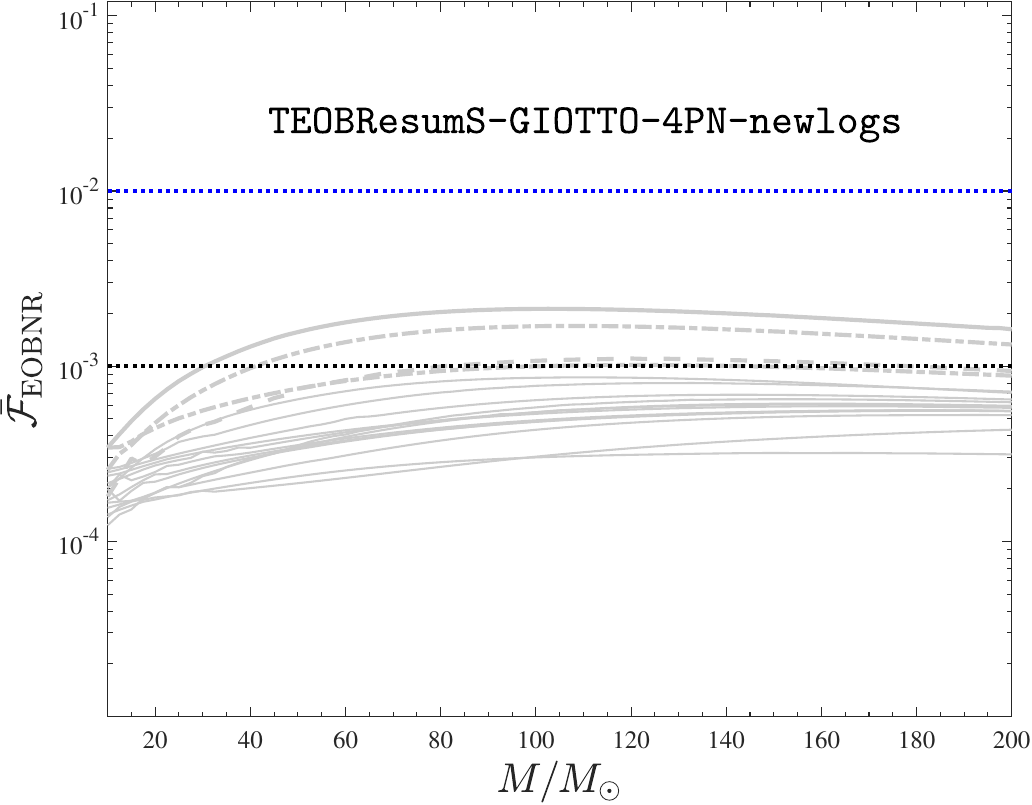}
	\caption{\label{fig:barF_giotto}EOB/NR $\ell=m=2$ unfaithfulness with the three flavors 
	of the \TEOBg{} model computed with alla SXS nonspinning datasets  available, with
	$1\leq q\leq 10$ and $q=15$. The full 4PN waveform information (in two different
	resummed fashions) yields a {\it less performant} model with respect to the standard
	version (leftmost panel) that relies on the Taylor-expanded $\rho_{22}^{\rm 3^{+2}PN}$ function.
	The solid, thin, grey lines correspond to $\bar{\cal F}^{\rm max}_{\rm EOBNR}<10^{-3}$.}
\end{figure*}

\section{Eccentric initial conditions}
\label{sec:eccentric_ics}
Determining the initial conditions for eccentric bound systems in the EOB formalism
is equivalent to finding the mapping between the desired initial eccentricity $e_0$, true anomaly $\zeta_0$ and 
some reference frequency $\Omega_0$ and the EOB dynamical parameters $(r_0, p_{r_{*}}^0, p_{\varphi,0})$. In previous 
iterations of \TEOBResumS{}, for simplicity and without loss of generality, the value of the anomaly $\zeta_0$ was fixed to either $0$ or $\pi$.
This implied that -- in order for all possible orbits to be covered during parameter estimation -- the initial eccentricity $e_0$ and the initial 
frequency $\Omega_0$ had to be treated as free parameters~\cite{Chiaramello:2020ehz,Nagar:2021gss,Bonino:2022hkj}. 
Further, the user-specified initial frequency $\Omega_0$ was
interpreted as the average frequency between periastron and apastron, $\Omega_{0, \rm avg} = (\Omega_{+} + \Omega_{-})/2.$ 
(see Appendix~C of Ref.~\cite{Nagar:2021gss} for details).
In this work, we implement an alternative approach to determine the initial conditions for eccentric orbits, which allows 
to fix the value of the true anomaly $\zeta_0$ to any desired value and allows for users to specify an initial orbit-averaged 
frequency $\bar{\Omega}_0$ as input parameters. This approach follows the one described in \cite{Ramos-Buades:2023yhy}, and relies on the following steps:
\begin{itemize}
\item[(i)] Given a set of initial conditions $(e_0, \zeta_0, \bar{\Omega}_0)$, we compute the istantaneous frequencies at apastron
and periastron $\Omega_{\pm}$ via the Newtonian expression:
\begin{equation}
\Omega_{\pm} = \frac{\bar{\Omega}_0 (1 \pm e_0)^2}{1-e_0^2}^{3/2} \, .
\end{equation}
From the istantaneous frequencies at apastron and periastron we then estimate $\Omega_{0, \rm avg}$.
\item[(ii)] Recalling that 
\be
r = \rho_0/(1+e_0\cos(\zeta_0)) \, ,
\ee 
we numerically find the initial semilatus rectum $\rho_0$ and radial momentum $p_{r_*}^0$ by imposing that 
the average frequency is the desired one:
\be
\begin{split}
	2 \Omega_{0, \rm avg}  &= \frac{d \hat{H}}{d p_\varphi} (r(\rho_0, e_0, \zeta_0=0), j_0(\rho_0, e_0), p_{r_*}^0=0) \\
                           &+ \frac{d \hat{H}}{d p_\varphi} (r(\rho_0, e_0, \zeta_0=\pi), j_0(\rho_0, e_0), p_{r_*}^0=0) \, .
\end{split}
\ee
and ensuring energy conservation at the point specified by $\zeta_0$:
\be
\begin{split}
\hat{H}_{\rm eff}(r(\rho_0, e_0, \zeta_0=0), j_0(\rho_0, e_0), p_{r_*}^0=0) = 
\\ \hat{H}_{\rm eff}(r(\rho_0, e_0, \zeta_0), j_0(\rho_0, e_0), p_{r_*}^0) \, .
\end{split}
\ee

In the equations above, $j_0$ is the value of angular momentum obtained by imposing energy conservation at apastron and periastron.
\end{itemize}

We note that these initial conditions are adiabatic, meaning that they do not incorporate effect of radiation reaction.
While this approximation is expected to not lead to significant errors for large eccentricities, in the quasi-circular limit 
it is known that non-adiabatic initial data can lead to some spurious eccentricity in the waveforms~\cite{Knee:2022hth, Shaikh:2023ypz}
Given that we find such spurious eccentricity to be of the order of $10^{-3}$, we expect this to be a minor effect
with respect to other differences between the two \TEOBg{} and \TEOBd{} models.


\section{Numerical relativity quasi-circular datasets}
\label{sec:nrdata}
In this Appendix we collect the details of the simulations employed 
in the paper to inform or validate the \TEOBd{} model.

The configurations employed to inform $a^6_c$ are collected in Table~\ref{tab:a6c},
that also report the first-guess values of $a_6^c$ shown in Fig.~\ref{fig:a6_fits}. Note that
the table lists the values for the three choices for the radiation reaction that we have explored,
that is: (i) using $\rho_{22}$ at $3^{+2}$~PN accuracy in Taylor-expanded form;(ii) using $\rho_{22}$
at 4PN accuracy, fully analytical, and resummed with a $(2,2)$ Pad\'e approximant;(iii) using $\rho_{22}$
at effective 4PN accuracy (still in Pad\'e resummed form) where the 4PN $\nu$-dependence is informed 
to NR simulations. 
The first-guess values for $c_3$, for either \daliAN{} and \daliNR{} are listed in the two rightmost columns
of Tables~\ref{tab:c3_eqmass} and ~\ref{tab:c3_uneqmass}.
Scattering angles are reported in Table~\ref{tab:chi_scattering} (for nonspinning configurations), again with
the three different analytical choices explored in the main text. Finally, the EOB/NR unfaithfulness for the
publicly available SXS simulations are listed in  Table~\ref{tab:SXS}.

  \begin{table}[t]
	\caption{\label{tab:c3_eqmass}First-guess values for $c_3$ for equal-mass, equal-spin configurations.
	They are  used to determine $c_3^{=}$ in Eq.~\eqref{eq:c3fit}.}
	\begin{center}
  \begin{ruledtabular}
	\begin{tabular}{lllc|ccc}
	  $\#$ & ID & $(q,\chi_1,\chi_2)$ & $\tilde{a}_0$ &$c_3^{\tt 4PNan}$ & $c_3^{\tt 4PNnr}$ & \\
	  \hline
  1 & BBH:1137 & $(   1, -0.   97, -0.   97)$ & $-0.97$ & 86.0  & 91.0\\ 
 2 & BBH:0156 & $(   1, -0.9498, -0.9498)$ & $-0.95$ & 84.5 & 90.4 \\ 
 3 & BBH:0159 & $(   1, -0.90, -0.   90)$ & $-0.90$ & 80.5  & 86.8 \\ 
 4 & BBH:2086 & $(   1, -0.80, -0.   80)$ & $-0.80$ & 73.5  & 79.5 \\ 
 5 & BBH:2089 & $(   1, -0.60, -0.   60)$ & $-0.60$ & 64   & 71.0\\ 
 6 & BBH:2089 & $(   1, -0.20, -0.   20)$ & $-0.60$ & 48   & 53.0\\
 7 & BBH:0150 & $(   1, +0.   20, +0.   20)$ & $+0.20$ & 29  & 37.0 \\ 
 8 & BBH:0170 & $(   1, +0.   4365, +0.   4365)$ & $+0.20$ & 23.5 & 29.0 \\ 
 9 & BBH:2102 & $(   1, +0.   60, +0.   60)$ & $+0.60$ & 18.0  & 23.5 \\ 
 10 & BBH:2104 & $(   1, +0.   80, +0.   80)$ & $+0.80$ & 12.5 & 15.5  \\ 
 11 & BBH:0153 & $(   1, +0.   85, +0.   85)$ & $+0.85$ & 11.5  & 14.5 \\ 
 12 & BBH:0160 & $(   1, +0.   90, +0.   90)$ & $+0.90$ & 10.3  & 11.0\\ 
 13 & BBH:0157 & $(   1, +0.   95, +0.   95)$ & $+0.95$ & 8.7   & 6.4\\ 
 14 & BBH:0177 & $(   1, +0.   99, +0.   99)$ & $+0.99$ & 7.0  & 6.0\\   
 \end{tabular}
  \end{ruledtabular}
  \end{center}
  \end{table}
 
  \begin{table}[t]
	\caption{\label{tab:c3_uneqmass}First-guess values for $c_3$ for the unequal-spin and
	unequal-mass configurations. They are  used to determine $c_3^{\neq}$ in Eq.~\eqref{eq:c3fit}.}
	\begin{center}
  \begin{ruledtabular}
	\begin{tabular}{lllc|ccc}
	  $\#$ & ID & $(q,\chi_1,\chi_2)$ & $\tilde{a}_0$ & $c_3^{\tt 4PNan}$ & $c_3^{\tt 4PNnr}$ \\
	  \hline
 15 & BBH:0004 & $(   1, -0.   50,  0.    0)$ & $-0.25$ & 55.5 &  56.4 \\ 
 16 & BBH:0005 & $(   1, +0.   50,  0.    0)$ & $+0.25$ & 35 & 34.6 \\ 
 17 & BBH:2105 & $(   1, +0.   90,  0.    0)$ & $+0.45$ & 27.7 & 27.5 \\ 
 18 & BBH:2106 & $(   1, +0.   90, +0.   50)$ & $+0.70$ & 19.1 & 20.7 \\ 
 19 & BBH:0016 & $( 1.5, -0.   50,  0.    0)$ & $-0.30$ & 56.2 & 56.5 \\ 
 20 & BBH:1146 & $( 1.5, +0.   95, +0.   95)$ & $+0.95$ & 14.35 & 12.0 \\ 
 21 & BBH:0552 & $(1.75,+0.80,-0.40)$& $+0.36$ & 29 & 30.5\\
 22 & BBH:1466 & $(1.90,+0.70,-0.80)$& $+0.18$ & 33 & 37.5 \\
 23 & BBH:2129 & $(   2, +0.   60,  0.    0)$ & $+0.40$ &  29.5 & 30.0 \\ 
 24 & BBH:0258 & $(2,+0.87,-0.85)$& $+0.296$ & 32 & 32 \\
 25 & BBH:2130 & $(   2, +0.   60, +0.   60)$ & $+0.60$ & 23 & 24.5 \\ 
 26 & BBH:2131 & $(   2, +0.   85, +0.   85)$ & $+0.85$  & 15.8 & 17.0 \\ 
 27 & BBH:1453 & $(2.352,+0.80,-0.78)$& $+0.328$ & 29 & 32.5 \\
 28 & BBH:2139 & $(   3, -0.   50, -0.   50)$ & $-0.50$ & 65.3 & 65.0 \\ 
 29 & BBH:0036 & $(   3, -0.   50,  0.    0)$ & $-0.38$ & 61 & 58 \\ 
 30 & BBH:0174 & $(   3, +0.   50,  0.    0)$ & $+0.37$ & 28.5 & 27.4 \\ 
 31 & BBH:2158 & $(   3, +0.   50, +0.   50)$ & $+0.50$ & 27.1 & 27.5 \\ 
 32 & BBH:2163 & $(   3, +0.   60, +0.   60)$ & $+0.60$ & 24.3 &  25.5\\ 
 33 & BBH:0293 & $(   3, +0.   85, +0.   85)$ & $+0.85$ & 16.0 & 18.0 \\ 
 34 & BBH:0292  & $(3,+0.73,-0.85)$& $+0.335$ & 30.6& 31.5 \\
 35 & BBH:1447 & $(3.16, +0.7398, +0.   80)$ & $+0.75$ & 19.2 & 21.0 \\ 
 36 & BBH:1452  & $(3.641,+0.80,-0.43)$& $+0.534$ & 25.6& 28.5 \\
 37 & BBH:2014 & $(   4, +0.   80, +0.   40)$ & $+0.72$ & 21.5 & 22.5 \\ 
 38 & BBH:1434 & $(4.368, +0.7977, +0.7959)$ & $+0.80$ & 19.8 & 19.8 \\ 
 39 & BBH:0111 & $(   5, -0.   50,  0.    0)$ & $-0.42$ & 54 & 53.5\\ 
 40 & BBH:0110  & $(   5, +0.   50,  0.    0)$ & $+0.42$ & 29.5 & 30.5 \\ 
 41 & BBH:1428  & $(5.516,-0.80,-0.70)$ & $-0.784$ & 80& 80 \\
 42 & BBH:1440 & $(5.64,+0.77,+0.31)$& $+0.70$ & 21.5 & 24.5\\
 43 & BBH:1432 & $(5.84, +0.6577, +0. 793)$ & $+0.68$ & 25 & 24.0 \\ 
 44 & BBH:1437 & $(6.038,+0.80,+0.15)$& $+0.7076$ & 21.5& 24.0 \\
 45 & BBH:1375  & $(   8, -0.   90,  0.    0)$ & $-0.80$ & 70 & 63.5 \\ 
 46 & BBH:1419 & $(8,-0.80,-0.80)$& $-0.80$ & 81.5 & 80 \\
 47 & BBH:0114 & $(   8, -0.   50,  0.    0)$ & $-0.44$ & 61 & 57.5 \\ 
 48 & BBH:0065 & $(   8, +0.   50,  0.    0)$ & $+0.44$ & 26.5 & 27.0 \\ 
 49 & BBH:1426 & $(   8, +0.4838, +0.7484)$ & $+0.51$ & 30.3 & 28.5 \\ 
  \end{tabular}
  \end{ruledtabular}
  \end{center}
  \end{table}

\bibliography{refs20241017.bib,local.bib}

\end{document}